\documentclass[floatfix,aps,prd,amsmath,amssymb,superscriptaddress,nofootinbib,eqsecnum,reprint,showpacs,twocolumn]{revtex4}
\usepackage{color}
\usepackage[pdftex]{graphicx}
\usepackage{subfig}
\usepackage[colorlinks=true, citecolor=blue,urlcolor=blue]{hyperref}
\usepackage{xcolor}

\newcommand{\leaveout}[1]{}
\newcommand{\MC}[1]{}
\newcommand{\LF}[1]{}
\newcommand{\MCImp}[1]{}
\newcommand{\LL}[1]{}

\newcommand{\nsDM}[1]{non-standard DM}
\newcommand{\nsDMacr}[1]{NSDM}

\newcommand{\indmode}{\Lambda}
\newcommand{\indmodeExp}{s \ell m\omega}
\newcommand{\indmodemmmo}{s,\ell,-m,-\omega}
\newcommand{\indmodems}{-s,\ell,m,\omega}
\newcommand{\horind}{+}

\newcommand{\AngVel}{\Omega_H}
\newcommand{\Temp}{T_H}

\newcommand{\expW}{\beta}

\newcommand{\Rin}{R^{\text{in}}_{\indmode}}
\newcommand{\Rup}{R^{\text{up}}_{\indmode}}

\newcommand{\RinupExp}{R^{\text{in/up}}_{\indmodeExp}}

\newcommand{\R}{R_{\indmode}}
\newcommand{\RExp}{R_{\indmodeExp}}

\newcommand{\RinS}{\mathcal{R}^{\text{in}}_{\indmode}}
\newcommand{\RupS}{\mathcal{R}^{\text{up}}_{\indmode}}
\newcommand{\RinupS}{\mathcal{R}^{\text{in/up}}_{\indmode}}

\newcommand{\Rininc}{B^{\text{inc}}_{\indmode}}
\newcommand{\Rinref}{B^{\text{ref}}_{\indmode}}
\newcommand{\Rinincref}{B^{\text{inc/ref}}_{\indmode}}
\newcommand{\Rupinc}{C^{\text{inc}}_{\indmode}}
\newcommand{\Rupref}{C^{\text{ref}}_{\indmode}}
\newcommand{\Rupincref}{C^{\text{inc/ref}}_{\indmode}}
\newcommand{\RinIncRef}{B^{\text{inc/ref}}_{\indmode}}
\newcommand{\RupIncRef}{C^{\text{inc/ref}}_{\indmode}}

\newcommand{\RinincS}{\mathcal{B}^{\text{inc}}_{\indmode}}
\newcommand{\RinrefS}{\mathcal{B}^{\text{ref}}_{\indmode}}
\newcommand{\RinincrefS}{\mathcal{B}^{\text{inc/ref}}_{\indmode}}
\newcommand{\RupincS}{\mathcal{C}^{\text{inc}}_{\indmode}}
\newcommand{\RuprefS}{\mathcal{C}^{\text{ref}}_{\indmode}}
\newcommand{\RupincrefS}{\mathcal{C}^{\text{inc/ref}}_{\indmode}}

\newcommand{\Amplif}{Z_{\indmode}}
\newcommand{\AmplifExp}{Z_{\indmodeExp}}
\newcommand{\Amplifmmmo}{Z_{\indmodemmmo}}
\newcommand{\Amplifms}{Z_{\indmodems}}

\newcommand{\kSub}{\tilde{\omega}}

\newcommand{\Real}{\text{Re}}
\newcommand{\Imag}{\text{Im}}

\newcommand{\W}{W_{\indmode}}
\newcommand{\Wf}{W_{\indmode}^f}
\newcommand{\Wsub}{\mathcal{W}_{\indmode}}

\newcommand{\Glm}{g_{\indmode}}

\newcommand{\aMST}[1]{a^{\nu}_{#1}}

\newcommand{\eigenl}{\lambda_{\indmode}}
\newcommand{\eigenlExp}{\lambda_{\indmodeExp}}
\newcommand{\eigenAS}{\lambdabar}

\newcommand{\oSR}{\omega_{\text{SR}}}
\newcommand{\oQNM}{\omega_{\ell m n}}
\newcommand{\oQNMmm}{\omega_{\ell,-m,n}}

\newcommand{\del}{\delta}

\newcommand{\delSR}{\delta_{SR}}
\newcommand{\q}[2]{q_{#1}^{#2}}
\newcommand{\e}{\chi_{s}}
\newcommand{\et}{ \chi_{-s}}
\newcommand{\rArbit}{p}

\begin{document}


\author{Marc Casals}
\email{mcasals@cbpf.br,marc.casals@ucd.ie}
\affiliation{Centro Brasileiro de Pesquisas F\'isicas (CBPF), Rio de Janeiro, 
CEP 22290-180, 
Brazil.}
\affiliation{School of Mathematics and Statistics, University College Dublin, Belfield, Dublin 4, Ireland.}

\author{Lu\'is F. Longo Micchi}
\email{luisflm@cbpf.br}
\affiliation{Centro Brasileiro de Pesquisas F\'isicas (CBPF), Rio de Janeiro,  CEP 22290-180,  Brazil.}
\affiliation{Centro de Matem\'atica, Computa\c{c}\~ao e Cogni\c{c}\~ao, Universidade Federal do ABC (UFABC), 09210-170 Santo Andr\'e, S\~ao Paulo, Brazil.}

\title{Spectroscopy of Extremal (and Near-Extremal) Kerr Black Holes}

\begin{abstract}
We 
investigate
linear, spin-field perturbations of 
Kerr black holes 
in the extremal limit
throughout the complex-frequency domain.
We  calculate  quasi-normal modes of extremal Kerr, as well as of near-extremal Kerr,
via a novel approach: using the method of Mano, Suzuki and Takasugi (MST).
We also show how, 
in  the extremal limit, 
a 
branch cut is formed at the superradiant-bound frequency, $\oSR$, via
a simultaneous accumulation of quasi-normal modes and totally-reflected modes.
For real frequencies, we calculate the superradiant amplification factor, which  yields
 the  amount of rotational energy that can be extracted from a black hole. In the extremal limit, this factor is the largest  and it displays a discontinuity at  $\oSR$   for some modes.
Finally, we find no 
exponentially-growing modes
nor branch points on the upper-frequency plane in extremal Kerr after a numerical 
investigation,
thus
 providing 
  evidence of the mode-stability of this space-time away from the horizon.
\end{abstract}

\date{\today}
\maketitle







\section{Introduction}

Extremal (i.e., maximally-rotating) Kerr black holes play a special role within theories of gravity.
From a theoretical point of view, 
 the Bekenstein-Hawking entropy
 for extremal black holes can be reproduced via a counting of microstates within String Theory~\cite{Strominger:1996sh}.
Furthermore,
extremal black holes
exhibit
a near-horizon enhanced symmetry~\cite{Bardeen:1999px},
which
has led to the Kerr/Conformal Field Theory correspondence conjecture~\cite{PhysRevD.80.124008}.
Extremal black holes also enjoy a special place in relation to the (weak) Cosmic Censorship conjecture~\cite{penrose1969gravitational}, in the sense that
they are the ``last frontier" between rotating black holes and  naked singularities.
From an observational point of view, 
 there is  evidence of astrophysical black holes which are highly spinning~\cite{Gou:2013dna,McClintock:2006xd,0264-9381-30-24-244004},
 for which exact extremality cannot be excluded.
Interestingly, 
high spins lead
to a distinct observational feature in the gravitational waveform 
when a particle is 
in a near-horizon inspiral
into a near-extremal   Kerr (NEK) black hole~\cite{Gralla:2016qfw} 
or  into an extremal Kerr black hole~\cite{Hadar:2014dpa,Compere:2017hsi}.

An important 
feature
of a black hole is its  quasi-normal mode (QNM) spectrum.
QNMs describe the 
exponentially-damped, 
characteristic ``ringdown" of a black hole when perturbed by a field~\cite{Vishveshwara:1970zz} (see, e.g.,~\cite{Berti:2009kk} for a review).
Such ringdown was actually observed in the
late-time regime of the 
 gravitational waveforms detected by the Laser Interferometer Gravitational-Wave Observatory~\cite{PhysRevLett.116.061102}.
Physically, QNMs are field modes which decay exponentially with time and possess no incoming radiation: they are purely ingoing into the event horizon and purely outgoing to radial infinity. Mathematically, QNMs correspond to
poles
in the complex frequency ($\omega\in \mathbb{C}$) plane of the Fourier modes
 of the retarded Green function of the equation obeyed by the field perturbation.
The real part of these QNM frequencies determines
the characteristic frequencies of vibration of the black hole, while the (negative\footnote{We consider  the behaviour in time $t$  of a field mode of frequency $\omega$ to be $e^{-i\omega t}.$}) imaginary part determines the exponential decay rate of the modes.

It has been observed that, as extremality is approached, QNMs accumulate towards a real 
frequency $\omega=\oSR$~\cite{detweiler1980black}, thus forming a branch cut (BC)
 in the extremal case itself~\cite{PhysRevD.64.104021,casals2018perturbations}.
This frequency $\oSR$ is in fact the largest frequency that a (boson) field wave can have in order to be able to extract 
 rotational energy from a rotating
black hole -- this phenomenon of extraction of rotational energy is known as superradiance~\cite{starobinskii1973amplification,zel1971generation}.
Although superradiance exists for any 
rotating black hole,
it is particularly interesting to study it in the extremal limit since 
the closer the black hole is to extremality,
the larger is the amount of rotational energy that can be extracted from it.

Apart from the superradiant BC 
stemming
from $\oSR$, which is present  in extremal Kerr only, there is another BC 
stemming 
from the frequency at the origin ($\omega=0$), which exists both in sub-extremal and extremal Kerr.
The contribution from 
the BC near the origin
to
the full field perturbation is a power-law decay at late-times, in sub-extremal Kerr~\cite{Leaver:1986} (see~\cite{Casals:Ottewill:2015,CDOW13,PhysRevD.94.124053,PhysRevLett.109.111101,Casals:2012ng} for higher-order contributions which include logarithmic terms)
as well as in extremal Kerr~\cite{casals2018perturbations}.
This power-law decay, originally observed by Price~\cite{Price:1971fb}, typically kicks in  
after the QNM exponential decay.

Both superradiance and BCs have important consequences for the stability properties of rotating black holes.
For example, superradiance is associated with 
exponentially-growing mode instabilities 
  of rotating black holes in certain settings.
 These 
 settings
include
(i)  {\it massive} field perturbations~\cite{damour1976quantum,zouros1979instabilities,PhysRevD.22.2323},
and (ii)
 massless field perturbations 
 when the  black hole  is
 either surrounded by a mirror~\cite{Press:1972zz} or in an asymptotically anti-de Sitter Universe~\cite{PhysRevD.70.084011}.
 In this paper, however, we shall consider {\it massless} field perturbations in Kerr space-time (so asymptotically flat and without a mirror),
 to which we shall restrict ourselves from now on.
 
 Sub-extremal Kerr is known to possess no 
exponentially-{\it growing} modes
obeying the physical boundary condition of no incoming radiation
 (i.e., modes corresponding to poles of the Green function but, 
unlike QNMs, with a frequency that has a {\it positive} imaginary part)~\cite{whiting1989mode}.
In fact, the full (i.e., non-modal) linear stability 
(decay of the field and its derivatives to arbitrary order)
of sub-extremal Kerr has been proven under 
{\it scalar} perturbations up to and including the event horizon~\cite{dafermos2016decay}.

As for extremal Kerr, 
Aretakis found that  transverse derivatives of the full scalar field in the axisymmetric case  diverge {\it on the horizon} of an extremal Kerr black hole~\cite{Aretakis:2012ei} (see~\cite{Aretakis:2011ha} for an earlier, similar result in the case of an extremal Reissner-Nordstr\"om black hole).
A similar result was
obtained for
 electromagnetic and gravitational perturbations in~\cite{lucietti2012gravitational}.
 In~\cite{casals2016horizon,Gralla:2017lto} it was shown that this blow-up  is due
  to the  extra, superradiant BC in extremal Kerr.
Refs.~\cite{casals2016horizon,Gralla:2017lto}
also showed that the superradiant BC leads to an even  ``stronger" blow-up in the non-axisymmetric case, although it is power-law in both the
axisymmetric and non-axisymmetric cases.
\MC{Is there a proof that such a blow-up does not occur in sub-extremal Kerr?}
Away from the horizon of extremal Kerr,
the decay of the full linear field in the {\it axisymmetric} case has been proven in~\cite{aretakis2012decay}
for the scalar field and in~\cite{dain2015bounds} for the gravitational field.
With regards to {\it generic} (i.e., including non-axisymmetric) perturbations away from the horizon, there is evidence of stability  coming from mode analyses.
Despite these important results for extremal Kerr  space-times, neither their linear stability (whether modal or non-modal)  away from the horizon nor a bound on the rate of the blow-up on the horizon have, to the best of our knowledge,  been proven yet  for
non-axisymmetric perturbations. 
In other words,
  there is no yet proof that in extremal Kerr there exist no  unstable  modes with   $m\neq 0$,
whether as poles or as branch points of the Green function in the upper frequency plane.
We note, however, that there exists numerical support for the non-existence of such modes:
 in~\cite{Burko:2017eky} by numerically solving the 
equation obeyed by
azimuthal-$m$ mode
 field perturbations, and 
in~\cite{detweiler1973stability,Richartz:2015saa} by numerically looking for 
poles of the Green function  in the upper complex-frequency plane 
 and not finding any.


In this paper we carry out a thorough investigation of massless, integer-spin field perturbations of extremal Kerr black holes in the complex-frequency plane.
We accompany this investigation with a similar one  in NEK, which helps understand better the results in extremal Kerr.
In particular, we numerically look both for 
poles
and  branch points of the retarded Green function in the {\it upper} frequency plane in extremal Kerr
 and report that we do not find any (in the case of 
 poles,
 this is in agreement with~\cite{detweiler1973stability,Richartz:2015saa}).
 We also give a simple analytical argument against the existence of poles in the upper plane.
The result of our investigation supports the mode stability of extremal Kerr away from the horizon and no  instability on the horizon other than
the power-law
of the Aretakis phenomenon.

With regards to the {\it lower} frequency plane, we calculate and tabulate QNM frequencies  for both  NEK and extremal Kerr black holes.
To the best of our knowledge, Richartz's~\cite{Richartz:2015saa}  is the only work in the literature where QNMs of extremal Kerr black holes have been calculated. We reproduce Richartz's QNM values, extend the precision to 16 digits and obtain new frequencies for higher overtones (i.e., for larger magnitude of the imaginary part of the frequency).
Furthermore, we provide values for QNMs in extremal Kerr which were missed by~\cite{Richartz:2015saa} in extremal Kerr as well as by analyses~\cite{Yang:2012pj,yang2013quasinormal} in NEK -- notably, for the important gravitational modes with $\ell=m=2$ and $3$ (the first one was observed -- although not tabulated -- in NEK in~\cite{Cook:2014cta}, the latter was not).
Similarly, in NEK, our QNM values extend the precision to 16 digits of those tabulated in~\cite{QNMBertiNew} and we  include values for extra modes  (particularly for the so-called zero-damping modes~\cite{Yang:2012pj}).
We also show the interesting way in which the extra superradiant BC forms in the extremal limit. Namely, it forms via an accumulation of, not only QNMs,
as had been previously observed,
 but also totally-reflected modes (TRMs).
Our calculation in NEK also serves to illustrate the intricate structure of QNMs which has already been observed in the literature~\cite{detweiler1980black,hod2008slow,Yang:2012pj,yang2013quasinormal,Zimmerman:2015rua} and to validate our method and results.

Finally, on the {\it real}-frequency line,
we calculate the superradiant amplification factor 
in  sub-extremal Kerr and,  for the first time to best of our knowledge, in 
extremal Kerr.
This factor allows us to quantify the  maximum amount of 
energy that can be extracted from a rotating black hole via superradiance.
We show that, in extremal Kerr, the  amplification factor is,  for some modes, discontinuous at the superradiant-bound frequency, as predicted by the asymptotic analyses in~\cite{starobinskii1973amplification,Starobinskil:1974nkd}.

To carry out our investigations
we used the semi-analytic method of Mano, Suzuki and Takasugi (MST), which was originally developed for sub-extremal Kerr in~\cite{Mano:Suzuki:Takasugi:1996,Mano:1996mf,Sasaki:2003xr} and  recently developed for extremal Kerr in~\cite{casals2018perturbations}.
To the best of our knowledge, this is the first time that the MST method has been used for calculating QNM frequencies for any black hole space-time\footnote{In Refs.~\cite{CDOW13,zhang2013quasinormal} the MST method was used for calculating QNM-related quantities such as excitation factors but not for calculating the QNM {\it frequencies} themselves (which were  calculated there via the Leaver method~\cite{Leaver:1985}).}.
In this paper we point to  advantages that this new approach  for calculating QNM frequencies has over the  standard Leaver method~\cite{Leaver:1985} in sub-extremal Kerr and its adaptation in~\cite{Richartz:2015saa} to extremal Kerr\MC{How can the expansion in \cite{Richartz:2015saa} about the ordinary point $r=2M$ converge everywhere when there's an irregular singular point at $r=M$? when he does convergence tests, he only takes into account the coefficients but not the whole summand?}.

\MC{make code and data available online?}

The layout of the rest of the paper is as follows.
In Sec.\ref{sec:perts} we introduce the basic quantities for linear spin-field perturbations of Kerr black holes.
In Sec.\ref{sec:method} we briefly describe the method used for obtaining our results: both the MST method and the method
for searching for
poles of the Green function.
There, we also give asymptotics near the two branch points in extremal Kerr (namely, $\omega=0$ and $\oSR$).
 In Sec.\ref{sec:super}  we calculate
the superradiant amplification factor in sub-extremal and  extremal Kerr.
In Sec.\ref{sec:QNMs&TRMs NEK} we calculate QNMs in NEK and in 
Sec.\ref{sec:BCs} we investigate the presence and formation of BCs in (sub-)extremal Kerr.
In Sec.\ref{sec:QNM&unstable ext} we calculate QNMs and  search for 
unstable modes  in extremal Kerr.
We conclude with some comments in Sec.\ref{sec:conclusions}.
In Appendix \ref{sec:special freq} we investigate various specific frequencies, including the so-called algebraically-special frequencies~\cite{1973JMP....14.1453W,chandrasekhar1984algebraically}. 
Finally, appendixes \ref{sec: app QNMs NEK} and \ref{sec:app QNMs extreme} contain tables of QNMs in, respectively, 
NEK and extremal Kerr.

We choose units such that $c=G=1$.


\section{Linear perturbations}\label{sec:perts}

A Kerr black hole is uniquely characterized by its mass $M$ and angular momentum per unit mass $a$.
Using Boyer-Lindquist coordinates $\{t,r,\theta,\varphi\}$, 
the Kerr metric admits two linearly-independent Killing vectors: $\partial_t$ (stationarity) and $\partial_{\varphi}$ (axisymmetry).
The Boyer-Lindquist radii of the  event horizon and the Cauchy horizon of the black hole
are respectively given by 
 $r_{+}= M+ \sqrt{M^2-a^2}$ and  $r_{-}= M- \sqrt{M^2-a^2}$.
 The angular velocity of the event horizon
 $\AngVel$  
 and
 the Hawking temperature $\Temp$ are
 \begin{equation}
 \AngVel\equiv \frac{a}{r_{\horind}^2+a^2},\quad
\Temp\equiv \frac{\left(M^2-a^2\right)^{1/2}}
{4\pi M r_{\horind}}.
\end{equation}
Clearly, for $a>M$ there is no event horizon and 
the Kerr metric would correspond to
a rotating naked singularity.
Thus, the maximal angular momentum that a rotating black hole can have is $a=M$, in which case it is called an extremal Kerr black hole.
In an extremal black hole, the
Boyer-Lindquist radii of the
 event  and Cauchy horizons coincide
  ($r_+=r_-=M$),
 the angular velocity is $\AngVel=1/(2M)$
  and the temperature is zero ($\Temp=0$).
 
In this paper we consider  linear massless-field perturbations with general integer-spin $s$ 
($=0$, $\pm 1$ and $\pm 2$ for, respectively, the scalar,  electromagnetic and gravitational field)
 of
Kerr black holes.
Teukolsky~\cite{Teukolsky:1973ha} managed to decouple the equations obeyed by such field perturbations.
Furthermore, 
 assuming a field
dependence on the time $t$ and azimuthal angle  $\varphi$ of the type $e^{-i\omega t+im\varphi}$, with
frequency $\omega\in \mathbb{C}$ and azimuthal number 
$m=-\ell,-\ell+1,\dots, \ell$,
Teukolsky
showed that the equations
 separate into two ordinary differential equations (ODEs): one for a radial factor $\RExp$ and
 the other for a polar-angle factor.
The  separation constant $\eigenlExp$  is the angular eigenvalue, which is
  partly labelled by  the multipole number $\ell=|s|, |s|+1, |s|+2,\dots$.
The polar-angle factor is a spin-weighted spheroidal harmonic~\cite{Berti:2005gp};
the radial factor we deal with in the following subsections.
 To reduce cluttering, henceforth
we shall use the subindex
$\Lambda\equiv \{s, \ell,m,\omega\}$.

The retarded Green function of the Teukolsky field equation serves to evolve  initial data 
to its future.
Similarly to the field, the Green function may be  decomposed into radial modes $\Glm$, thus involving an integral over (just above) the real frequency line.
Leaver~\cite{Leaver:1986} 
carried out a spectral decomposition of the Green function in Schwarzschild space-time by
deforming this integral into the complex frequency plane -- for a similar decomposition in Kerr, see, e.g.,~\cite{PhysRevD.94.124053,Yang:2013shb}.
In  this paper we use this spectral decomposition in Kerr space-time.

\subsection{Extremal Kerr}\label{sec:ext Kerr}
 
 Specifically  in the extremal Kerr  case,
  the radial factor $\R$ of the perturbations obeys the ODE
\begin{align}\label{eq:radial ODE}
&
\left(x^{-2s}\frac{d}{dx} \left(x^{2s+2}\frac{d}{dx}\right) + 
 \frac{k^2}{4x^2} + \frac{k^2}{x} + \frac{k(m-is)}{x} -
 \eigenl
\right.  \nonumber \\ & 
+(k+m)\Big( k+ (k+m) (x+1)^2+2 i s x \Big)
\bigg)
\R(x) =0.
\end{align}
Here, 
we have defined a shifted radial coordinate 
$x \equiv r-M$ 
and a shifted frequency 
$k\equiv \omega-m\cdot \AngVel(a=M)$.
This second-order, linear 
ODE
possesses two irregular singular points: at infinity
 ($x=\infty$)
 and
at the horizon 
($x=0$).
Henceforth,
and following Leaver~\cite{Leaver:1985,Leaver:1986,Leaver:1986a}, 
 we shall set $M=1/2$, so that, in particular, 
$k=\omega-m$.

Eq.\eqref{eq:radial ODE} admits two linearly independent solutions.
The ones used to construct the retarded Green function are $\Rin$, which is purely-ingoing into the horizon, and $\Rup$, which is purely-outgoing to infinity.
Specifically, these ``ingoing" and ``upgoing" solutions are respectively defined by the following boundary conditions:
%
\begin{align}\label{eq:Rin bc}
\Rin \sim 
\left\{\begin{array}{l l}
\displaystyle
e^{ik/(2x)}\frac{e^{-i\omega\ln x}}{x^{2s}},&  x\to 0^+,
\vspace{0.4cm}
 \\
\displaystyle
\Rininc \frac{e^{-i\omega ( x+ \ln x) }}{x}+
\Rinref \frac{e^{i\omega ( x+ \ln x) }}{x^{1+2s}},& x\to \infty,
\end{array}
\right.
\end{align}
and
\begin{align}
\label{eq:Rup bc}
&
\Rup  \sim
\nonumber \\ &
\left\{\begin{array}{l l}
\displaystyle
\Rupref\ e^{ik/(2x)}\frac{e^{-i\omega\ln x}}{x^{2s}}
+\Rupinc\ e^{-ik/(2x)}e^{i\omega\ln x}
,&  x\to 0^+,
\vspace{0.4cm}\\
\displaystyle
 \frac{e^{i\omega ( x+ \ln x) }}{x^{1+2s}},& x\to \infty,
\end{array}
\right.
\end{align}
where
$\RinIncRef$ and $\RupIncRef$
are incidence/reflection coefficients of, respectively, the ingoing and upgoing solutions.
Using these solutions, we can define the following constant ``Wronskian":
\begin{align} \label{eqextwronskian}
\W \equiv 
\Delta^{s+1}
 \left( \Rin\dfrac{d\Rup}{dx} - \Rup\dfrac{d\Rin}{dx}\right) = 
  2i\omega\Rininc.
\end{align}

The radial modes of the retarded Green function can then be expressed as
\begin{equation}\label{eq:radialgreenfunction}
\Glm(x,x') = - \frac{\Rin(x_<)\Rup(x_>)}{\W} ,
\end{equation}
where $x_<\equiv \text{min}(x,x')$ and $x_>\equiv \text{max}(x,x')$.

Clearly, any 
zeros of the Wronskian correspond to
poles of the Green function modes.
From Eqs.\eqref{eq:Rin bc} and \eqref{eqextwronskian}, such poles possess $\Rininc=0$ and are thus simultaneously purely-ingoing into the horizon and purely-outgoing to infinity, i.e., they possess no incoming radiation.
If $\text{Im}(\omega)<0$, such poles are the so-called QNM frequencies $\oQNM$ 
 where $n=0,1,2,\dots$
is the overtone number and it 
 increases with 
the magnitude of $\text{Im}(\oQNM)$.
Clearly, each QNM decays exponentially with time.
If $\text{Im}(\omega)> 0$, on the other hand,
such modes would grow exponentially with time and
would lead to a mode instability of the space-time
(if $\text{Im}(\omega)=0$, one could say that they are marginally unstable modes).

Apart from poles, the Green function modes may also possess branch points if  $\Rin$ and/or $\Rup$ possess any.
It is well-known that $\Rup$ 
possesses a branch point at $\omega=0$ in both sub-extremal and extremal Kerr~\cite{Leaver:1986,Leaver:1986a,PhysRevLett.84.10,PhysRevD.94.124053,casals2018perturbations}. 
This branch point is due to the irregular character of the 
singularity of the radial ODE at $r=\infty$. It is also known that $\Rin$ 
 possesses a branch point at 
$\omega=m$ 
(i.e., $k=0$) in extremal Kerr~\cite{Leaver:1986,casals2018perturbations}. 
This extra branch point is due to the  irregular character of the
singularity of the  ODE at $r=r_{\horind}$.
These branch points at $\omega=0$ and 
$m$
 in the radial solutions in extremal Kerr carry over to the Wronskian $\W$
 as well as to the Green function modes $\Glm$
\footnote{
We note that, apart from these BCs which appear due to the irregular singular points in the radial ODE,
 the Wronskian and the Green function modes also possess BCs coming in from the angular eigenvalue~\cite{oguchi1970eigenvalues,BONGK:2004}. These angular BCs, however,
 do not possess any physical significance~\cite{Hartle:Wilkins:1974,PhysRevD.94.124053} and so we do not consider them in this paper.}.

Physically, the contribution to the field perturbation from the BC  to leading order near $\omega=0$ is known to decay  at late-times as a power-law, both in sub-extremal~\cite{Leaver:1986,PhysRevLett.84.10,PhysRevD.94.124053} and extremal Kerr~\cite{casals2018perturbations}.
In its turn, the contribution to the field perturbation of extremal Kerr from the BC  to leading order near  $k=0$ also decays as a power-law off the horizon~\cite{casals2018perturbations}, while it
gives rise to the Aretakis phenomenon on the horizon~\cite{casals2016horizon,Gralla:2017lto}.
It has also been observed in~\cite{Gralla:2017lto} 
that the Aretakis phenomenon may be viewed as a consequence of the enhanced symmetry which the 
near-horizon geometry of extremal Kerr (NHEK) 
possesses: the isometry group of NHEK is $\text{SL}(2,\mathbb{R})\times U(1)$~\cite{Bardeen:1999px}, as opposed to the two-dimensional isometry group of the full Kerr geometry which is formed from the Killing vectors
 $\partial_t$ and $\partial_{\varphi}$.

\subsection{Sub-extremal Kerr}

Since we will also be showing  results in sub-extremal Kerr, we conclude this section by introducing the basic quantities that we
need in the sub-extremal case.
One can define the following  linearly-independent solutions of the radial Teukolsky equation in sub-extremal Kerr:
\begin{align}\label{eq:Rin bc sub}
\RinS \sim 
\left\{\begin{array}{l l}
\displaystyle
\Delta^{-s}e^{-i \kSub r_*}, & r\rightarrow r_{\horind}, \\
\displaystyle
r^{-2s-1} \RinrefS e^{i \omega r_*} +r^{-1} \RinincS e^{-i \omega r_*}, & r \rightarrow \infty,
\end{array}
\right.
\end{align}
and
\begin{equation}
\label{eq:Rup bc sub}
\RupS  \sim
\left\{\begin{array}{l l}
\displaystyle
\RupincS e^{i \kSub r_*}+\RuprefS\Delta^{-s}e^{-i \kSub r_*}, & r\rightarrow r_{\horind},
\\
\displaystyle
r^{-2s-1} 
e^{i \omega r_*}, & r \rightarrow \infty,
\end{array}
\right.
\end{equation}
where
$\kSub\equiv \omega-m\AngVel$,
$\Delta\equiv (r-r_{\horind})(r-r_-)$
and 
\begin{equation}
r_{*}\equiv  
r + \dfrac{r_{+}\ln\left(r-r_{+}\right) -r_{-}\ln\left(r-r_{-}\right)}{r_{+}-r_{-}}.
\end{equation}
These ingoing and upgoing solutions are the equivalent 
of Eqs.\eqref{eq:Rin bc} and \eqref{eq:Rup bc} in extremal Kerr.

In sub-extremal Kerr, one may define a Wronskian 
similarly to  Eq.\eqref{eqextwronskian}:
\begin{equation}\label{eq:sub wronskian}
\Wsub\equiv \Delta^{s+1}\left(\RinS \frac{d\RupS}{dr}-\RupS \frac{d\RinS}{dr}\right)
=2i\omega \RinincS.
\end{equation}
In sub-extremal Kerr, $\RupS$ and  $\Wsub$  possess a branch point at $\omega=0$ but, since $r=r_+$ is a regular singular
point of the radial ODE,
 $\Rin$ and $\Wsub$  do  not  possess a branch point at $\omega=m\Omega_H$.


\section{Method and asymptotics}\label{sec:method}

We now describe the method  which we use to obtain
our results.
In the first and second subsections we briefly review the MST method in, respectively, extremal and sub-extremal Kerr.
 In the  third subsection we explain how we carried out the search for poles in the  complex frequency plane.
 In the last subsection we give asymptotics of the Wronskian in extremal Kerr near the branch points.
For details of the MST method in sub-extremal and extremal Kerr, we refer the reader to, respectively,~\cite{Sasaki:2003xr,PhysRevD.94.124053} and~\cite{casals2018perturbations}, and references
therein.
We note that the MST method is developed for 
$\text{Re}(\omega)\ge 0$
 but  we may use
Eq.\eqref{eq:symm W ext} below
(and its extremal Kerr counterpart)
to cover the whole plane.


\subsection{MST method in extremal Kerr}\label{eq:MST ext}

The MST method in extremal Kerr
essentially consists of expressing the solutions of the radial ODE \eqref{eq:radial ODE} as infinite series of confluent hypergeometric functions, with
 the same series coefficients
$\aMST{n}$ (see Eq.\eqref{eq:rec rln an} below) for both the ingoing and upgoing solutions.
This allows for obtaining the following expression for the Wronskian~\cite{casals2018perturbations}:
\begin{widetext}
\begin{equation}\label{eq:IncI}
\W
=
\frac{\sin(\pi(\nu+i\omega))}{\sin(2\pi\nu)}
\frac{
\left(\bar K_{\nu}(-ik)^{-2\nu-1}+
\mathcal{\bar C}
\bar K_{-\nu-1}\right)
 2^{1+s-i\omega}\omega^{\nu+s}(i\omega)^{1-\nu-i\omega}e^{-3\pi \omega/2}e^{-\pi i}
\sum_{n=-\infty}^{\infty}
(-1)^n  \aMST{n}}
{
(-ik)^{s+i\omega-\nu-1} e^{-i \pi \et /2} e^{- i \pi (\nu + \frac12)} \sum_{n=-\infty}^{\infty} \frac{\Gamma(\q{n}{\nu} + \e )}{\Gamma(\q{n}{\nu} - \e)} \aMST{n}},
\quad \text{Re}(\omega)>0,
\end{equation}
\end{widetext}
where
\begin{align}\label{eq:R+ nu -> -nu-1}
&
\mathcal{\bar C}
\equiv
 ie^{-2\pi i \nu}(i\omega)^{2\nu}\omega^{-2\nu},
\end{align}
\begin{equation}\label{eq:Knu}
\bar K_{\nu}\equiv
(2\omega)^{-\nu-1}e^{i \pi s}\frac{\sum_{n=\rArbit}^{\infty} C_{n,n-\rArbit}}{\sum_{n=-\infty}^\rArbit D_{n,\rArbit-n}},
\end{equation}

\MC{Do we know if {\it all} the QNMs that we find come from the zeros of $\left(\bar K_{\nu}(-ik)^{-2\nu-1}+\mathcal{\bar C}\bar K_{-\nu-1}\right)$?}
with  $\rArbit$ an arbitrary integer,
\begin{equation}\label{eq:Dnj}
D_{n,j}\equiv  \frac{\Gamma(\q{n}{\nu} + \e )\Gamma(1-2 \q{n}{\nu})(\q{n}{\nu} + \e)_j}{\Gamma(\q{n}{\nu}- \e)\Gamma(1-\q{n}{\nu}+\e)(2 \q{n}{\nu})_j\, j!}
 \aMST{n}(-2 i\omega)^{n+j},
\end{equation}
\begin{equation}\label{eq:Cnj}
C_{n,j}\equiv
\frac{\Gamma(\q{n}{\nu} + \e)\Gamma(2\q{n}{\nu}-1)(1-\q{n}{\nu}+ \et)_j}{\Gamma(\q{n}{\nu}-\e)\Gamma(\q{n}{\nu} + \et)(2-2\q{n}{\nu})_j\, j!}
 \aMST{n}(-ik)^{j-n},
\end{equation}
and
\begin{align} 
\e \equiv s-i\omega,
\quad 
\q{n }{\nu} \equiv n+\nu+1.
\end{align} 
Here, $(z)_j\equiv \Gamma(z+j)/\Gamma(z)$ denotes the Pochhammer symbol.
The coefficients $\aMST{n}$ satisfy the following three-term bilateral recurrence relations:
\begin{equation}\label{eq:rec rln an}
\alpha_n \aMST{n+1}+
\beta_n\aMST{n}+
\gamma_n \aMST{n-1}=0,\quad n\in \mathbb{Z},
\end{equation}
where
\begin{subequations}
\begin{align}\label{eq:rec.eq.coeffs}
\alpha_n \equiv& \frac{
\omega k
(\q{n}{\nu}+\e)(\q{n}{\nu} - \et)}{\q{n}{ \nu} (2 \q{n}{ \nu} +1 )},\\
\beta_n \equiv & ( \q{n}{ \nu} -1 ) \q{n}{\nu} +
2 \omega^2
 -s(s+1) -
 \nonumber
  \\ &
\eigenl - 
m \omega
- \frac{ \omega k\,  \e \et}{ (\q{n}{ \nu} -1 ) \q{n}{ \nu}  }, 
\nonumber
 \\ 
\gamma_n \equiv & \frac{
\omega k
(\q{n}{ \nu} -1 -\e)(\q{n}{ \nu} -1 + \et)}{(\q{n}{\nu}-1)(2\q{n}{ \nu} -3)}.
\nonumber
\end{align}
\end{subequations}
Finally, $\nu$ is the so-called renormalized angular momentum parameter.
Its value is chosen so that the solution $\aMST{n}$ of the recurrence relation in Eq.\eqref{eq:rec rln an} is minimal (i.e., $\aMST{n}$ is
the unique -up to a normalization- solution of the recurrence relation which is subdominant with respect to the other solutions) both as $n\to \infty$
and as $n\to -\infty$.
In practise, the value of  $\nu$ may be found by imposing the condition:
\begin{align}\label{eq:eq for nu}
R_n L_{n-1} = 1,
\end{align}
where 
$R_n\equiv \aMST{n}/\aMST{n-1}$ and $L_n\equiv \aMST{n}/\aMST{n+1}$.
We note that Eqs.\eqref{eq:rec rln an} and \eqref{eq:eq for nu}  satisfied by $\aMST{n}$ and $\nu$ agree with the extremal limit of their
sub-extremal counterparts, which are given in Eqs.(123) and (133) in~\cite{Sasaki:2003xr}.
The choice of $n$ in Eq.\eqref{eq:eq for nu} is arbitrary and henceforth we choose it to be $n=1$.

Although $\nu$ has been introduced here via the MST method, it is in fact a rather fundamental parameter.
For example, it yields the monodromy of the upgoing solution around $r=\infty$~\cite{castro2013black}.
Also, 
$\nu_c(\nu_c+1)$ is the eigenvalue of the Casimir operator of the  $\mathfrak{sl}(2,\mathbb{R})$ 
factor in the
algebra of NHEK~\cite{Gralla:2015rpa}\MC{only true for $s=0$?}, where we have defined $\nu_c\equiv \nu(k=0)$.
As we shall see, $\nu$
plays a  pivotal role in the physics of extremal black holes and so here we describe some of its properties.
Firstly, 
$\nu$
is either real-valued or else complex-valued with a real part that is equal to 
a half-integer number~\cite{fujita2005new,Sasaki:2003xr}).
This property readily follows for $\nu_c$ from the analytical result shown in~\cite{casals2018perturbations} that
\begin{equation}\label{eq:nu_c}
\nu_c=-\frac{1}{2}\pm \sqrt{
\lambda_{s\ell mm}-m^2+\left(s+\frac{1}{2}\right)^2},
\end{equation}
and the fact that $\lambda_{s\ell mm}\in\mathbb{R}$;
one is free to choose the sign in Eq.\eqref{eq:nu_c}.
It is  thus convenient to define
\begin{equation}\label{eq:delSR}
\delSR^2\equiv 
-\left(\nu_c+\frac{1}{2}\right)^2
\in\mathbb{R},
\end{equation}
 following~\cite{starobinskii1973amplification,Starobinskil:1974nkd}.
Clearly, $\delSR^2<0$ if $\nu_c$ is real-valued and $\delSR^2>0$ otherwise.
As we shall see throughout the paper, various properties of  quantities will depend on the sign of $\delSR^2$ -- we collect these properties in Table \ref{tab}.
In Sec.\ref{sec:QNMs NEK props} we discuss the modes for which $\delSR^2$ is positive and for which it is negative.
 Numerically, we have observed that $\nu_c$ is non-integer except in the axisymmetric case $m=0$, for which it is 
$\lambda_{s\ell 00}=\ell(\ell+1)-s(s+1)$, and so
either $\nu_c=\ell$ or $\nu_c=-\ell-1$, and $\delSR^2=-(\ell+1/2)^2$.
Lastly, it follows from the symmetries of the angular equation that
 $\nu_c$ is invariant under $s\to -s$ and, separately, under $m\to -m$:
\begin{equation}\label{eq:symms nu_c}
\left.\nu_c\right|_s=\left.\nu_c\right|_{-s},\quad
\left.\nu_c\right|_m=\left.\nu_c\right|_{-m}.
\end{equation}



\begin{widetext}
\begin{table*}[t]
 \begin{center}
 \begin{tabular}{| c | c   | c   |  c | c | c  |} 
  \hline
$\delSR^2$  & $\nu_c$ & $\mathcal{F}_s^2$
& $\nu_c(\nu_c+1)$
& $\Amplif$ for $a=M$ near $k=0$ & $\exists$ DMs in NEK \\ [0.5ex] 
 \hline
 \hline
$<0$ &  $\in\mathbb{R}$ & $<0$ & $>0$ & continuous and monotonous & Yes  \\ 
\hline
 $>0$ &\ $=-1/2+i\cdot \mathbb{R}$ \ &  $>0$  &   $<0$ & discontinuous and oscillatory  & No
   \\
 \hline
\end{tabular}
\end{center}
\caption{
Various properties of quantities depending on the sign of $\delSR^2$ in Eq.\eqref{eq:delSR}:
renormalized angular momentum parameter at $k=0$, $\nu_c$  in Eq.\eqref{eq:nu_c};
 $\mathcal{F}_s^2$ in Eq.\eqref{eq:cond DM}; the eigenvalue $\nu_c(\nu_c+1)$ of the Casimir operator of the  $\mathfrak{sl}(2,\mathbb{R})$ in NHEK;
amplification factor $\Amplif$ in Eq.\eqref{eqZdef};
DMs in NEK (but note that it does not apply to \nsDMacr\ s).
 }
\label{tab}
\end{table*}
\end{widetext}

Let us here indicate how we performed the practical calculation of the renormalized angular momentum $\nu$ and 
the angular eigenvalue $\eigenl$.
An alternative to calculating  $\nu$ via Eq.\eqref{eq:eq for nu} 
is by
using the monodromy method described in~\cite{castro2013black}.
Ref.\cite{castro2013black} provides the weblink~\cite{monodromysite} to a {\rm MATHEMATICA} code
which we used
 for calculating $\nu$
\MC{prove analytically how MST's $\nu$ is related to monodromy's $\alpha_{irr}$?}.
As for the calculation of
$\eigenl$, we used the 
\textit{Mathematica} function \texttt{SpinWeightedSpheroidalEigenvalue} in the  toolkit in~\cite{BHPToolkit} 
(for $s=0$, one may  also use the
corresponding in-built  {\rm MATHEMATICA} function).

We end up this subsection
by noting that, to the best of our knowledge,  the MST method has never been used before for calculating QNM frequencies themselves.
One of the most standard methods for calculating QNMs is the continued fraction method which Leaver  introduced in~\cite{Leaver:1985}.
This method was 
later  also used
 in, e.g.,~\cite{QNMBertiNew,Yang:2012pj,yang2013quasinormal}, for obtaining QNM frequencies in sub-extremal Kerr  space-time.
Ref.~\cite{Richartz:2015saa} adapted Leaver's method to the case of extremal Kerr.
We note that this adaptation is not guaranteed to work at $\omega=m$.
In its turn, the infinite series in the MST expression \eqref{eq:IncI} for the Wronskian in extremal Kerr converges faster the closer the frequency is to $\omega=0$
or to $\omega=m$~\cite{casals2018perturbations}.
Therefore, the MST method in extremal Kerr is probably more suitable near 
$\omega=m$
 than the adaptation of Leaver's method in~\cite{Richartz:2015saa}.


\subsection{MST method in sub-extremal Kerr}\label{sec:subext}

The asymptotic radial coefficients in  sub-extremal Kerr may also be obtained via MST expressions.
Specifically, the incidence coefficient $\RinincS$ 
may be obtained by dividing Eq.(168) by Eq.(167) in~\cite{Sasaki:2003xr}
and the reflection coefficient $\RinrefS$ by dividing Eq.(169) by Eq.(167) in~\cite{Sasaki:2003xr}.
An important point to note is that, in the resulting expressions, both  $\RinincS$ and $\RinrefS$ contain an explicit overall factor 
$\Gamma(1-s-2\textit{i}\epsilon_{+})$, which comes from Eq.(165)~\cite{Sasaki:2003xr}, where
\begin{align}
\epsilon_{+}\equiv 
\frac{1}{2}\left(1+\frac{1}{\kappa}\right)\tilde{\omega},
\quad
\kappa\equiv \sqrt{1-(2a)^2}.
\end{align}
Simple poles of this $\Gamma$-factor, therefore, correspond to simple poles of both $\RinincS$ and $\RinrefS$ (unless such a pole is somehow cancelled out by other factors in $\RinincS$ and/or $\RinrefS$ -- such potential cancellation
 is possible but certainly not apparent from the MST expressions). 
Therefore, the modes corresponding to these poles are, in principle, totally-reflected modes (TRMs).
Such potential poles in $\RinincS$ carry over to the Wronskian, and we remove them ``by hand" by defining the following ``Wronskian factor":
\begin{align} \label{eq:Wronsk fac}
W^{f}_{\indmode} \equiv 
 \dfrac{
2^{i\omega} \kappa^{2s}
 e^{i\kappa \epsilon_{+}(1+2\log\kappa/(1+\kappa))}e^{i\omega(\ln\omega - \frac{1-\kappa}{2})}}
 { e^{-\frac{\pi}{2}\omega} e^{i\omega \kappa}e^{\frac{\pi}{2}i(\nu+2-s)}\Gamma(1-s-2i\epsilon_{+})}
\Wsub,
\end{align}
where, for calculational convenience, we have also included some
 extra factors.
As can be expected, the removal of the above $\Gamma$-factor is rather convenient for  practical purposes,  
as we shall explicitly see in Sec.\ref{sec:calc QNMs NEK}.


\subsection{Search for poles}

The expression in Eq.\eqref{eq:IncI} (as well as its sub-extremal counterpart) 
for the Wronskian 
is only valid for 
$\text{Re}(\omega)\ge 0$. 
In order to investigate the region $\text{Re}(\omega)<0$, one may use a 
 symmetry that follows from the angular equation,
\begin{equation}\label{eq:symm eigen}
\eigenlExp=\lambda_{s,\ell,-m,-\omega^*}^*
\end{equation}
in order to obtain the following
radial symmetry
\begin{equation}\label{eq:symm rad}
\RinupExp=R^{\text{in/up}^*}_{s,\ell,-m,-\omega^*},
\end{equation}
and similarly for $\RinupS$.
This symmetry implies that the Wronskian
satisfies:
\begin{equation}\label{eq:symm W ext}
W_{s,\ell,-m,-\omega^*}=-W^*_{s,\ell,m,\omega},
\end{equation}
and similarly for $\Wsub$,
and that the QNM frequencies satisfy:
\begin{equation}\label{eq:symm QNM}
\oQNM=-\oQNMmm^*.
\end{equation}
Furthermore, by virtue of the so-called Teukolsky-Starobinsky identities~\cite{teukolsky1974perturbations,Chandrasekhar}, 
which relate radial 
solutions with spin $s$
to radial
solutions with spin ``$-s$", the QNM frequencies \MC{or, more generally, poles?} are the same for spin $s$ and for spin ``$-s$", as long as the frequency is not an algebraically-special
frequency~\cite{MaassenvandenBrink:2000ru} -- see App.\ref{sec:AS} for a description and calculation of algebraically-special
frequencies.

As mentioned above, both QNMs and exponentially unstable modes are 
poles of the Green function modes and
so
zeros of the Wronskian. However, in practise, we  looked for minima 
--instead of zeros--
of the absolute value of the Wronskian.
From the minimum modulus principle of complex analysis, 
if a function  is analytic in a certain region,
a point is a zero of that function  within that region if and only if it is a local minimum of the absolute value of the function (e.g.,~\cite{Rudin}).
The Wronskian is not analytic everywhere in the complex frequency plane: it has poles and branch points. However, poles can clearly not correspond to minima of $|\W|$. In their turn, branch points of $\W$ could correspond to local minima of $|\W|$ which are not zeros, but they could   easily be  discarded by spotting the appearance of a discontinuity 
stemming from such points.


The minimization routine that we chose to use is the Nelder-Mead method \cite{NM}.
We obtained the initial guesses for
minima of $|\W|$
in the Nelder-Mead method in the following way.
First, we calculated $|\W|$ over a grid of frequencies in a region of the complex plane.
We chose the 
grid
 stepsizes 
$\Delta\omega_r$  and $\Delta\omega_i$
in,
respectively, 
$2M\text{Re}(\omega)$ and $2M\text{Im}(\omega)$,
large enough so that we could manage to cover in practise the region chosen, yet small enough so as to try to not miss any 
minima.
More specifically, we picked  $\Delta\omega_r=0.01$  for all cases shown later except for  $a=0.998M$, $s=-2$, $\ell=2$, $m=1$, for which we picked $\Delta\omega_r=0.005$;
for $\Delta\omega_i$ we picked values between $0.0005$ and $0.01$, depending on the case (App.\ref{sec:Hod} is also an exception -- we state there the values chosen). 
Finally, we picked the frequencies $\omega_k$, $k=0,1,2,\dots$, in the grid which yield (local) minima of $|\W|$ among its values on the grid.
Each one of these frequencies $\omega_k$ is an initial guess for a
minimum  of $|\W|$.

For each guess $\omega_k$, the  Nelder-Mead method requires three initial points. As these initial points we chose: $\omega_k+\zeta\, e^{j 2 \pi i/3}$, $j=1$, $2$ and $3$,
and we chose a value of $\zeta>0$ which is smaller than $\min\{\Delta\omega_r,\Delta\omega_i\}$.
We then applied the  Nelder-Mead method for each $\omega_k$ with the corresponding three initial points
and we required
 16 digits of precision in both the imaginary and real parts of the zeros of $|\W|$.

We similarly applied the above procedure  to the Wronskian factor $W^{f}_{\indmode}$ (Eq.\eqref{eq:Wronsk fac}) in sub-extremal Kerr.


\subsection{Wronskian near the branch points  in extremal Kerr }\label{sec:MST branch pts}

As mentioned in Sec.\ref{sec:ext Kerr}, the Wronskian  possesses branch points at $\omega=0$ and 
$k=0$
in extremal Kerr. 
As mentioned in Sec.\ref{eq:MST ext}, the MST method is particularly suited near these points and so we use it here to give
the analytical behaviour of the Wronskian near these points.
The asymptotics we give follow readily from the MST expressions in~\cite{casals2018perturbations}
and
 will help explain various features which we shall see in the next sections.

First, near the origin, and for $m\neq 0$, it is
\begin{equation}\label{eq:W omega->0}
\W \sim 
c_0
\, \omega^{-\ell+s-1}+ \omega^{-\ell+s}\left(
c_1+c_l
\, \ln(-i\omega) \right),\quad
\omega\to 0,
\end{equation}
for some coefficients $c_0$, $c_1$ and $c_l$
which do not depend on $\omega$
(although they generically depend on $\ell$ and $m$).
 In its turn, 
to leading order near the superradiant-bound frequency, it is
\begin{align}\label{eq:W k->0}
&
\W \sim 
c_k (-ik)^{-s-im} \cdot
\\ &
\left(
S_{\nu_c}
\, \left( -ik\right)^{-\nu_c}+ 
e^{i\pi (1/2-\nu_c)}
S_{-\nu_c-1}
\, \left( -ik\right)^{\nu_c+1}
\right), 
\quad k\to 0,
\nonumber
\end{align}
where, for $m>0$ (for $m<0$ one may use Eq.\eqref{eq:symm W ext})
\begin{equation}\label{eq:Sdef}
S_{\nu_c} \equiv (2m)^{-\nu_c}  \frac{ \Gamma(2 \nu_c +1) \Gamma(s - \nu_c - im) }{\Gamma(-2 \nu_c-1)\Gamma(\nu_c + 1 - im -s ) },
\end{equation}
and
\begin{align}
&
c_k\equiv 
e^{\pi i(s+\nu_c)/2}e^{-\pi m/2}(2m)^{s-im}
\cdot \\ &
\frac{\sin(\pi(\nu_c+im))}{\sin(2\pi\nu_c)}
 \frac{\Gamma(\nu_c+1-s + im) }{\Gamma(\nu_c + 1 +s- im) }.
 \nonumber
\end{align}
In the case $m=0$, the superradiant-bound frequency
is located at
 the origin and we have
\begin{align}\label{eq:W k->0,m=0}
&
\W \sim 
c_{\ell} 
\, \omega
\left(
S_{\ell}
\, \left( -i\omega\right)^{-2\ell-1}+
S_{-\ell-1}
\, \left( -i\omega\right)^{2\ell+1}
\right), 
\, \omega\to 0,
\end{align}
approached with $\text{Re}(\omega)\ge 0$,
where
\begin{equation}\label{eq:Sdef m=0}
S_{\ell} \equiv (-1)^{\ell}  \frac{ \Gamma(2 \ell +2) \Gamma(2\ell+1) }{\Gamma^2(\ell+1-s) },
\end{equation}
and
\begin{align}
&
c_{\ell}\equiv 
2^{s+1}i\,
 \frac{\Gamma(\ell+1-s)}{\Gamma(\ell + 1 +s) }.
 \nonumber
\end{align}

The asymptotics in Eqs.\eqref{eq:W omega->0} and \eqref{eq:W k->0} manifestly show that 
$\omega=0$ and 
$k=0$ are branch points of the Wronskian\footnote{The leading order Eq.\eqref{eq:W k->0,m=0} does not show a BC for $m=0$ but a BC is expected to appear at a higher order.} and a BC is taken to run vertically ``down" from each one of these points.

The  asymptotics in Eq.\eqref{eq:W k->0}  show that, for $m>0$, 
 \begin{equation}\label{eq:W k->0,ext}
 |\W|
 =O\left(k^{-s+\expW}\right), 
 \quad \text{as}\ k\to 0,
 \end{equation}
 with 
  \begin{equation}
 \expW\equiv \frac{1}{2}-\left|\frac{1}{2}+\text{Re}(\nu_c)\right|.
  \end{equation}
 Clearly, it is
 $\expW=1/2$ in the case   $\delSR^2\ge 0$ and 
 $\expW<1/2$ in the case   $\delSR^2<0$.
 From Eq.\eqref{eq:W k->0,m=0} it follows that, for $m=0$,
\begin{equation}\label{eq:W k->0,ext,m=0}
 \W=O\left(\omega^{-2\ell}\right), 
 \quad \text{as}\ \omega\to 0.
 \end{equation}


After describing the method that we used and giving the asymptotics near the branch points, we turn to the results of our calculations.




\section{Superradiance}\label{sec:super}

Superradiance is the phenomenon, originally observed in~\cite{zel1971generation,starobinskii1973amplification,Starobinskil:1974nkd}, whereby a field wave which is incoming from radial infinity and is partially reflected may extract rotational energy from a rotating
black hole.
It can be shown that, for a field mode with frequency $\omega\in\mathbb{R}$ and azimuthal number $m\in\mathbb{Z}$, superradiance occurs if and only if the condition 
$ \omega\cdot (\omega-\oSR) <0$
 is satisfied, where
$\oSR \equiv m\AngVel$ (which is  equal to $m$ in  extremal Kerr).
In order to  ``quantify" superradiance, it is useful to define the so-called amplification factor:
\begin{equation} \label{eqZdef}
\Amplif \equiv \dfrac{dE_{out}}{dt} \left(\dfrac{dE_{in}}{dt} \right)^{-1} -1 , 
\end{equation}
where $E_{in}$ and $E_{out}$ are the energy of, respectively, the incident and the reflected part of 
the wave.

For our purposes, it is  convenient to write the  amplification factor in terms of the Wronskian. This is readily achieved by using the expressions
in~\cite{brito2015superradiance}
and Eqs.5.6--5.8 in~\cite{teukolsky1974perturbations}, which relate the  coefficients at the horizon with the coefficients
at infinity\MCImp{\footnote{Eqs.5.6--5.8 in~\cite{teukolsky1974perturbations}  are obtained in sub-extremal Kerr but straightforwardly carry over to the corresponding extremal coefficients (i.e., with $\Rinincref$ and $\Rupincref$
instead of $\RinincrefS$ and $\RupincrefS$, respectively).I don't see this anymore - when we looked at this we neglected terms in the Wronskian with $O(1/x)$ wrt terms with $O(1)$ but this was an evaluation of the Wronskian for $x\to 0$, not for $x\to \infty$, since it referred to the extra factor $x^{-im}$ at the horizon. Therefore the terms with $O(1/x)$ should be {\it dominant} over the terms with $O(1)$?}}.
 The result is
\begin{equation} \label{eqZforsbyW}
\Amplif = \left\lbrace \begin{matrix} -\dfrac{ 8 \omega Mk\, r_{\horind}}
{|\Wsub|^{2}}, 
\quad\quad\quad\quad\quad\quad\quad\quad\quad \text{for} \quad s=0,
 \\ -\dfrac{2\omega^{3}}{Mkr_{\horind}
  |\Wsub|^2
 },  \quad\quad\quad\quad\quad\quad\quad\quad \text{for} \quad s=1, \\ - \dfrac{4 \omega^{5}}{k(2Mr_{\horind})^{3}\left(k^{2}+4 
\pi^2 \Temp^2\right)}
 \dfrac{1}
 {|\Wsub|^{2}},
 \quad \text{for} \quad s=2,  \\ \end{matrix} \right.
\end{equation}
with $\omega\in\mathbb{R}$.
We wrote the amplification factor in terms of the sub-extremal Wronskian $\Wsub$, but $\Amplif$  in  extremal Kerr is also given by Eq.\eqref{eqZforsbyW} 
with $\Wsub$ replaced by the extremal $\W$.
From the symmetries 
in Eq.\eqref{eq:symm rad}
it readily follows that $\AmplifExp=\Amplifmmmo$, $\forall \omega\in\mathbb{R}$.
Also, the amplification factor is independent of the sign of $s$: $\AmplifExp=\Amplifms$~\cite{brito2015superradiance}.

Refs.\cite{starobinskii1973amplification,Starobinskil:1974nkd}  (see also~\cite{teukolsky1974perturbations})
 obtained  asymptotic expressions for $\Amplif$ in 
 the following
  two regimes:
(i)  for  $M\omega\ll 1$  and\footnote{See~\cite{brito2015superradiance} for a further simplification in the limit
$\omega \ll \oSR$ 
of the
 original expression in~\cite{starobinskii1973amplification,Starobinskil:1974nkd}  for  $M\omega\ll 1$.}  
 $\omega \ll \oSR$ 
 when $a\leq M$;
(ii) for 
$\omega \to m$
 in the case $\delSR^2<0$ and for $|\alpha| \ll m^{-4} \text{max}(1,|\alpha|^{2})$ in the case $\delSR^2>0$, where
$\alpha \equiv (1-\omega/m)$,
 when $a=M$ (these latter asymptotics have been extended
to NEK in~\cite{teukolsky1974perturbations})\MC{Can we reproduce the leading-order of these two asymptotics from the expressions in~\cite{casals2018perturbations}?}.

\begin{figure}[hp!]
\begin{center}
   \includegraphics[width=.4\textwidth]{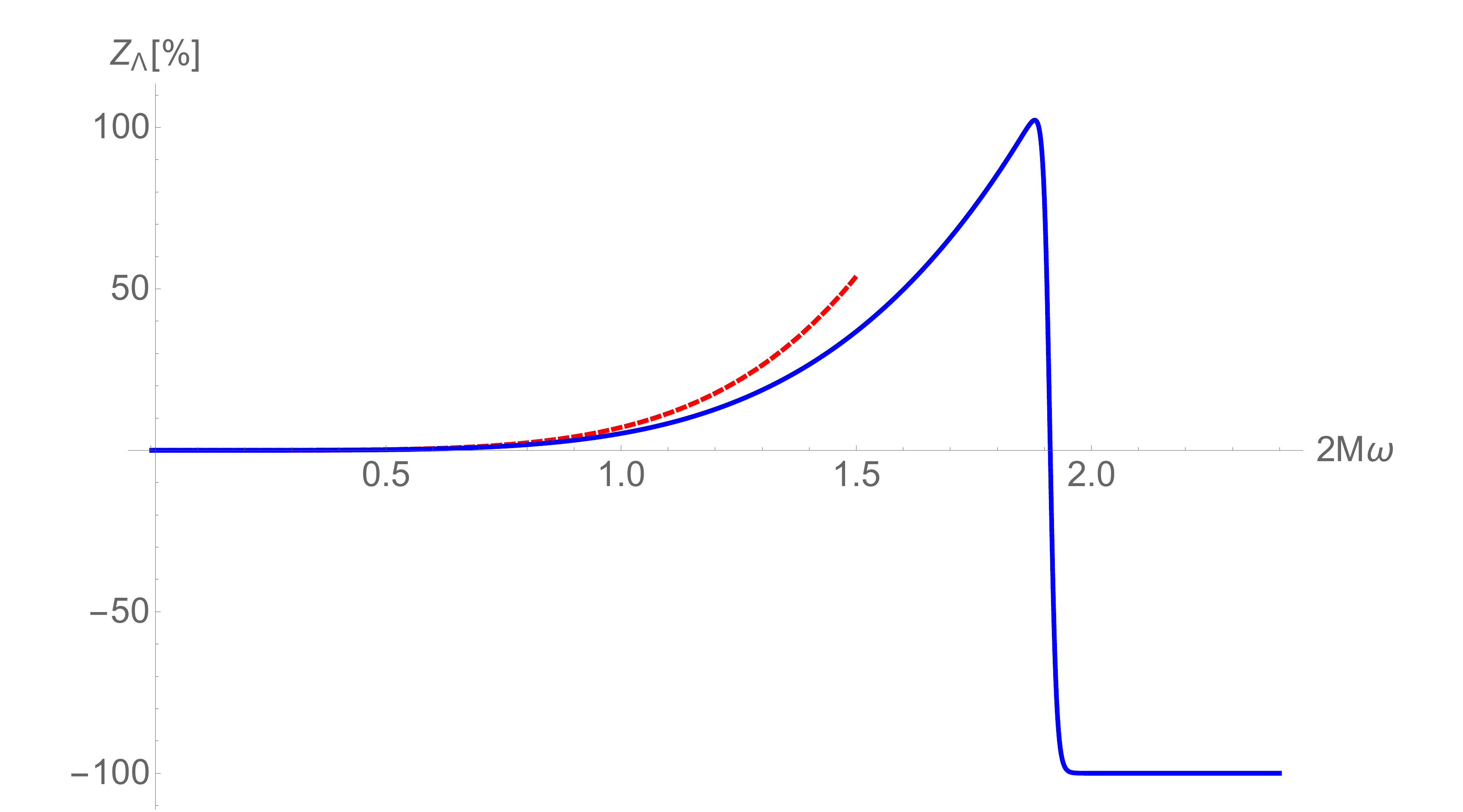}
   \end{center}
\caption{
\MC{The vertical axis is a bit dim?}
Amplification factor $\Amplif$ (in percentage form)  in Eq.\eqref{eqZforsbyW} as a function of the frequency for
the mode  $s=\ell=m=2$, $a=0.999M$.
The blue solid curve is our 
calculation using the MST method,
explained in Sec.\ref{sec:subext}.
The red dashed curve is the
approximation
(i) mentioned in Sec.\ref{sec:super}
(i.e., approximation for  $M\omega\ll 1$ and 
$\omega \ll \oSR$).
}
 \label{figZsub}
\end{figure}

\begin{figure}[hp!]
   \begin{center}
         \includegraphics[width=.4\textwidth]{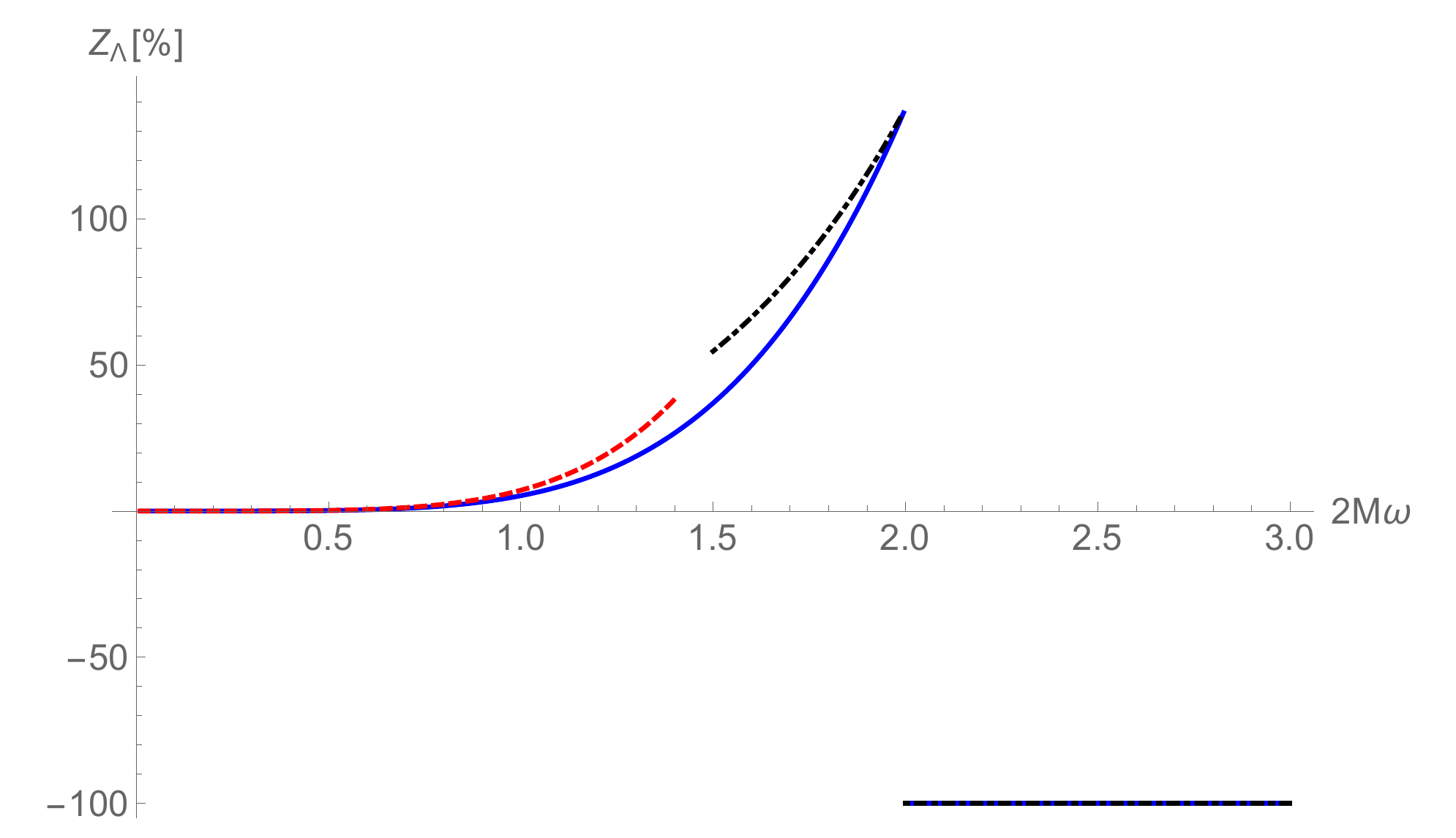}                          
   \end{center}
\caption{
\MC{the axes in this plot are  poor}
Amplification factor $\Amplif$ (in percentage) in    Eq.\eqref{eqZforsbyW} in  extremal Kerr   as a function of the frequency.
The mode  is $s=\ell=m=2$,  for which $2M\oSR=2$  and   $\delSR^2>0$.
 The solid blue curve is our 
 calculation using the MST method,  
   Eq.\eqref{eq:IncI}. 
 The red dashed curve is the
approximation
(i) mentioned in Sec.\ref{sec:super}
 (i.e., approximation for  $M\omega\ll 1$ and 
 $\omega \ll \oSR$)
 with the choice $a=0.9999M$. The dashed black curve
 (which overlaps with the blue one for $M\omega>1$)
  corresponds to the asymptotics in~\cite{Starobinskil:1974nkd} for $|\alpha| \ll m^{-4} \text{max}(1,|\alpha|^{2})$ for extremal Kerr.
}
\label{figZextreme}
\end{figure}

\begin{figure}[hp!]
   \begin{center}
      \subfloat[$s=0,\ell=3,m=2$.]{
\label{figZextreme1}
                     \includegraphics[width=.4\textwidth]{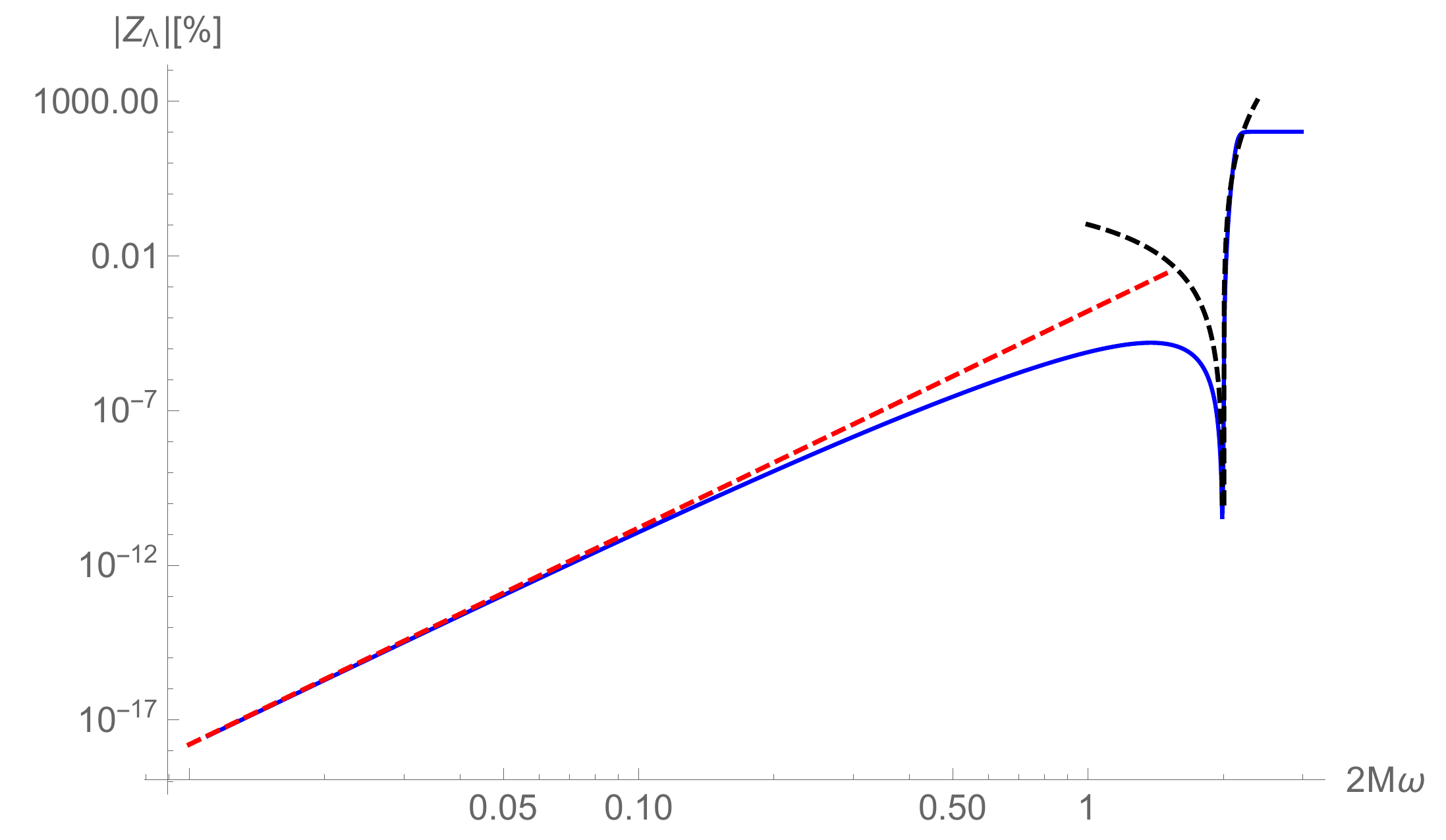}                             
   }\quad
      \subfloat[$s=\ell=m=1$]{
            \label{figZextreme2}
                  \includegraphics[width=.4\textwidth]{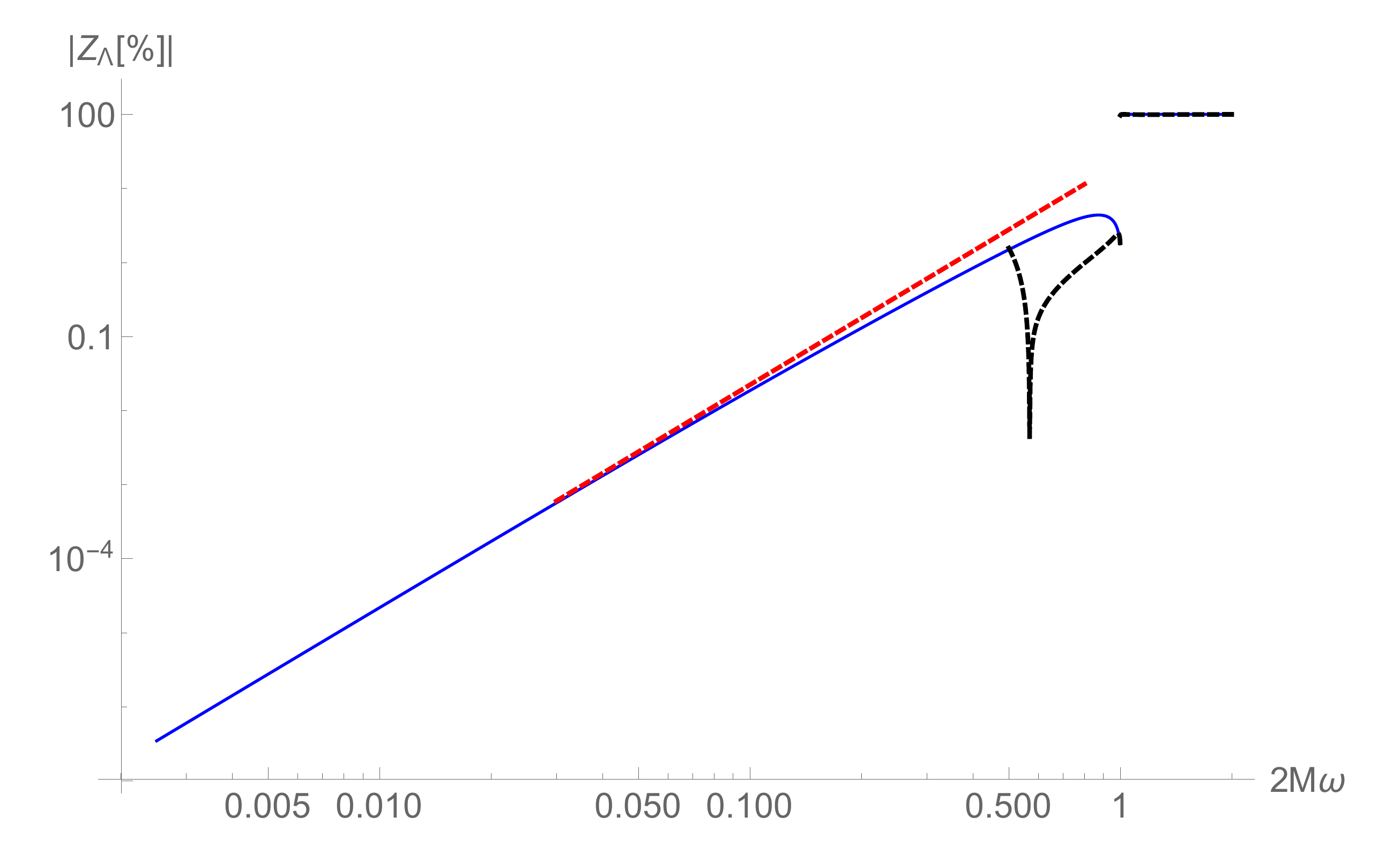}                          
                     }\quad
      \subfloat[$s=\ell=m=1$]{
      \label{figZextreme2Z1}
         \includegraphics[width=.4\textwidth]{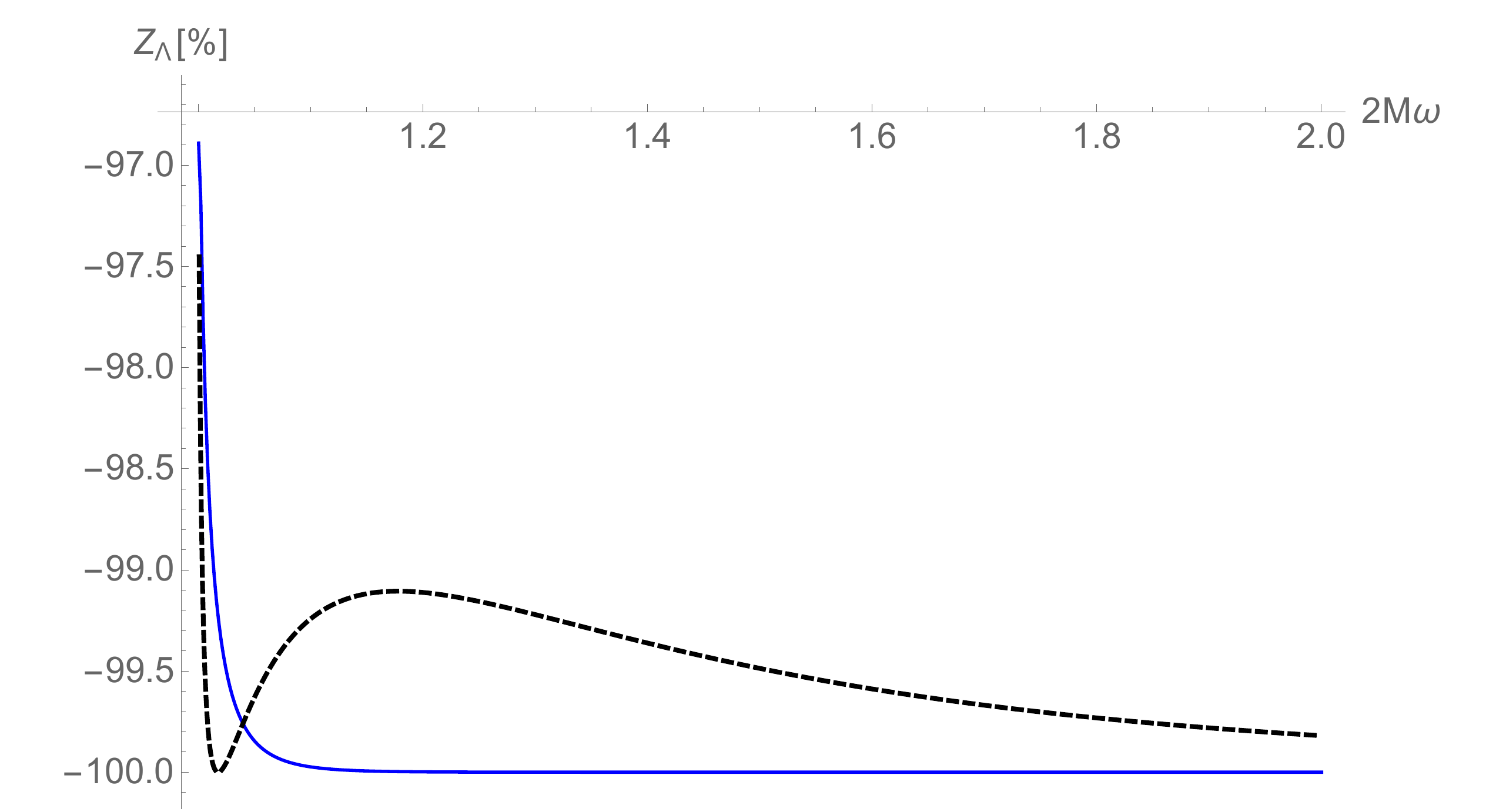}         
   }\quad
         \subfloat[$s=\ell=m=1$]{
      \label{figZextreme2Z2}
         \includegraphics[width=.4\textwidth]{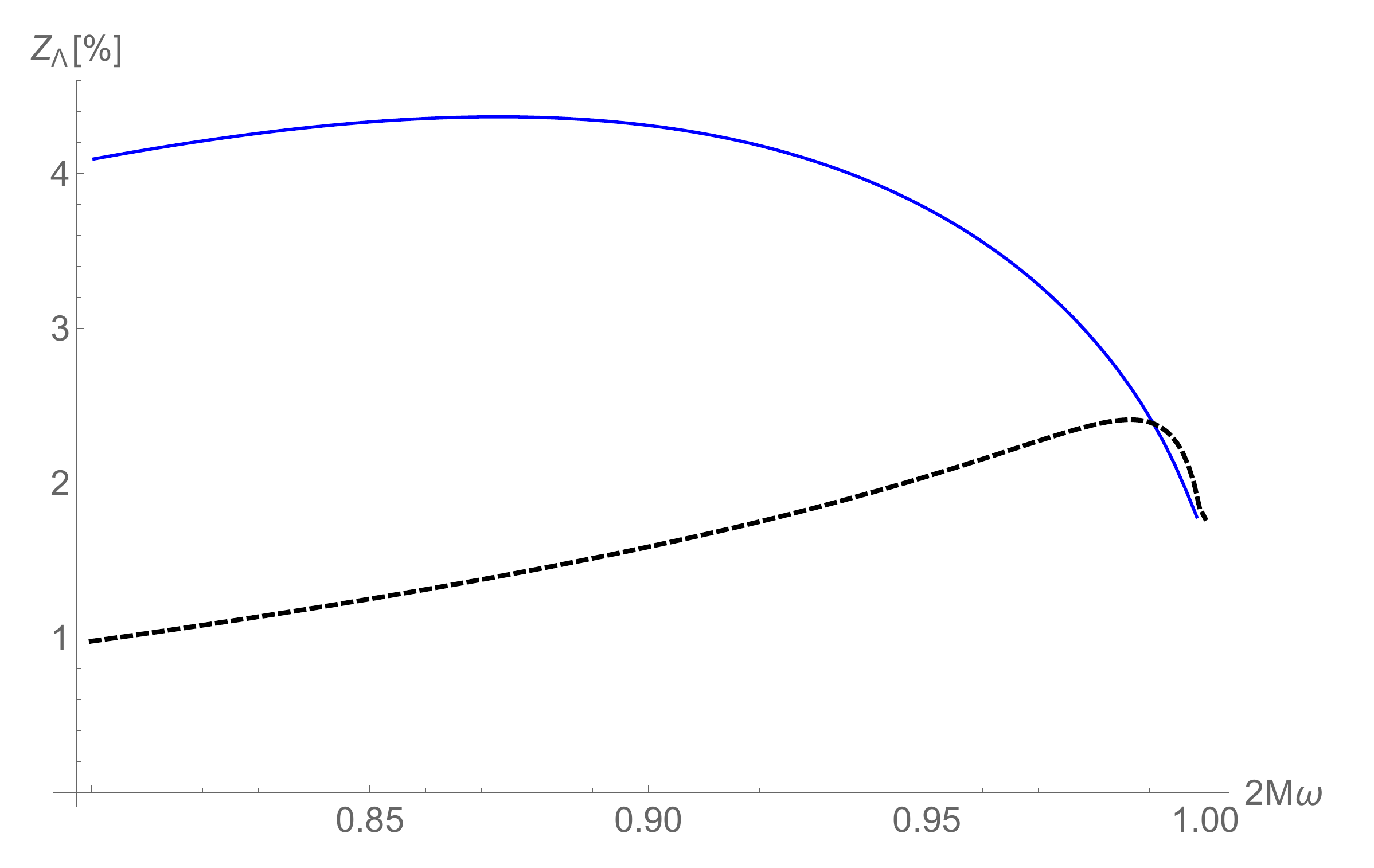}         
   }
   \end{center}
\caption{Amplification
 factor $\Amplif$ (in percentage) in  extremal Kerr   as a function of the frequency.
Same colour and style coding for the curves as in Fig.\ref{figZextreme}.
Modes: (a) log-log plot for $s=0$, $\ell=3$, $m=2$, for which $2M\oSR=2$ and  $\delSR^2<0$;  
(b) log-log plot  for $s=\ell=m=1$, for which $2M\oSR=1$ and  $\delSR^2>0$;  
(c) and (d) are linear and zoomed-in (near $k=0$, from the right and the left, respectively) versions of (b).
}
\label{figZextreme s=1}
\end{figure}

Using the MST method described in the previous section, we calculated the Wronskian and, via Eq.\eqref{eqZforsbyW}, the amplification factor.
First, as a check, we reproduced Fig.12 in~\cite{brito2015superradiance}  for the case of $s=0$, $\ell=m=1$, $a=0.99M$.
We then studied new cases.
In Figs.\ref{figZsub}--\ref{figZextreme s=1} we plot the exact, numerical values of $\Amplif$ in sub-extremal and extremal Kerr and compare them against the asymptotics in (i) and (ii)
mentioned in the above paragraph.
We find  good agreement between our  values and the two asymptotic expressions in their corresponding regimes of validity.

The asymptotics 
of the extremal Wronskian 
in Eq.\eqref{eq:W k->0,ext}
 lead to two distinct behaviours of the amplification factor in
Eq.\eqref{eqZforsbyW} in  extremal Kerr depending on whether $\delSR^2$ is positive or negative.
In the case that $\delSR^2< 0$, we have that $\Amplif\to 0$ as $k\to 0$ and $\Amplif$ is continuous at $k=0$.
This agrees with the asymptotics in (ii), which also show that, in this case, $\Amplif$ 
varies monotonically.
We exemplify the case  $\delSR^2< 0$ in Fig.\ref{figZextreme s=1}(a)\MC{Zoom into $\oSR$ in Fig.\ref{figZextreme s=1}(a) to show monotonicity}.
In the case that $\delSR^2\ge  0$, on the other hand, Eq.\eqref{eq:W k->0,ext} implies that $\Amplif$ goes like $\left(k/|k|\right)^N$, with $N=1,-1,-3$ for $s=0,1,2$, respectively.
This implies that $\Amplif$ has a discontinuity at $k=0$ in extremal Kerr when  $\delSR^2\ge  0$.
This agrees with the asymptotics in (ii), which also show that, in this case, $\Amplif$ presents an infinite number of oscillations between:
(a) two positive values as 
$ \omega\to m^-$,
 and (b) between a negative value
and ``$-1$" as  
$ \omega\to m^+$.
We exemplify the  case $\delSR^2\ge  0$ in Figs.\ref{figZextreme} and \ref{figZextreme s=1}(b)--(d)
\MCImp{But we can't see the infinite oscillations in the plots?}.
\MC{I was going to write ``With regards to the oscillations,~\cite{starobinskii1973amplification} already noted that they are connected to quasi-stationary bound states near the event horizon of NEK, which were investigated in~\cite{?}", but (i) I don't understand the connection; (ii) I don't even understand if these bound states are meant to be ZDMs (as one would expect from quasi-boundedness but ZDMs are always present)
or DMs (which are actually present for $\delSR^2<0$, not $\delSR^2>0$ as it is for the oscillations, and on the top of that ZDMs are associated with the peak away from the horizon, not with the near-horizon geometry as claimed above)?}.
With regards to the discontinuity, we can see its formation in Fig.\ref{figZsub} and its presence in
Fig.\ref{figZextreme}: as extremality is approached, the slope of   $\Amplif$ near  
$\omega=\oSR$ 
increases until a discontinuity is reached in the actual extremal limit.

Whereas the amplification factor in subextremal Kerr has been calculated in an exact numerical manner in, e.g.,~\cite{Starobinskil:1974nkd,teukolsky1974perturbations,brito2015superradiance},
to the best of our knowledge,
this is the first work where
this is achieved
in extremal Kerr.
The amplification factor  for a rotating Kerr
black hole is the largest  in extremal Kerr, such as in the cases that we plot in  Figs.\ref{figZextreme} and \ref{figZextreme s=1}.
In extremal Kerr, we obtain that the largest (percentage) value of $\Amplif$
 is approximately equal to
4.3640\%
  for $s=\ell=m=1$
   and   to
 137.61\%  for $s=\ell=m=2$ 
 (cf. the values in~\cite{Starobinskil:1974nkd,brito2015superradiance}).

In this  section we  considered $\omega$ real; from now on we shall consider $\omega$ to be generally complex.


\section{Modes in near-extremal Kerr}\label{sec:QNMs&TRMs NEK}

In this section 
 we turn to QNMs (Sec.\ref{sec:QNMs NEK}) and TRMs (Sec.\ref{sec:TRMs}) in NEK and so we consider $\text{Im}(\omega)< 0$
 (we remind the reader that, in subextremal Kerr, no exponentially-unstable modes exist~\cite{whiting1989mode}, and so no poles of the Green function
 modes exist for $\text{Im}(\omega)>0$).


\subsection{QNMs in NEK}\label{sec:QNMs NEK}

In order to understand the limit to extremal Kerr,  we  describe in the first subsection the  properties of QNMs in NEK which were derived
in~\cite{detweiler1980black,PhysRevD.31.290,hod2008slow,Yang:2012pj,yang2013quasinormal,Zimmerman:2015rua,hod2013purely,Leaver:1985}.
In the second subsection, we  present our numerical calculation of these modes, showing agreement with results in the literature.
 This agreement validates our implementation
of  the MST and Nelder-Mead methods for calculating
 QNMs. 
We shall turn to QNMs in extremal Kerr in Sec.\ref{sec:QNM ext}.
We note that in that section we will find QNMs in the extremal case for mode parameters (namely, spin-2 field with $\ell=m=2$ and $3$) which the above analyses in NEK missed to find.


\subsubsection{Properties of QNMs in NEK}\label{sec:QNMs NEK props}

It was shown in~\cite{Yang:2012pj} that
 two
families of QNMs  branch off from the same family of QNMs
as $a$ approaches extremality:
zero-damping modes (ZDMs)
and damped modes (DMs).
As $a\to M$, 
the  ZDM frequencies tend  to $\oSR$, and so their imaginary part tends to zero, as their name suggests.
The imaginary part of DM frequencies, on the other hand, tends to a finite value as $a\to M$.
 We next give some properties of these two families.
 
 Let us start with the ZDMs.
These modes are present for all values of $\ell$ and $m$.
 ZDMs are associated with the near-horizon geometry.
For example, it can be shown that, in the  eikonal limit  $\ell\gg 1$, ZDMs reside on the extremum which  the   potential  of the radial equation  (obtained by suitably transforming the original Teukolsky equation to one with a real potential in the case of non-zero spin)
possesses on the event horizon for all $\ell$ and $m$.
The  frequencies of ZDMs
 have the following asymptotics
~\cite{hod2008slow,Yang:2012pj,yang2013quasinormal,hod2013purely} (which were partly based on, and corrected, an original derivation in~\cite{detweiler1980black})\footnote{Ref.\cite{yang2013quasinormal} gives a correcting term to  Eq.\eqref{eq:ZDM delta2<0} above, which, for fixed $a$, may become significant when
$\delSR^2<0$ and $|\delSR|$ is small \MC{that's what \cite{yang2013quasinormal} say above Eq.3.26 but above Eq.3.27 they say $|\del|\sim 1$?} (although this correcting term is small for $a$ sufficiently close to $M$).}
\footnote{For $m=0$, Eq.\eqref{eq:ZDM delta2<0} and~\cite{PhysRevD.60.107504} agree when taking into account that
$\delSR=(\ell+1/2)i$ in this axisymmetric case.}
\begin{equation}\label{eq:ZDM delta2<0}
\oQNM\sim 
\oSR
+2\pi \Temp \left(m-\delSR-i\, \left(n+\frac{1}{2}\right)\right),
\quad a\to M,
\end{equation}
where 
$\delSR$ is taken to be the principal square root in Eq.\eqref{eq:delSR} (so a positive number if  $\delSR^2>0$ 
and with a positive imaginary part if $\delSR^2<0$)
and $n=0,1,2,\dots$
Eq.\eqref{eq:ZDM delta2<0}  yields two slightly different behaviours for the leading order, as $a\to M$, of the
 imaginary part of the frequencies: it is ``$-2\pi \Temp(n+1/2)$" for the modes with $\delSR^2>0$, and it is even more negative
 for the modes with $\delSR^2<0$.

As $a\to M$, the temperature goes to zero and so the imaginary part of the frequencies in 
Eq.\eqref{eq:ZDM delta2<0} goes to zero.
The property that $\omega=\oSR$ is an accumulation point for the ZDM family of QNMs in the extremal limit was originally shown analytically in~\cite{detweiler1980black} 
and
corroborated numerically  later in~\cite{Leaver:1985}.
Despite it being an accumulation point for QNMs, which possess no incoming radiation (i.e.,  $\Rininc=0$),
the mode at $\omega=\oSR$ itself
does possess  incoming radiation (i.e.,  $\Rininc \neq 0$)
 in extremal Kerr.
 This can be seen from a basic conservation-of-energy argument which, in fact, also applies
 to all real-frequency non-superradiant modes
 in either extremal or sub-extremal Kerr~\cite{detweiler1973stability,teukolsky1974perturbations,casals2016horizon}.
In sub-extremal Kerr, this result has been extended to the superradiant regime: 
the only mode with $\omega\in\mathbb{R}$  and no incoming radiation   is the trivial mode~\cite{Andersson:2016epf}.
This then implies the non-existence of exponentially-growing modes in sub-extremal Kerr~\cite{Andersson:2016epf,Hartle:Wilkins:1974},
 a result  which had been previously proven in~\cite{whiting1989mode} in a different way.

The accumulation of ZDMs near the superradiant bound frequency leads to certain physical features.
Away from the horizon, it leads to a {\it temporary} power-law decay of the field 
at early times~\cite{PhysRevD.64.104021,yang2013quasinormal}, which then
gives way to the characteristic QNM exponential decay, before ending up
in a power-law decay due to the origin BC~\cite{Price:1971fb}.
Near the horizon, the accummulation of ZDMs leads to a {\it transient growth}  of the field~\cite{gralla2016transient}.
Also, this accummulation leads to a distinct observational feature in the gravitational waveform in a near-horizon inspiral~\cite{Gralla:2016qfw,Compere:2017hsi}, as we mentioned in the Introduction.

Let us now turn to DMs.
Although we are not aware of an actual proof,  the exact calculations in the literature and in this paper  seem to suggest that DMs satisfy the following properties:
\begin{itemize}
\item[(i)]
they have $|\text{Re}(\oQNM)| > |\oSR|$;
\item[(ii)] they originate from lower overtones within the family of QNMs at smaller $a$.
\end{itemize}

 Refs.~\cite{PhysRevD.31.290,yang2013quasinormal}  obtained large-$\ell$ WKB asymptotics for the values of the frequencies of  DMs.
In this  eikonal limit, it has been shown that DMs reside near the maximum of  the radial  potential 
outside the horizon, whenever such a maximum exists.
Still in the eikonal limit,
 the condition for the existence of such a maximum and, equivalently,  for the existence of DMs, is
$\mu< \mu_c$,
 where 
$ \mu\equiv |m|/(\ell+1/2)$  
 and
 $\mu_c\approx 0.74$.
For general  $\ell$ and 
$m$,
the 
 condition for the existence of  a maximum outside the horizon is
\begin{equation}\label{eq:cond DM}
\mathcal{F}_s^2\equiv \delSR^2+1/4<0.
\end{equation}
Refs.\cite{Yang:2012pj,yang2013quasinormal} thus
suggest
 that, in general, 
DMs are
 present 
 if and only if Eq.\eqref{eq:cond DM} is satisfied.
 On the other hand, Fig.3 in~\cite{Cook:2014cta} shows that for $-s=\ell=m=2$ there {\it is} a QNM with finite imaginary part, even though it satisfies  $\mathcal{F}_s^2>0$, as can be readily checked.
This QNM 
   possesses the properties  opposite to (i) and (ii) above.
Furthermore, this QNM 
is not the only one with  finite imaginary part for which Eq.\eqref{eq:cond DM} is not satisfied,
and which has  the property opposite to,
at least,
 (i) above -- 
we shall see  in Sec.~\ref{sec:QNM ext} that is the case in extremal Kerr  for $s=-2$ and $\ell=m=3$.
Therefore,  from the exact calculations in the literature and in this paper, 
it seems that there  exist QNMs with negative imaginary part always when Eq.\eqref{eq:cond DM} is satisfied, and also, at least in some cases, when Eq.\eqref{eq:cond DM} is not
satisfied. 
In this paper,  we shall continue to refer to the former ones (i.e., those that necessarily exist when $\mathcal{F}_s^2<0$)  as DMs and we shall refer to the latter ones  (i.e., those which may exist when $\mathcal{F}_s^2>0$) as 
``\nsDM\ s" (\nsDMacr\ )\footnote{We note that it has been argued in~\cite{Hod:2015swa,Hod:2016aoe} that DMs  may exist even for $\mathcal{F}_s^2<0$: see Eq.\eqref{eq:Hod freq}.
However, a search for these specific suggested modes was carried out in~\cite{Zimmerman:2015rua,Richartz:2015saa} and they were not found.
In  App.\ref{sec:Hod} we report a similar negative search for such modes. We also note that the $s=-2$ QNMs for $\ell=m=2$ and $3$ would lie to the left of the BC, whereas 
the suggested modes in Eq.\eqref{eq:Hod freq} lie to the right of the BC.} \MC{better idea for a name? eg, might be confusing that later we also talk about ``standard QNMs"}.
It also seems 
(although we are not aware of an actual proof)
that DMs satisfy properties (i) and (ii) above whereas \nsDMacr\ s satisfy the opposite properties.

We note that
for $m<0$ and $\text{Re}(\omega)>0$
 -- or, equivalently, for $m>0$ and $\text{Re}(\omega)<0$, by virtue
of the symmetry \eqref{eq:symms nu_c} --
there exist the 
standard QNMs, with a finite imaginary part but without being ``partnered up with"
ZDMs (i.e., these standard QNMs and the ZDMs did not branch off from the same family of modes at smaller $a$).

We finish this subsection by relating various conditions which we have mentioned.
 Firstly, it has been checked numerically that 
 $\delSR^2<0 \leftrightarrow \mathcal{F}_s^2<0$.
 This has been checked  for a large set of modes
 with $s=0$ and ``$-2$" in~\cite{Yang:2012pj,yang2013quasinormal} and for a set of modes with $|s|=0$, $1$ and $2$ by us.
 Secondly,~\cite{Yang:2012pj,yang2013quasinormal} show that 
the equivalence
 $\mu<  \mu_c\leftrightarrow \mathcal{F}_s^2<0$
  holds
 for most
 modes that they checked with $s=0$  and ``$-2$" but does {\it not} hold for a few modes.
Similarly, we checked that, for $|s|=1$, 
the equivalence
$\mu<  \mu_c\leftrightarrow \mathcal{F}_s^2<0$
holds
for most  modes but
does not hold
for some modes (such as $\ell=m=1$ and
$\ell=13$, $m=10$).
Finally, we note that $\mathcal{F}_s^2<0$ is equivalent to
$\nu_c(\nu_c+1)>0$.
That is, DMs exist
if and only if the eigenvalue of the Casimir operator of the  $\mathfrak{sl}(2,\mathbb{R})$ in NHEK is positive
(although \nsDMacr\ s may exist even if it is non-positive).




\subsubsection{Calculation of QNMs in NEK}\label{sec:calc QNMs NEK}

\begin{figure}[hp!]
   \begin{center}
   \subfloat[$s=-2,\ell=2,m=0,a=0.998M$]{
\label{figNEKa}
   \includegraphics[width=.4\textwidth]{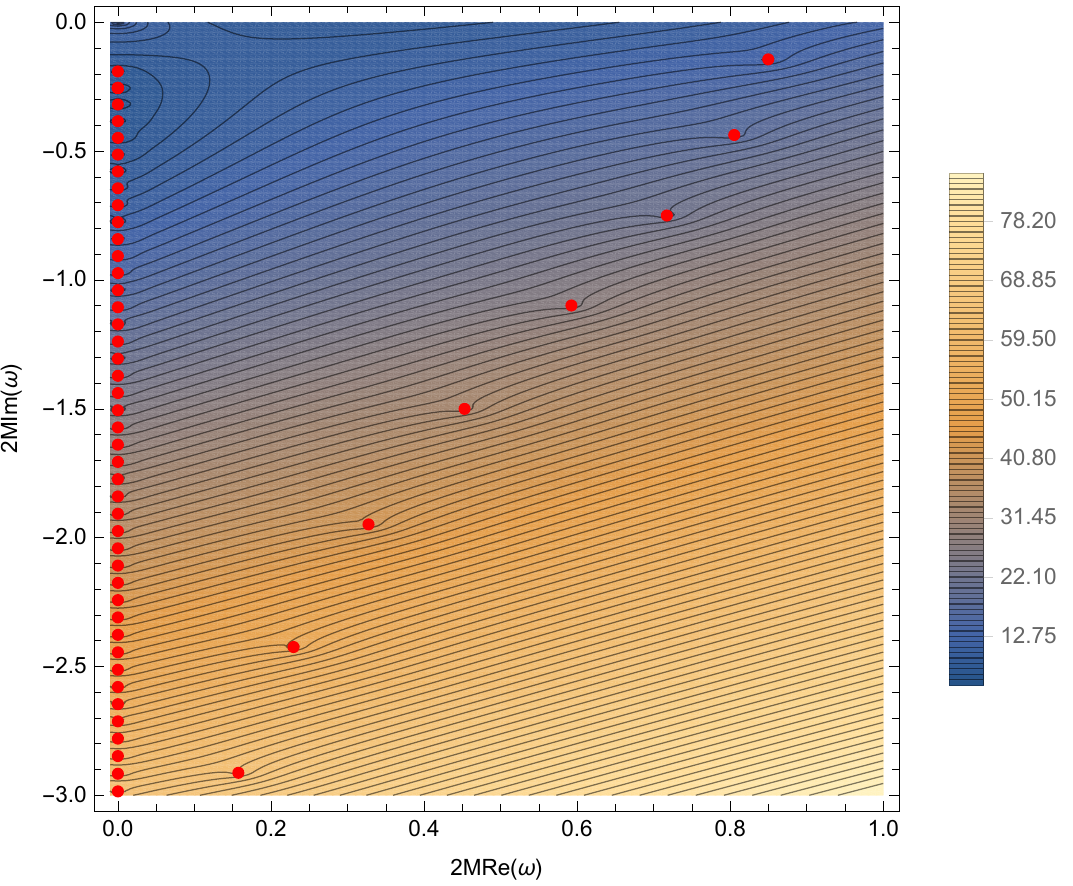}} \quad   
  \subfloat[$s=-2,\ell=2,m=1,a=0.998M$]{
\label{figNEKc}
   \includegraphics[width=.4\textwidth]{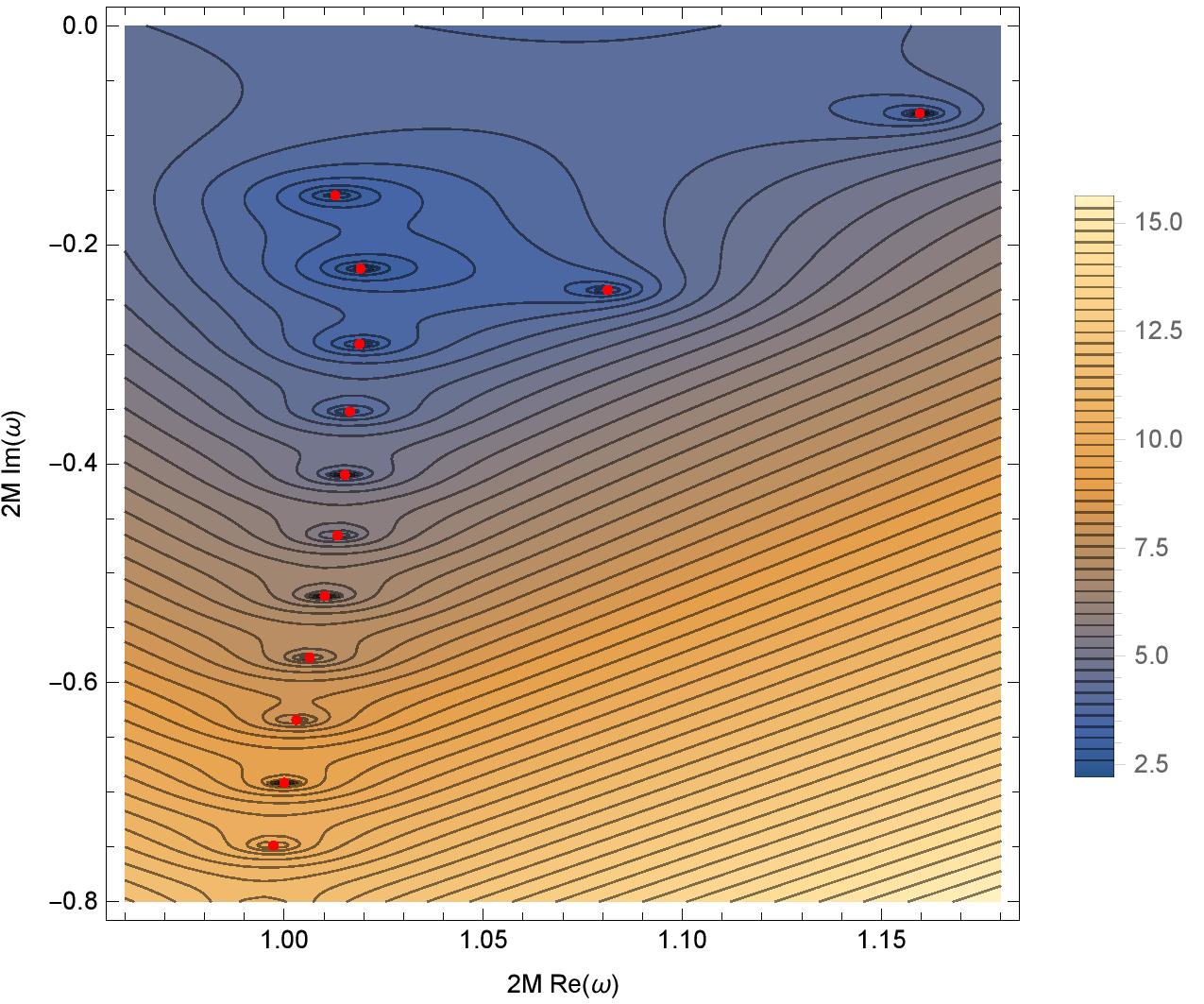}} \quad
  \subfloat[$s=-2,\ell=2,m=,a=0.9999M$]{
\label{figNEKd}
   \includegraphics[width=.4\textwidth]{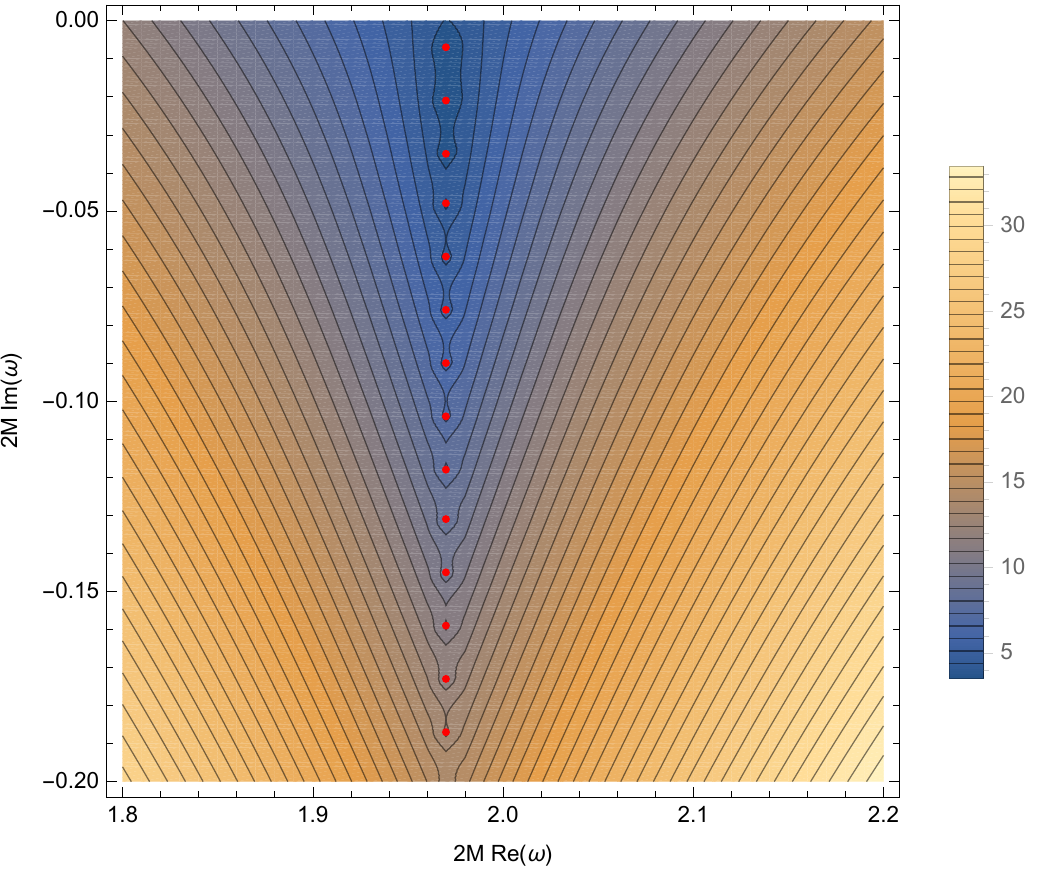}}
   \end{center}
\caption{
Contour plots of $\log_{10}|\Wf|$ in NEK  for $s=-2$, $\ell=2$ and $m=0,1,2$. The superradiant bound frequencies for these modes are 
$2M\oSR=0$ (for $m=0$), $\approx 0.938$ (for $m=1$) and $\approx 1.972$ (for $m=2$).
The QNMs
 in Tables \ref{table:QNMNEKsm2l2m2}-- \ref{table:QNMNEKsm2l2m1} are indicated with red dots\MC{but there's also extra ones here?}.
 N.B.: the top two plots 
  correspond
to Fig.7 in~\cite{yang2013quasinormal}.
}
\label{figQNMNEK}
\end{figure}

Let us now turn to our calculation of QNMs
in NEK.
The contourplots
of the Wronskian factor \eqref{eq:Wronsk fac} (to be precise, of $\log_{10}|\Wf|$)
 in Fig.\ref{figQNMNEK} shows the accumulation of ZDMs near $\omega = \oSR$.
 The only  
 case
 with  $\mathcal{F}_s^2>0$ in that figure
 is that in 
 Fig.\ref{figQNMNEK}(c).
 In this case,
 within the analyzed region of the complex plane, we found no DMs
-- as expected --
  and no \nsDMacr\ s.
The top 
two
 plots correspond to plots in Fig.7 in~\cite{yang2013quasinormal} (Fig.\ref{figQNMNEK}(c)
has a slightly larger value of $a$ than a plot
 in Fig.7 in~\cite{yang2013quasinormal}).

We note that Fig.7 in~\cite{yang2013quasinormal} shows contour plots of Leaver's continued fraction.
Apart from the QNMs, their plots show poles in the continued fraction, which can be ``very close" to the QNMs. In our Fig.\ref{figQNMNEK}, on the other hand, we plot the Wronskian factor $\Wf$
and it does not display poles near the QNM frequencies. This is due to having defined $\Wf$ in Eq.\eqref{eq:Wronsk fac}  by
 removing a $\Gamma$-factor in $\W$  which, as it seems, contains its poles --  see Secs.\ref{sec:subext} and \ref{sec:TRMs}. This can be seen as an advantage of our approach, since the poles 
 in Leaver's continued fraction
 become arbitrarily close to $\oQNM$, as can be observed at the bottom left panel of Fig.7~\cite{yang2013quasinormal}. The presence of these poles in Leaver's continued fraction may lead to numerical issues when approaching the extremal limit.

In Appendix \ref{sec: app QNMs NEK} we tabulate various QNM frequencies in NEK.
We calculated these QNMs  to 16 digits of precision using the  method described in Sec.\ref{sec:method}.
Fig.~\ref{figQNMNEK} shows these NEK QNMs  \MC{Luis: all of them?}.
Our NEK QNM values
for $s=-2$, $\ell=2$
 agree with those in~\cite{QNMBertiNew} to the following 
 typical
  number of digits of precision:
7
(worst was 4 digits) for $m=2$; 
10
 (worst was 7 digits) for $m=1$; 
5
(worst was 3 digits) for  $m=0$.
As stated in~\cite{QNMBertiNew}, however, the QNM values contained there are unreliable in NEK (``roughly, when $a/M\ge 0.999$"), where no error bars are given.
We note that, for the cases that  we calculated the QNMs for, Ref.~\cite{QNMBertiNew}  seems to provide values only for DMs\footnote{Refs.~\cite{Yang:2012pj,yang2013quasinormal}, on the other hand, do include ZDMs in the figures for all cases, although their values are not tabulated there.}. The exception to that is the case where no DMs exist,
which is that in Table \ref{table:QNMNEKsm2l2m2}, for which~\cite{QNMBertiNew} does provide the ZDM values. In our App.\ref{sec: app QNMs NEK}, on the other hand, we provide  DMs as well as ZDMs for all cases where they exist.

For $m=0$ we noticed that the values of the real part of the ZDM frequencies were below the precision we used in the calculation every time as we increased that precision (to even 32 digits).
We thus do not provide these values in the $m=0$ Table \ref{table:QNMNEKsm2l2m0}.
\MC{Is the real for $m=0$ part meant to be {\it exactly} zero? Compare our values for $m=0$ QNMs against \eqref{eq:ZDM delta2<0} and against Eq.12 in~\cite{PhysRevD.60.107504} and see which one we agree better with.
Why does~\cite{PhysRevD.60.107504} say ``we do not  find QNMs on or near the NIA in the vicinity of the numerical results reported by Yang et al" when Yang et al. do not give any numerical values?}

\subsection{TRMs in NEK}\label{sec:TRMs}

Apart from the QNMs, there is another interesting set of modes worth considering.
Totally reflected modes (TRMs) correspond, physically,
to waves with no transmission and, mathematically, 
 to a singularity in both  $\RinincS$ and $\RinrefS$ while $\RinincS/\RinrefS$ is finite\MC{check definition}.
Therefore,  the Wronskian is singular at a TRM frequency.

Ref.\cite{Keshet:2007be} 
derived  the following 
exact expression for
TRM frequencies in sub-extremal Kerr
\begin{equation}\label{TRM}
\omega =
\oSR
- 2\pi i \Temp(n-s+1),
\end{equation}
\MCImp{So in extremal Kerr ($\Temp=0$) there's only one TRM?
$k=0$ being a TRM agrees with \eqref{eq:W k->0,ext} and the numerics for some cases but not all of them? perhaps \eqref{TRM} is just not valid in extremal Kerr?... }
where $n = 0,1,2,\dots$
These frequencies coincide with the poles
 of the factor 
$\Gamma(1-s-2i\epsilon_{+})$ in Eq.\eqref{eq:Wronsk fac}, which we removed from the Wronskian precisely with the intention 
of avoiding its poles.
Note that the difference between the TRMs and the QNMs, which asympote as in 
\eqref{eq:ZDM delta2<0},
is only to higher order in the asymptotics  (for the specific case of $m=0$ in NEK, this had already been noticed  in~\cite{hod2013purely}).
\MC{From \eqref{eq:ZDM delta2<0} and \eqref{TRM} it seems that QNMs and TRMs don't quite pile up together when $\delSR^2<0$?}
 Fig.\ref{figBCextreme} shows that the TRMs and QNMs are coming closer together as we increase the value of $a$. When $a=M$ an infinite amount of both types of modes  accumulate and mix in such a way that a  finite discontinuity (BC) is formed. 
 We investigate this feature in the next section.

\MC{Luis was saying something like: Richartz says $\omega=m$ is a TRM (check this) and we're seeing that numerically (ie, we see a spike there). If this makes sense, let's mention it}

\MC{Can we calculate the TTM's (ie, $Bref/Binc=0$)? eg, see numerically from which factor in Eqs.4.30,4.31 thesis do the TTMs come from?
Keshet and Neitzke'08 only give WKB for TTM's in their Eq.54; but at the end of their Sec.II they say that TTMs pile up to $\omega=\oSR$ in the extremal limit...}



\section{Branch cuts}\label{sec:BCs}

In this section we first investigate the presence of  BCs in the complex frequency plane and afterwards
 the formation of the superradiant BC.


\subsection{Search for branch cuts}\label{sec:search BCs}

In order to start the investigation of the presence of BCs we plot both the absolute value and the phase of the Wronskian when taking a  loop around a certain frequency.
That is, we calculated the Wronskian for $\omega=\omega_0+Re^{i\phi}$,
given some frequency $\omega_0\in\mathbb{C}$, radius $R>0$ away from it, and varying the phase  $\phi: 0\to 2\pi$.
The discontinuity of the Wronskian at some $\phi$ would indicate a BC, possibly stemming from $\omega_0$ (branch point).

\begin{figure}[hp!]
   \begin{center}
   {
\label{figbcWabs}
   \includegraphics[width=.45\textwidth]{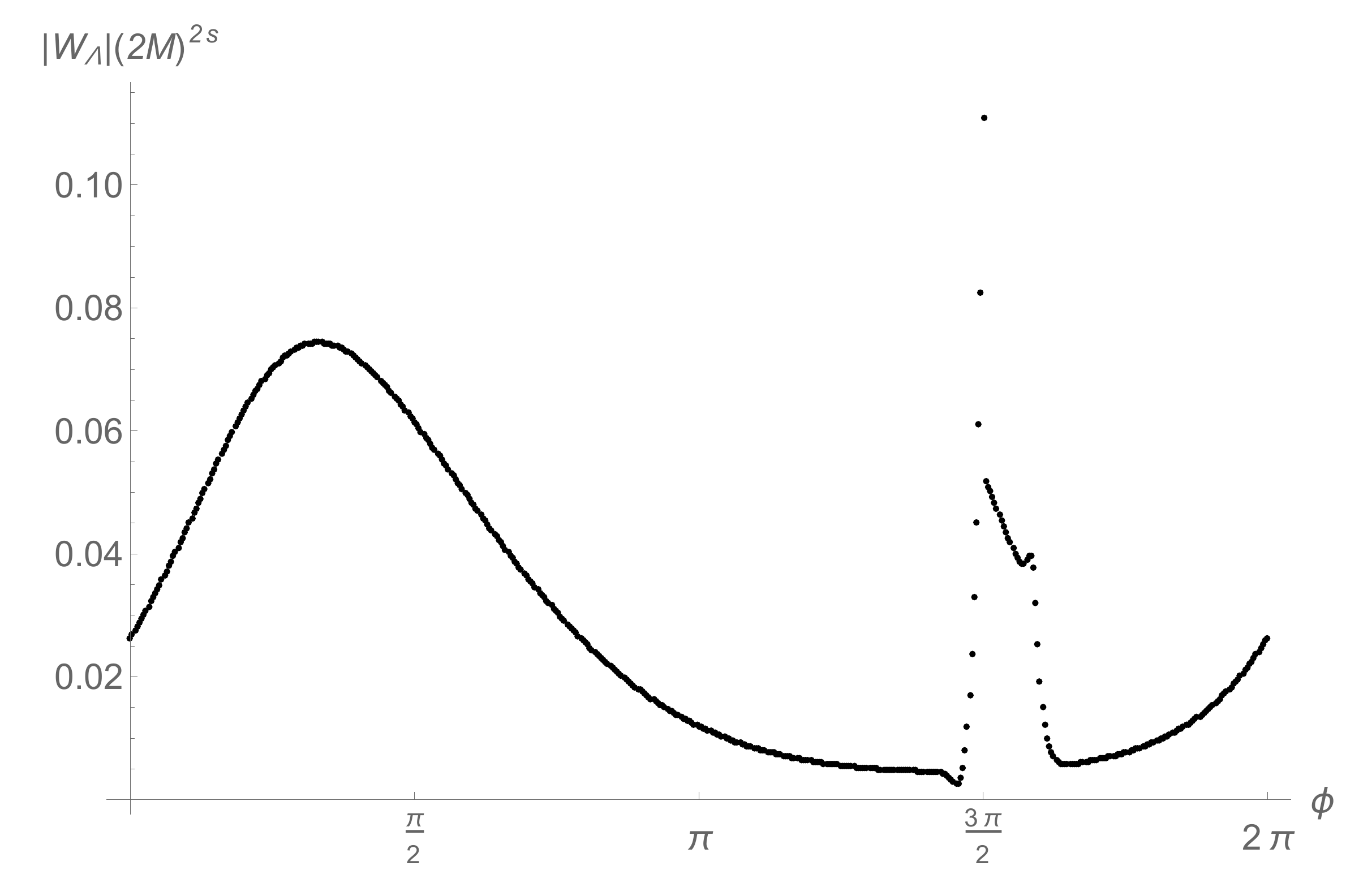}} \quad
{
\label{figbcWarg}
   \includegraphics[width=.45\textwidth]{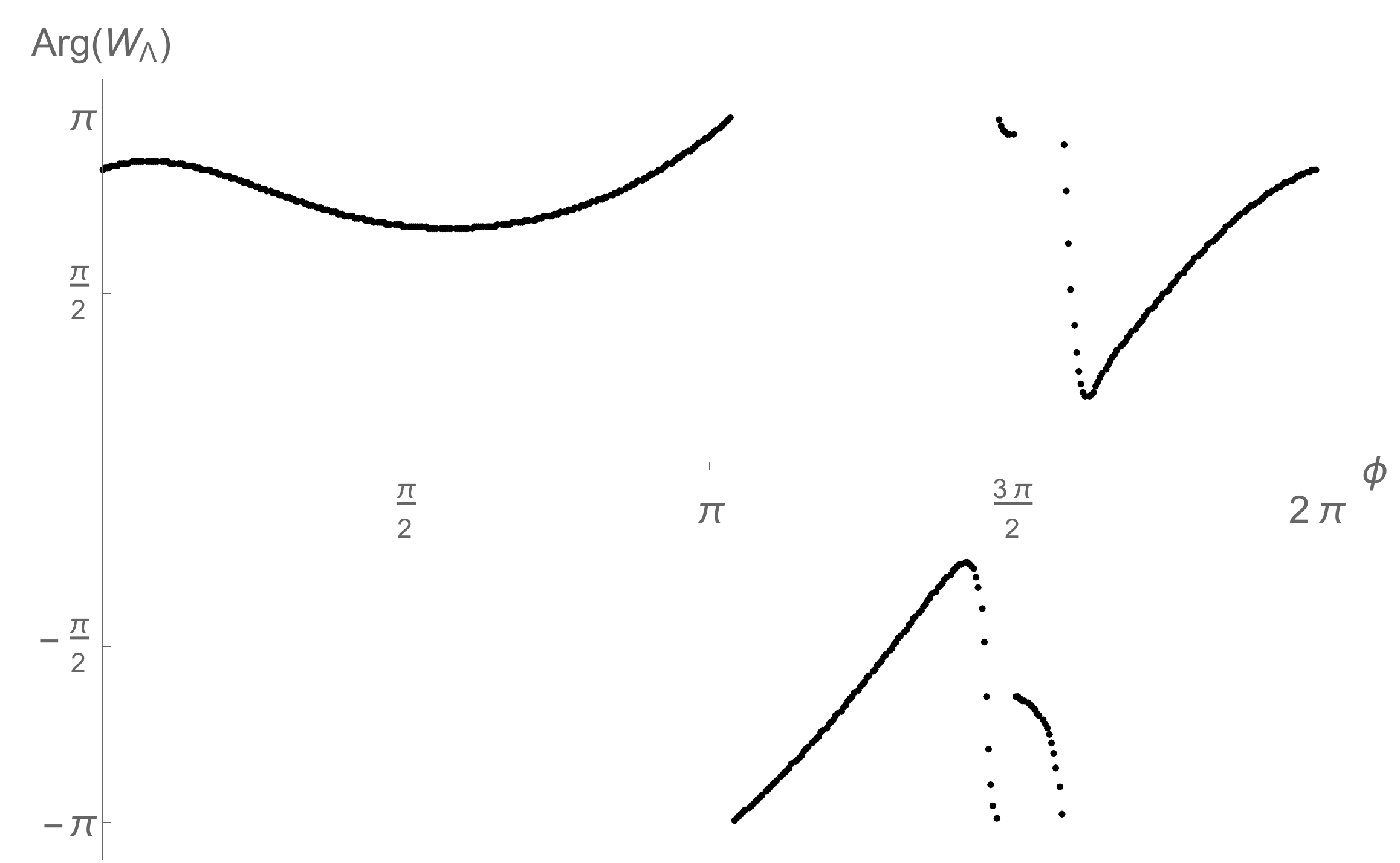}}
   \end{center}
\caption{
\MC{mention that axes labels are really  $|\Wsub|$ and  $\arg(\Wsub)$}
Plot of the absolute value (top) and argument (bottom) of
 $\Wsub$  for $s=\ell=m=2$ and $a=0.5M$
 on a loop 
$\omega=e^{i\phi}/M$ around the origin, which encircles $M\oSR\approx 0.2679$ and $M\omega_{HW}\approx 0.2679+0.2320i$.
There is a discontinuity near (or at) $\phi=3\pi/2$,
 corresponding to the well-known BC down from the origin.
 A steep structure is seen in the fourth quadrant, which is related to the formation of the superradiant BC in the extremal case. 
 }
 \label{figbcW}
\end{figure}

We performed such loops in many instances.
For example, in Fig.\ref{figbcW} we plot the Wronskian for $s=\ell=m=2$ in subextremal Kerr when going around the origin: 
$\omega_0=0$ and $R=1/M=2$.
For this mode, the superradiant bound frequency is $M\oSR\approx 0.2679$ and the particular Hartle-Wilkins frequency (see App.\ref{sec:HW}) is $M\omega_{HW}\approx 0.2679+0.2320i$.
Therefore both of these frequencies
lie inside the circle that we take around the origin.
The only  discontinuity that we find is near (or at) the  phase $\phi = 3\pi/2$. This discontinuity 
corresponds to the well-known BC  from $\omega=0$ down the negative imaginary axis
 ~\cite{Leaver:1986,PhysRevLett.84.10,PhysRevD.94.124053}
\footnote{We note that we did find discontinuities in the MST coefficients $a_{n}$ which do not correspond to the BCs from 
 $\omega=0$ or  $\omega=\oSR$
  -- see~\cite{MSc-Longo}. However, these discontinuities can be traced back to  discontinuities of $\nu$
  which can be ruled out on ``physical" quantities such as the Wronskian, on account of the symmetries of the MST equations.
  This is corroborated by our plots of the Wronskian.}. 
We also note that somewhere in the fourth quadrant
there is a  steep structure 
(though not an actual discontinuity),
which will result in the superradiant BC in the extremal limit.
We also did similar plots of the Wronskian around the origin for other modes and we found the same qualitative features:
(i) there is a BC down from $\omega=0$; (ii) there is an indication that an extra BC   from $\oSR$ is forming as $a$ approaches $M$.

 
We carried out a similar  search for BCs in the extremal case\MC{ie, going around the origin and enclosing $\omega_{HW}=\oSR$ also for  $s=\ell=m=2$?}.
We found BCs down from $\omega=0$ and $\omega=\oSR$ and no other BCs.
We anticipated the existence of the BC down from $\omega=\oSR$ in Sec.\ref{sec:perts}.
In the next subsection we show how this BC  is formed.


\subsection{Formation of superradiant BC}\label{sec:superrad BC}

Using asymptotics for the QNMs as $a\to M$, indications were found in~\cite{detweiler1980black,PhysRevD.64.104021} that the ZDMs
 in  NEK accumulate towards the superradiant bound frequency so as to try to form a new BC down from $\omega=\oSR$. 
We already saw this  accumulation of ZDMs in Fig.\ref{figQNMNEK}. 

\begin{figure}[hp!]
   \begin{center}
   \subfloat[$s=0,\ell=3,m=2,a=0.95M$]{
\label{figBCa095}
   \includegraphics[width=0.47\textwidth]{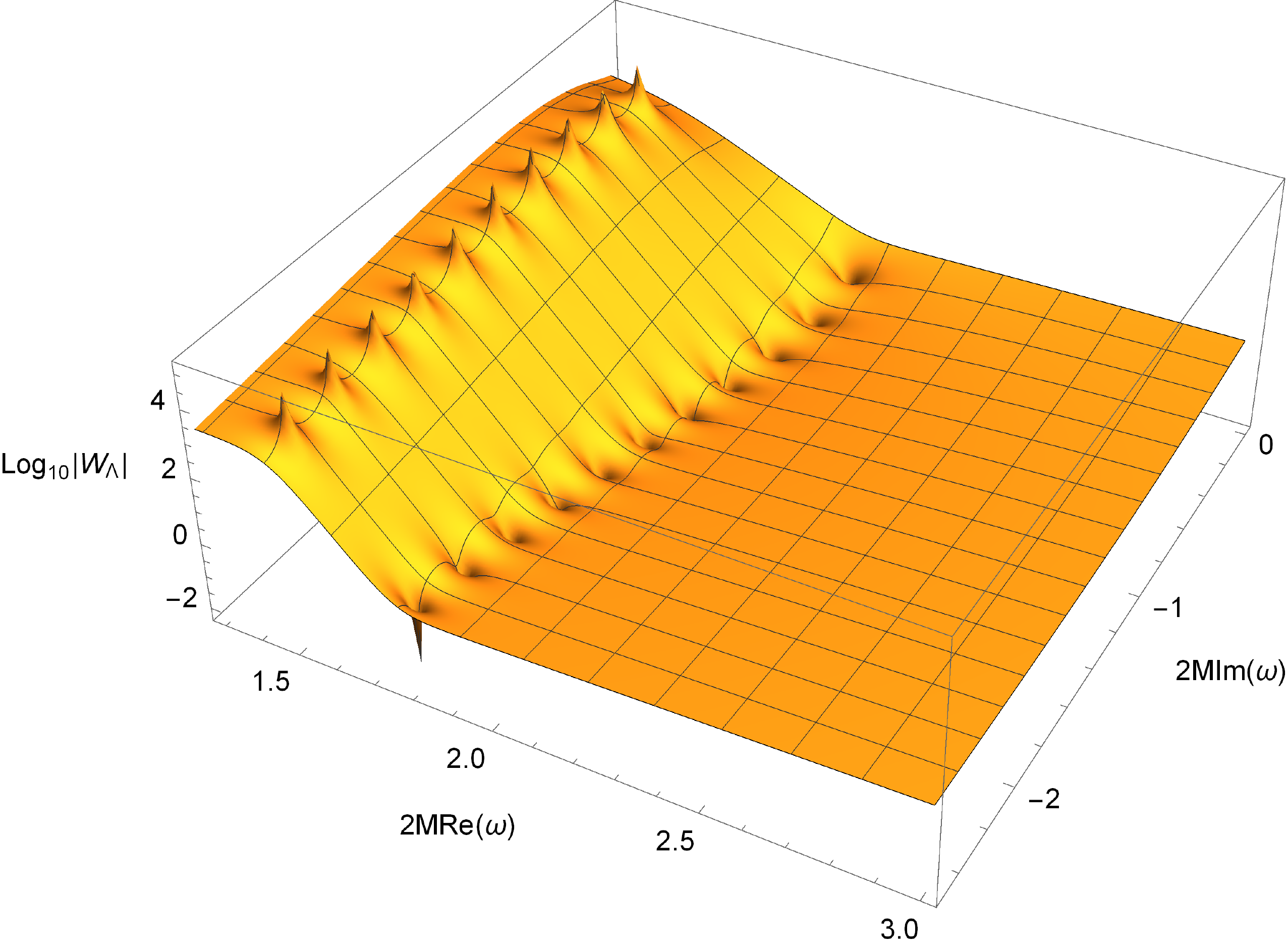}} \quad
   \subfloat[$s=0,\ell=3,m=2,a=0.99M$]{
\label{figBCa099}
   \includegraphics[width=.47\textwidth]{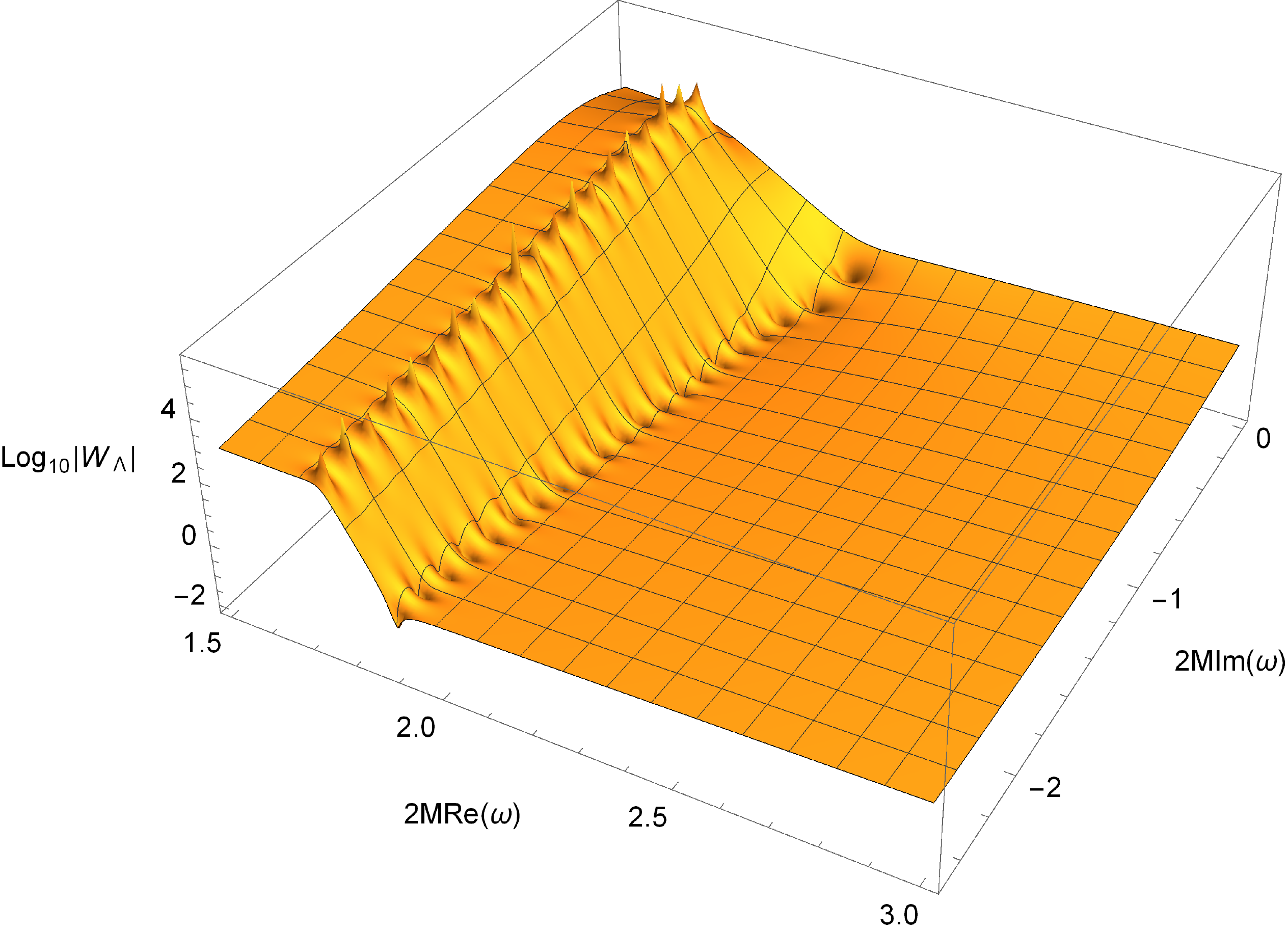}} \quad
   \subfloat[$s=0,\ell=3,m=2,a=M$]{
\label{figBCa1}
   \includegraphics[width=.47\textwidth]{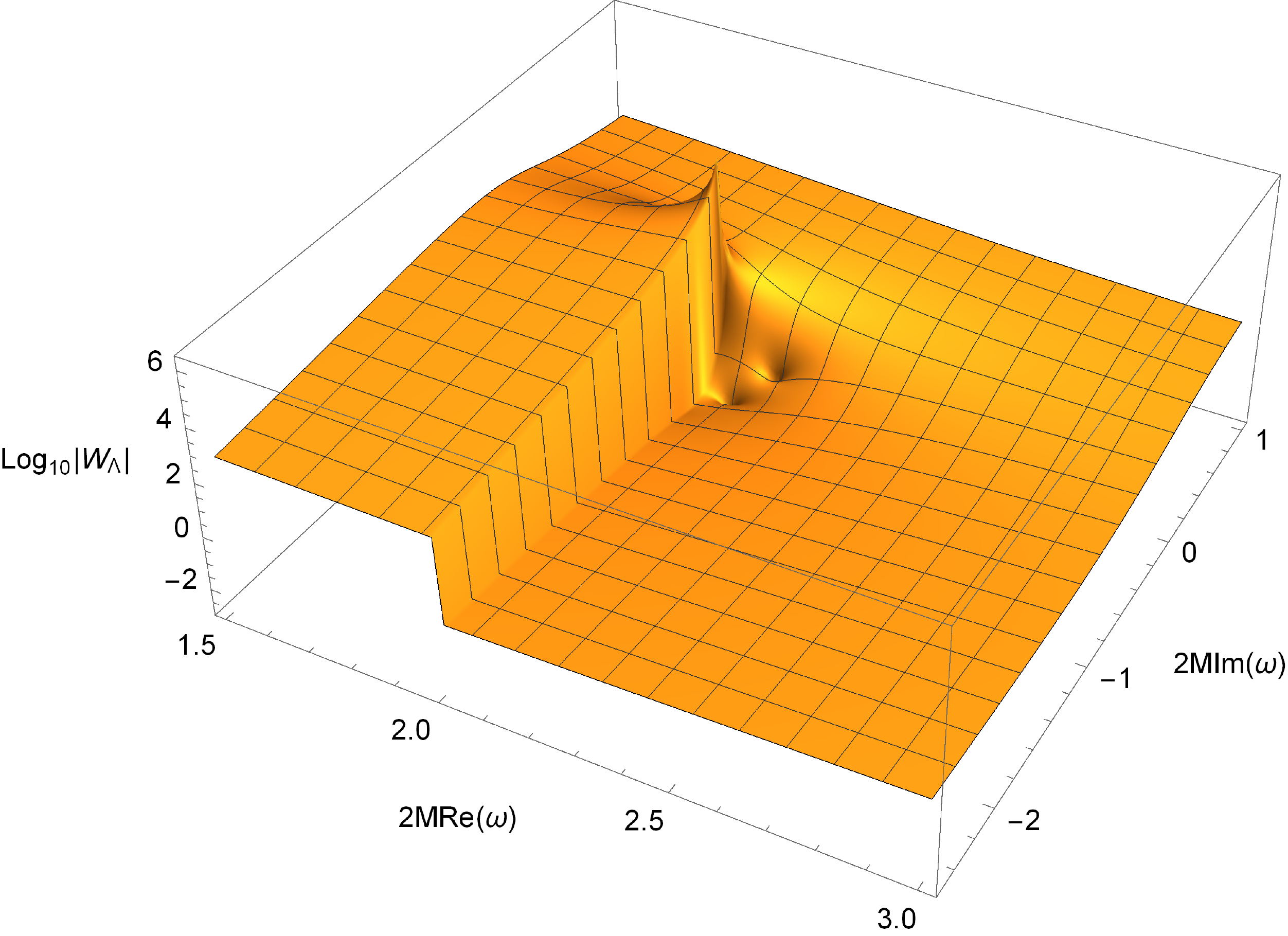}}
   \end{center}
\caption{
\MC{Maybe include the asymptotics of \eqref{TRM} (and of \eqref{eq:ZDM delta2<0}), but maybe better in a contour plot...?}
Plots of $\log_{10}\left|\Wsub\right|$ for $a=0.95M$ and $0.99M$ and of $\log_{10}\left|\W\right|$ (so for $a=M$). 
The mode is
 $s=0,\ell=3,m=2$, for which it is $2M\oSR\approx$ 1.447 (for $a=0.95M$), $\approx$1.735 (for $a=0.99M$), =2 (for $a=M$).}
\label{figBCextreme}
\end{figure}

The above indication of formation of a superradiant BC is illustrated more clearly in the 3D plots of the absolute value of the Wronskian in Fig.\ref{figBCextreme}, where we increase $a=0.95M \to 0.99M \to M$.
Fig.\ref{figBCextreme} clearly shows how, in NEK, a series of TRMs  (i.e., poles of the Wronskian) appear near a series of ZDMs (which are QNMs, i.e., zeros of the Wronskian), thus yielding a steep structure in the numerically-calculated Wronskian. 
We already spotted this  steep structure
in the fourth quadrant in Fig.~\ref{figbcW}\MC{Explain further}. 
As $a$ increases, the series of TRMs and the series of ZDMs are seen to approach each other, in agreement with Eqs.\eqref{eq:ZDM delta2<0}
and \eqref{TRM}.
This approach ends up yielding a BC discontinuity stemming from the (branch) point $\omega = \oSR$ in the actual limit $a=M$.

This superradiant BC in extremal Kerr can be seen in Figs.\ref{figQNMextremes0},  
\ref{figQNMsm2l2m0aM} and \ref{figQNMsm2extreml3},
 which contain a contourplot version of Fig.\ref{figBCa1} as well as similar contourplots  of the absolute value of the Wronskian for other modes.
 The superradiant BC is also manifest in the
phase of the Wronskian in extremal Kerr as shown in Fig.\ref{fig:phase W} for a sample of modes.
The BC is clear in Fig.\ref{fig:phase s0l3m2} for $m=2$; in Fig.\ref{fig:phase sm2l2m0} for $m=0$ the BC can be readily inferred from the symmetry \eqref{eq:symm W ext}\footnote{\label{ft:BC m=0}For $m=0$, due to the symmetry \eqref{eq:symm W ext}, the BC is only in the phase of the Wronskian, not in its absolute value -- this is similar to what happens 
to all modes in Schwarzschild space-time~\cite{Casals:Ottewill:2015}.}; 
Fig.\ref{fig:phase sm2l2mm1} for $m=-1$ has no BC for $\text{Re}(\omega)>0$ but we include it for completeness.
With red dots we indicate the QNM frequencies. One may see that the variation of the values of the phase along a loop around a QNM is precisely that corresponding to a simple zero of $\W$, and 
so to a simple pole of $1/\W$, ie, a QNM.
We investigate poles in extremal Kerr in the next section.

\section{QNMs and search for unstable modes in extremal Kerr}\label{sec:QNM&unstable ext}

In this section we investigate poles of the Green function modes (i.e., zeros of the Wronskian) in extremal Kerr.
These poles may correspond to either QNMs (if they lie in the  lower complex-frequency half-plane) or to exponentially-unstable
modes (if they lie in the upper half-plane).


\subsection{QNMs in extremal Kerr}\label{sec:QNM ext}

We here turn to the QNMs in extremal Kerr.
The condition in Eq.\eqref{eq:cond DM}  (which was given for NEK but carries over to extremal Kerr) differentiates between two different regimes:
presence of DMs when $\mathcal{F}_s^2<0$; 
absence of DMs when $\mathcal{F}_s^2>0$.
As indicated in Sec.\ref{sec:QNMs NEK props}, however, there may exist \nsDMacr\ s when  $\mathcal{F}_s^2>0$.

\begin{figure}[hp!]
   \begin{center}
   \subfloat[$s=0,\ell=3,m=2,a=M$]{
\label{figQNMs0l3m2}
   \includegraphics[width=.5\textwidth]{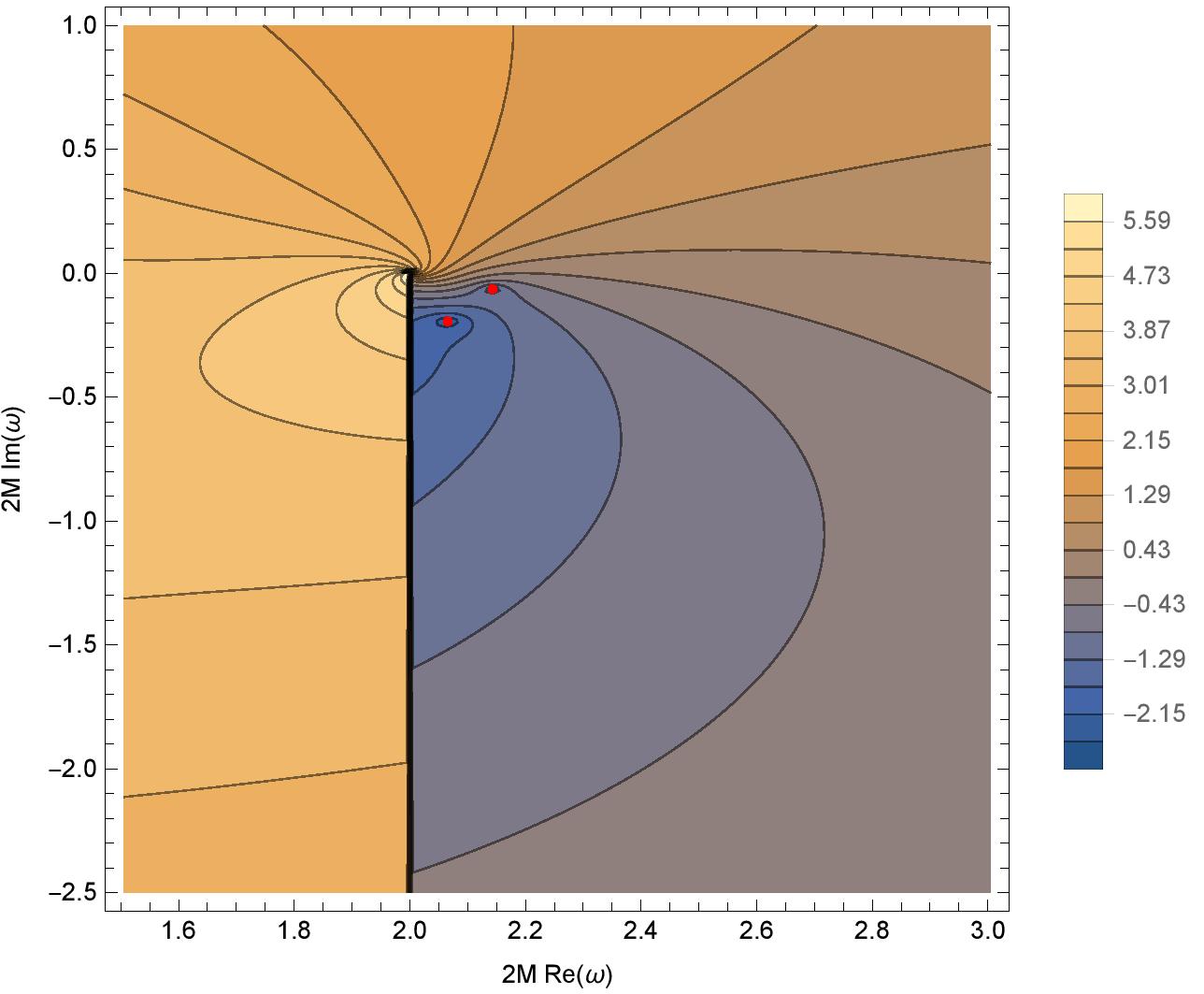} } \quad
      \subfloat[$s=\ell=m=1$, $a=M$]{
      \label{figQNMs1l1m1}
   \includegraphics[width=.5\textwidth]{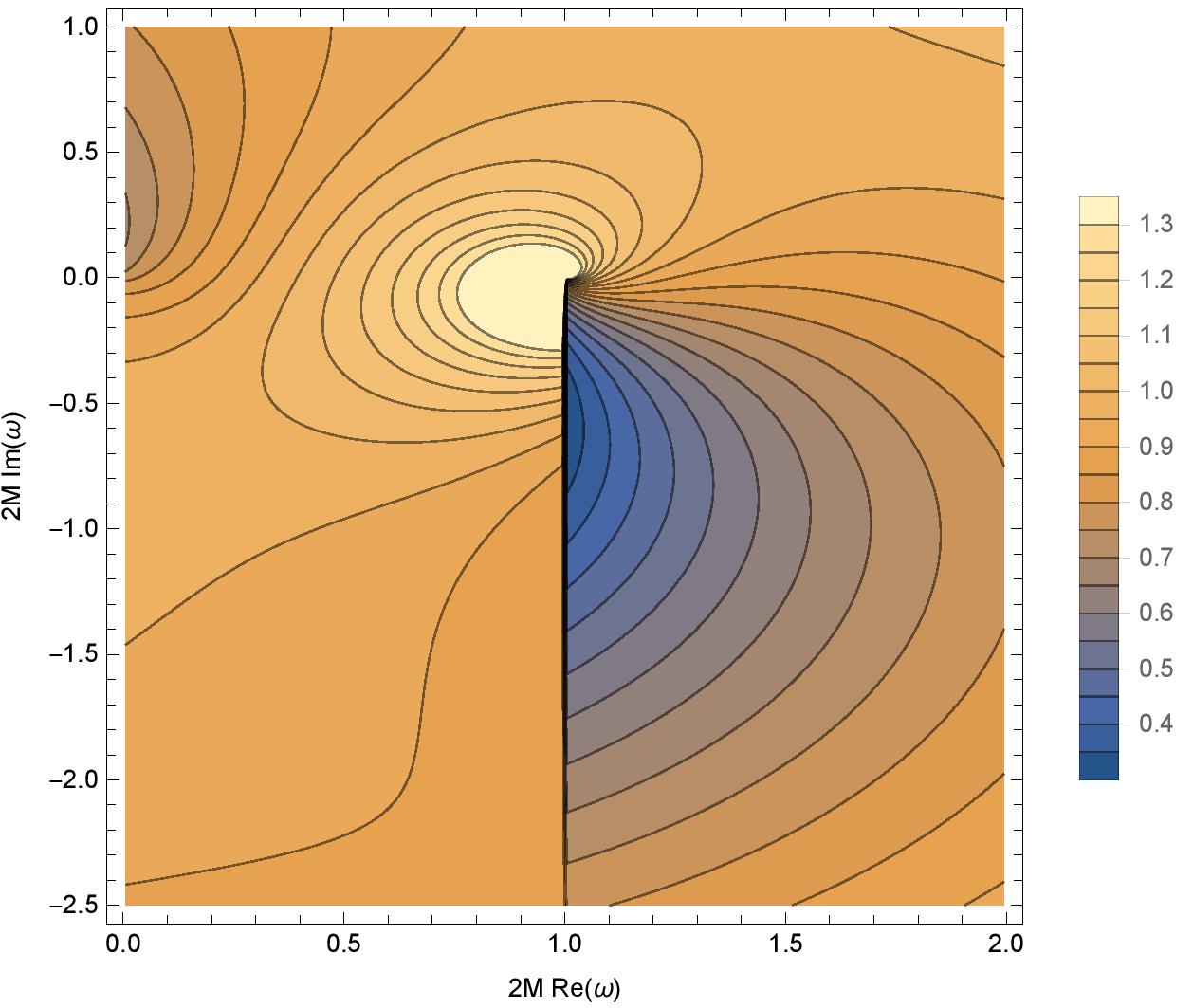}}
   \end{center}
\caption{
Contour plots of $\log_{10}\left|\left(2M\right)^{2s}\W\right|$ in the complex-frequency plane in extremal Kerr.  The modes are: (a) $s=0,\ell=3,m=2$, for which $2M\oSR=2$ and $\mathcal{F}_s^2<0$, in agreement with the presence of DMs which we observe (two to the right of the BC),
 and (b) $s=\ell=m=1$, for which $2M\oSR=1$  and $\mathcal{F}_s^2>0$, in agreement with the absence of DMs which we observe.
The QNMs 
in Table \ref{table:QNMs0l3m2a1} are indicated with red dots.
}
\label{figQNMextremes0}
\end{figure}

In Figs.\ref{figQNMextremes0}, 
\ref{figQNMsm2l2m0aM},
\ref{figQNMsm2extreml3} and 
\ref{fig:QNM s=-2,l=2,m=0,1,a=M,3D}
 we 
  plot the absolute value of the full Wronskian\footnote{There is no need to plot a Wronskian factor -- instead of the full Wronskian -- because here the TRMs (which complicated the numerics in NEK) are absent  (see Eq.\eqref{TRM}).}  (\ref{eqextwronskian})   in the complex-frequency plane
  for $m\geq 0$.
In 
Fig.\ref{figQNMs0l3m2}
  we can see, for $s=0$ and $\ell=m=2$, for which $\mathcal{F}_s^2<0$, two isolated 
QNMs frequencies appearing to the right of the BC, which correspond to DMs.
In 
Fig.\ref{figQNMs1l1m1}
  for $s=\ell=m=1$, for which $\mathcal{F}_s^2>0$, there are no DMs
\footnote{In this case, the condition 
$\mu< \mu_c$
for presence of DMs does not work, but that is not a problem since this condition is in principle only valid in the eikonal limit.}
nor, within the analyzed region, \nsDMacr\ s. 

\begin{figure}[hp!]
   \begin{center}
   \subfloat[$s=0,\ell=3,m=2,a=M$]
   {
\label{fig:phase s0l3m2}
      \includegraphics[width=.5\textwidth]{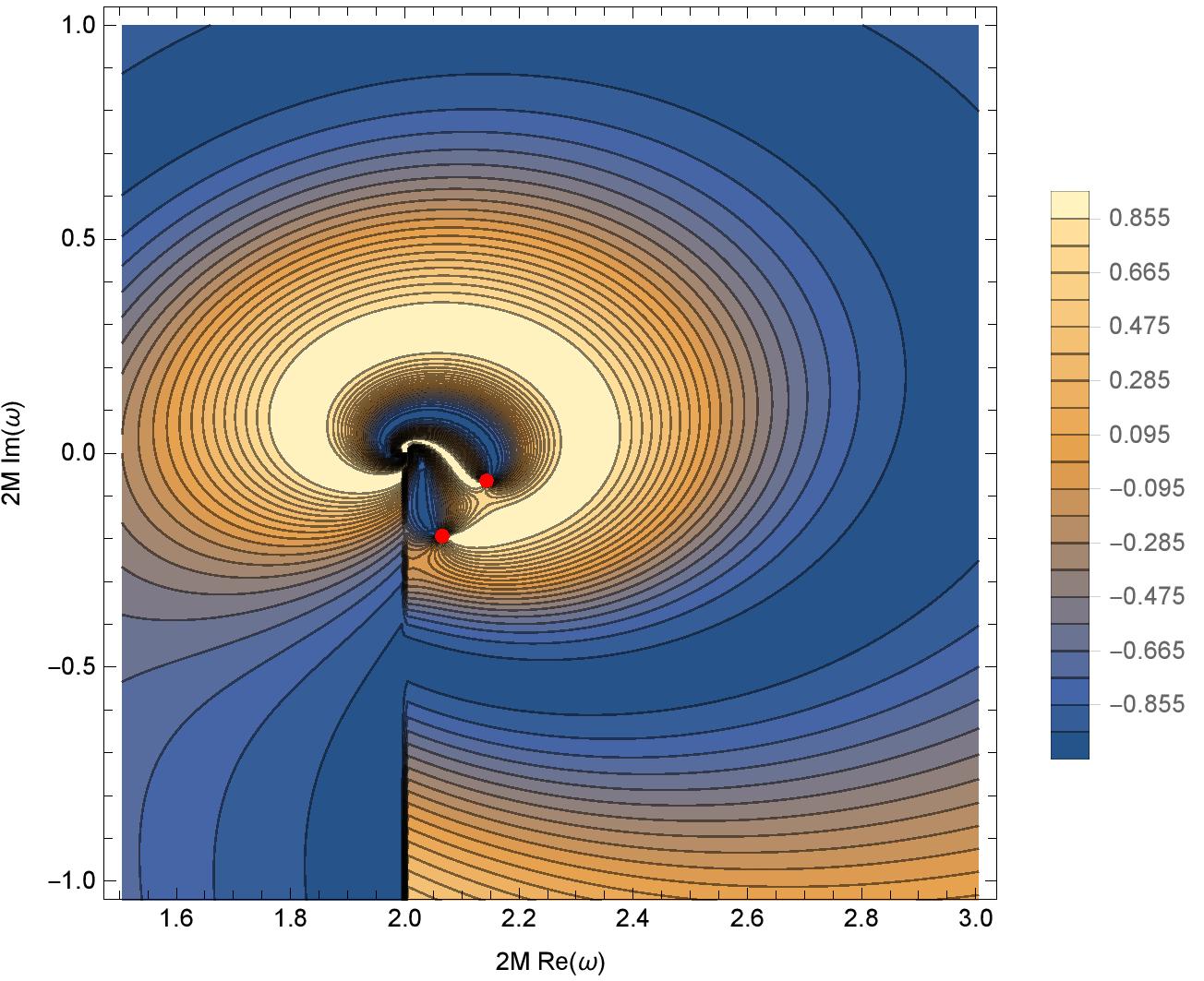}}\quad
   \subfloat[$s=-2,\ell=2,m=0,a=M$]
   {
\label{fig:phase sm2l2m0}
     \includegraphics[width=.5\textwidth]{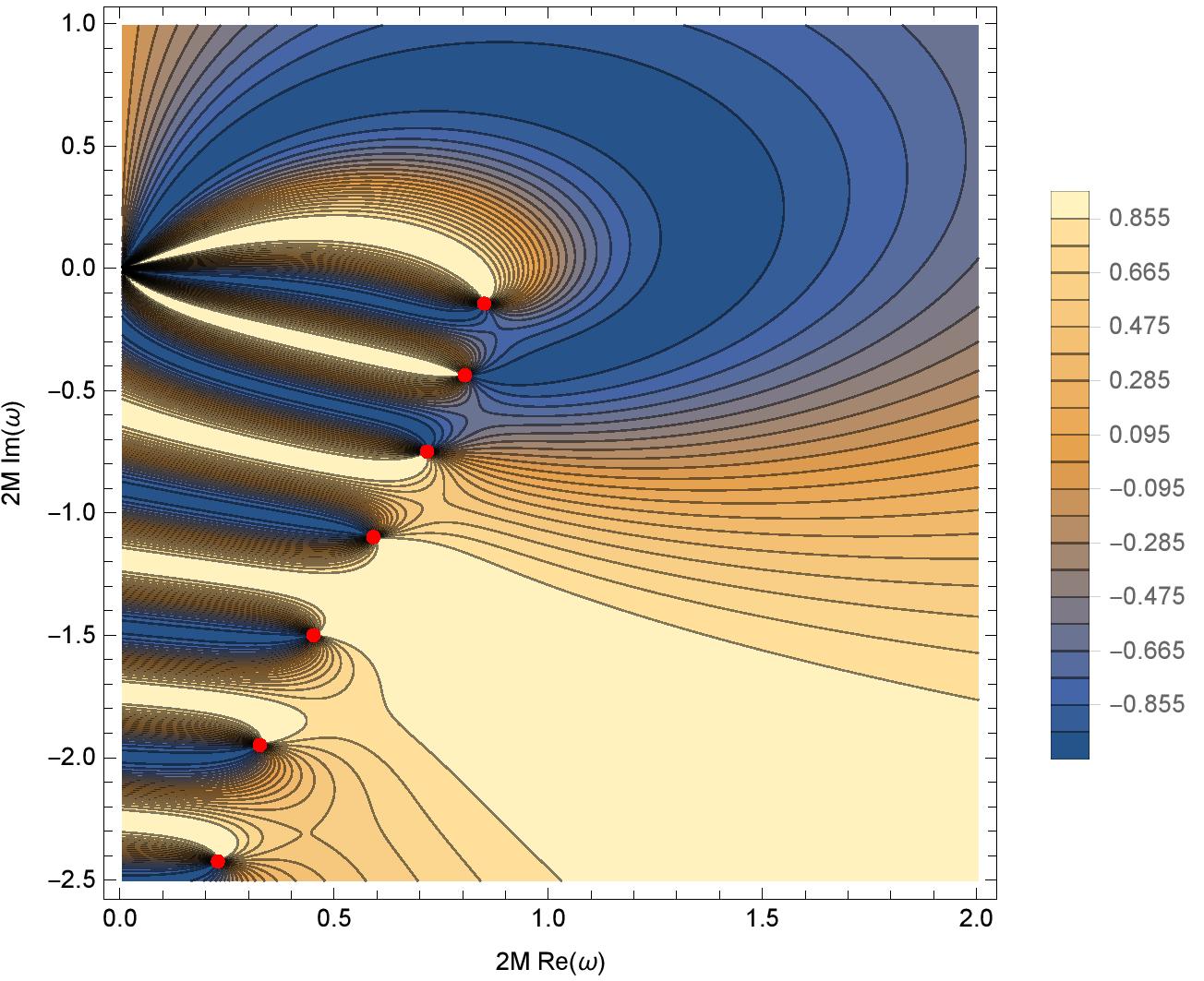}}   \quad
   \subfloat[$s=-2,\ell=2,m=-1,a=M$]
   {
\label{fig:phase sm2l2mm1}
      \includegraphics[width=.5\textwidth]{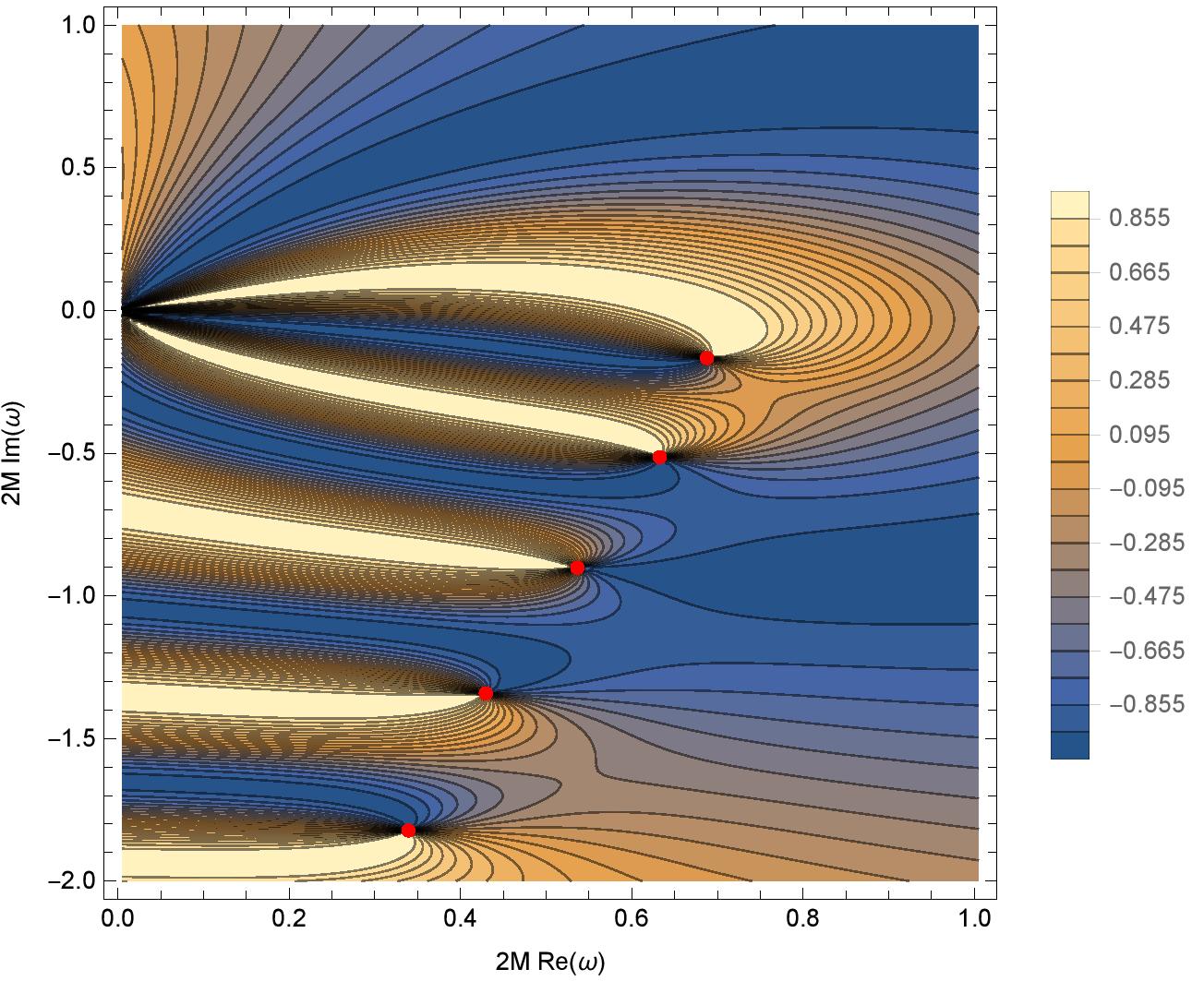}}    
   \end{center}
\caption{
Contour plots of 
$\sin\left(\arg\left(\W\right)\right)$
in the complex-$\omega$ plane in extremal Kerr.
The QNMs in Tables \ref{table:QNMs0l3m2a1}, \ref{table:QNMsm2l2m0a1} and \ref{table:QNMsm2l2mm1a1} are indicated with red dots.
\MCImp{also include the $-s=\ell=m=2$ one?}
}
\label{fig:phase W}
\end{figure}

\begin{figure}[hp!]
   \begin{center}
   \subfloat[$s=-2,\ell=2,m=0,a=M$]
   {
   \includegraphics[width=.46\textwidth]{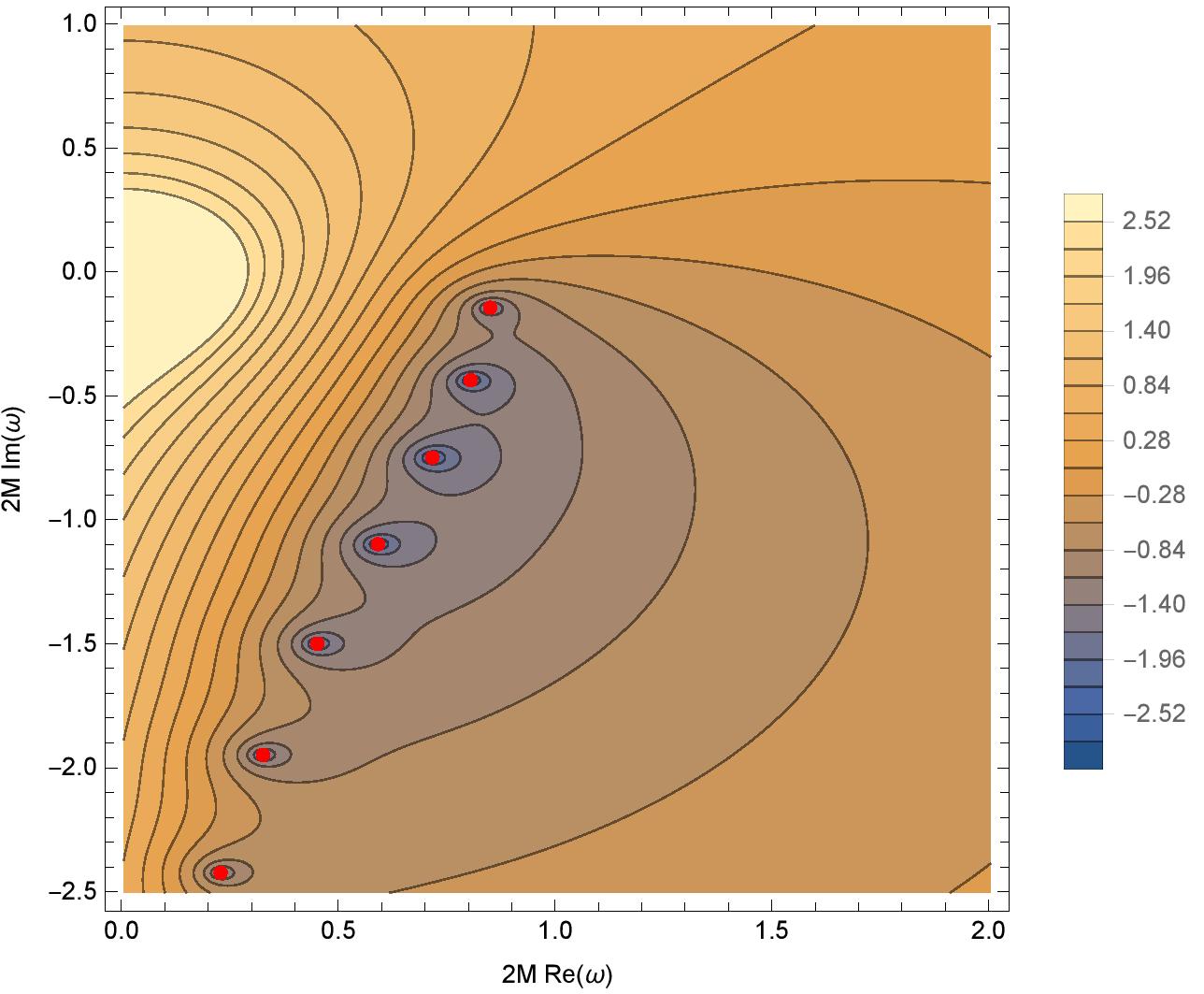}} \quad      
          \subfloat[$s=-2,\ell=2,m=1,a=M$]
      {
\label{figQNMsm2l2m1aM}
            \includegraphics[width=.46\textwidth]{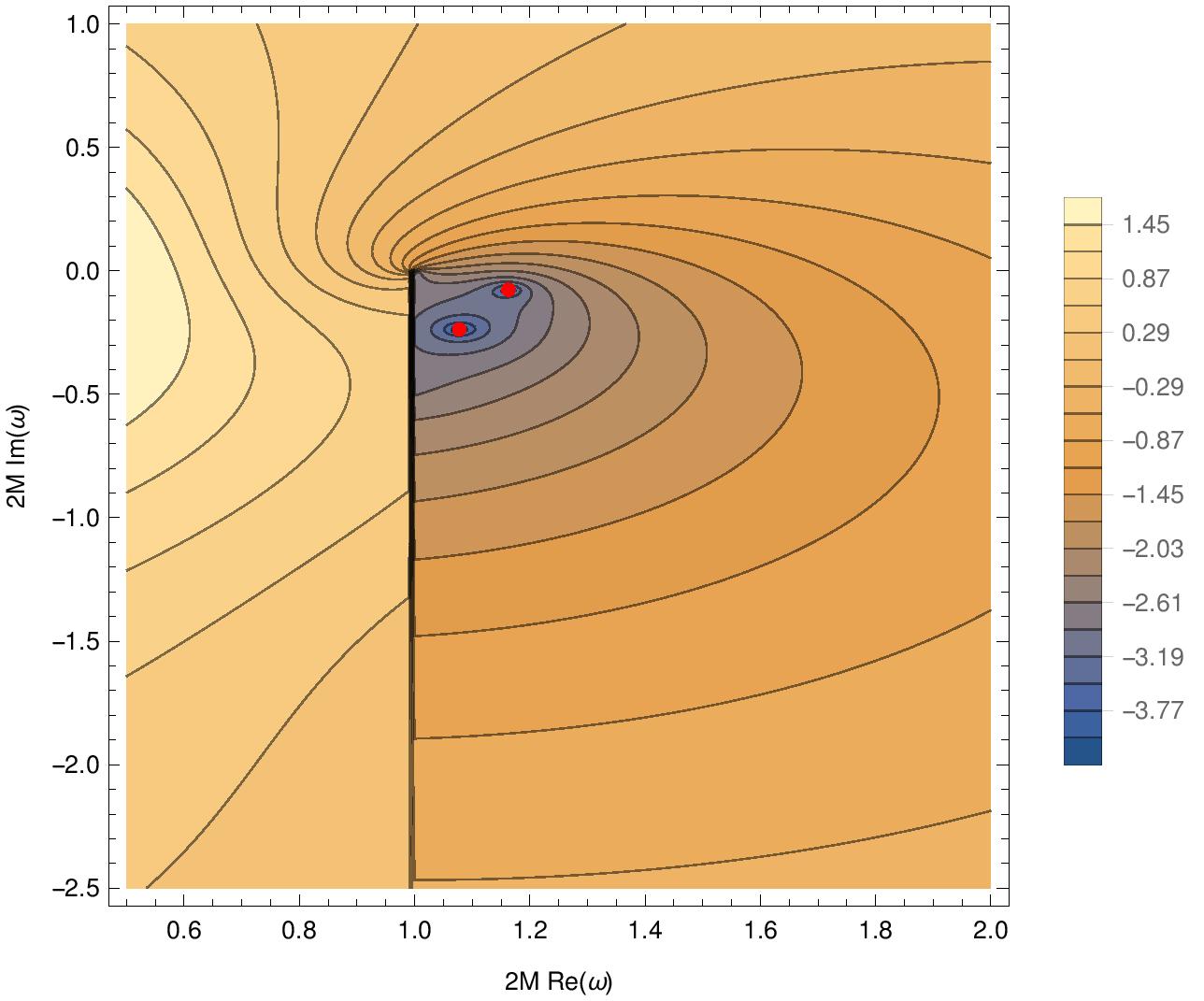}} \quad
   \subfloat[$s=-2,\ell=2,m=2,a=M$]
   {
\label{figQNMsm2l2m2aM}  
         \includegraphics[width=.46\textwidth]{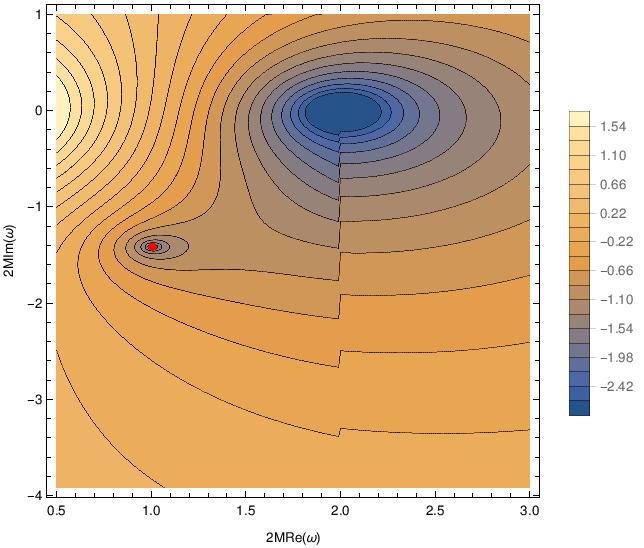}} \quad  
         \end{center}
\caption{Contour plot of $\log_{10}\left|\left(2M\right)^{2s}\W\right|$ in the complex-$\omega$ plane
 in extremal Kerr
 for  $s=-2$, $\ell=2$, $m=0$, $1$ and $2$.
For
$m=0$ and $1$
it is
$\mathcal{F}_s^2<0$
and so DMs  appear.
For $m=2$  it is
 $\mathcal{F}_s^2>0$
 and so DMs do not appear but  
 a \nsDMacr\  \ does appear. 
The QNMs in Tables \ref{table:QNMsm2l2m0a1}, \ref{table:QNMsm2l2m1a1}  and  \ref{table:QNMsm2l2m2a1} are indicated with red dots.}
\label{figQNMsm2l2m0aM}
\end{figure}

\begin{figure}[hp!]
   \begin{center}
      \includegraphics[width=.5\textwidth]{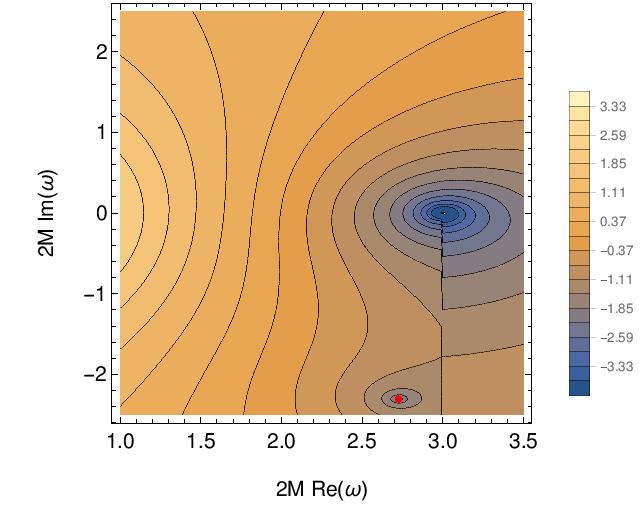}
   \end{center}
\caption{
Contour plot of $\log_{10}\left|\left(2M\right)^{2s}\W\right|$ in the complex-$\omega$ plane
for $s=-2$, $\ell=m=3$
 in extremal Kerr.
 It is $\mathcal{F}_s^2>0$
and so DMs do not appear, but there is the presence of a \nsDMacr\ .
The QNM in Table \ref{table:QNMsm2l3m3a1} is indicated with a red dot.
}
\label{figQNMsm2extreml3}
\end{figure}

In Figs.\ref{figQNMsm2l2m0aM}  we investigate the modes  $\ell=2$ and $0\leq m\leq2$ for $s=-2$.
It is $\mathcal{F}_s^2<0$ for $m=0$, $1$ and $\mathcal{F}_s^2>0$ for $m=2$.
Correspondingly, we observe DMs (again to the right of the BC, i.e., satisying property (i) in Sec.\ref{sec:QNMs NEK props}) for $m=0$, $1$ 
and no DMs for $m=2$ in the analyzed region of the complex-$\omega$ plane.
In the case $m=2$, however, we do observe a \nsDMacr\ , which is just the extremal limit of the corresponding ($n=5$) QNM in Fig.3 in~\cite{Cook:2014cta}.
 For 
 $m=0$ there is an accumulation of QNMs at $\omega=\oSR=0$  in NEK (see Fig.\ref{figNEKa}), leading to a BC along the negative imaginary axis
for the phase (but not the absolute value, as noted in footnote \ref{ft:BC m=0}) of the Wronskian  in extremal Kerr (Fig.\ref{fig:phase sm2l2m0}).
 For 
 $m=1$  we can  see 
 a BC that is formed from the accumulation in NEK of QNMs and TRMs - see Sec.\ref{sec:TRMs}. 

The previous plots are contour plots. In Fig.\ref{fig:QNM s=-2,l=2,m=0,1,a=M,3D} we plot the cases in 
Fig.\ref{figQNMsm2l2m0aM}
as 3-D plots instead. These show not only the cut and DM structure  but also the behaviour of the Wronskian as $k\to 0$: zero for $m=1$ and $2$
and divergent for $m=0$.
The vanishing for  $m=1$ and $2$ is in agreement with
Eq.\eqref{eq:W k->0,ext}, since it is $s=-2$ and, respectively,  $\nu_c\approx 1.42$ and $ -1/2+2.05i$ for these modes.
The divergence for $m=0$ is clearly in agreement with Eq.\eqref{eq:W k->0,ext,m=0}.
Similarly, Fig.\ref{figBCa1} shows that the Wronskian diverges as $k\to 0$ in this case, also in agreement with
Eq.\eqref{eq:W k->0,ext}, since it is  $s=0$ and $\nu_c\approx 1.71$.

Another mode for which we observe a \nsDMacr\ \ is the case $s=-2$, $\ell=m=3$ which we plot in Fig.\ref{figQNMsm2extreml3}.
This is a mode with  $\mathcal{F}_s^2>0$ and it lies to the left of the BC (i.e., the opposite of property (i) in Sec.\ref{sec:QNMs NEK props}).

In Fig.\ref{figQNMsm2Negatmextrem} we deal with the distinct case of negative $m$:
modes
 for $s=\ell=2$ and $m=-1$ and $-2$.
In these cases and for $\text{Re}(\omega)>0$, there are no BCs  (their superradiant BCs stem from $\oSR<0$) and there are the ``standard"
QNMs.
The QNMs in this figure are to be compared with the corresponding ones in Fig.3 in~\cite{Leaver:1985} --  visual agreement is found.

In Appendix \ref{sec:app QNMs extreme} we tabulate various QNM frequencies in extremal Kerr,
 calculated  using the MST method of Sec.\ref{sec:method} to 16 digits of precision.
These extremal Kerr QNMs are to be compared with the data in~\cite{Richartz:2015saa}. 
Our extremal Kerr QNM values agree with those in~\cite{Richartz:2015saa} to all digits of precision given in~\cite{Richartz:2015saa}; the frequencies in~\cite{Richartz:2015saa} are given up to a maximum of 7 digits of precision, whereas we give them up to 16.
We note that~\cite{Richartz:2015saa} only provides QNMs for overtones $n=0$ and $1$, whereas in App.\ref{sec:app QNMs extreme} we provide new QNM values up to 
higher overtones (up to $n=6$).
Importantly, we give the value for the most astrophysically relevant $s=-2$ QNM: $\ell=m=2$ in table \ref{table:QNMsm2l2m2a1},
as well as for $\ell=m=3$ in table \ref{table:QNMsm2l3m3a1}, neither of which was previously given in the literature, to the best of our knowledge.

\begin{figure}[hp!]
   \begin{center}
   \subfloat[$s=-2,\ell=2,m=0,a=M$]{
\label{fig3DQNMsm2l2m0aM}
   \includegraphics[width=.7\textwidth]{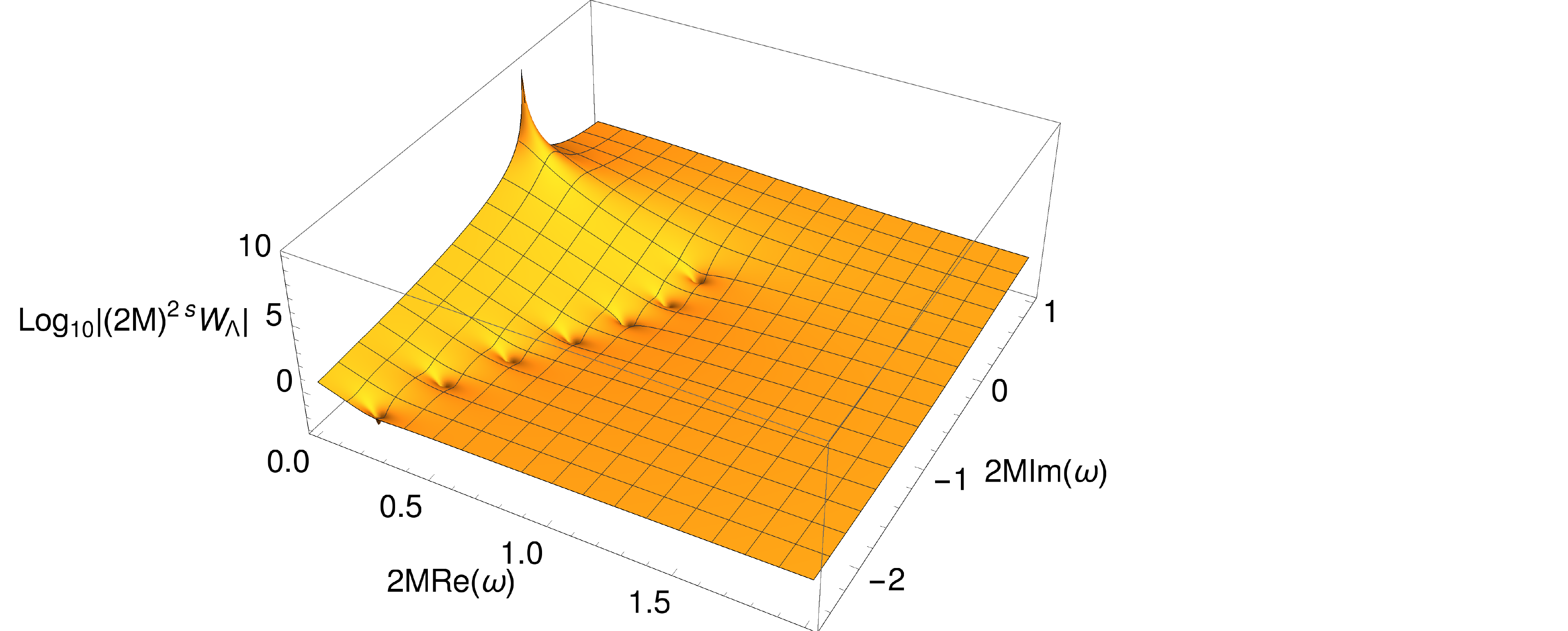}} \quad       
      \subfloat[$s=-2,\ell=2,m=1,a=M$]{
\label{fig3DQNMsm2l2m1aM}
      \includegraphics[width=.5\textwidth]{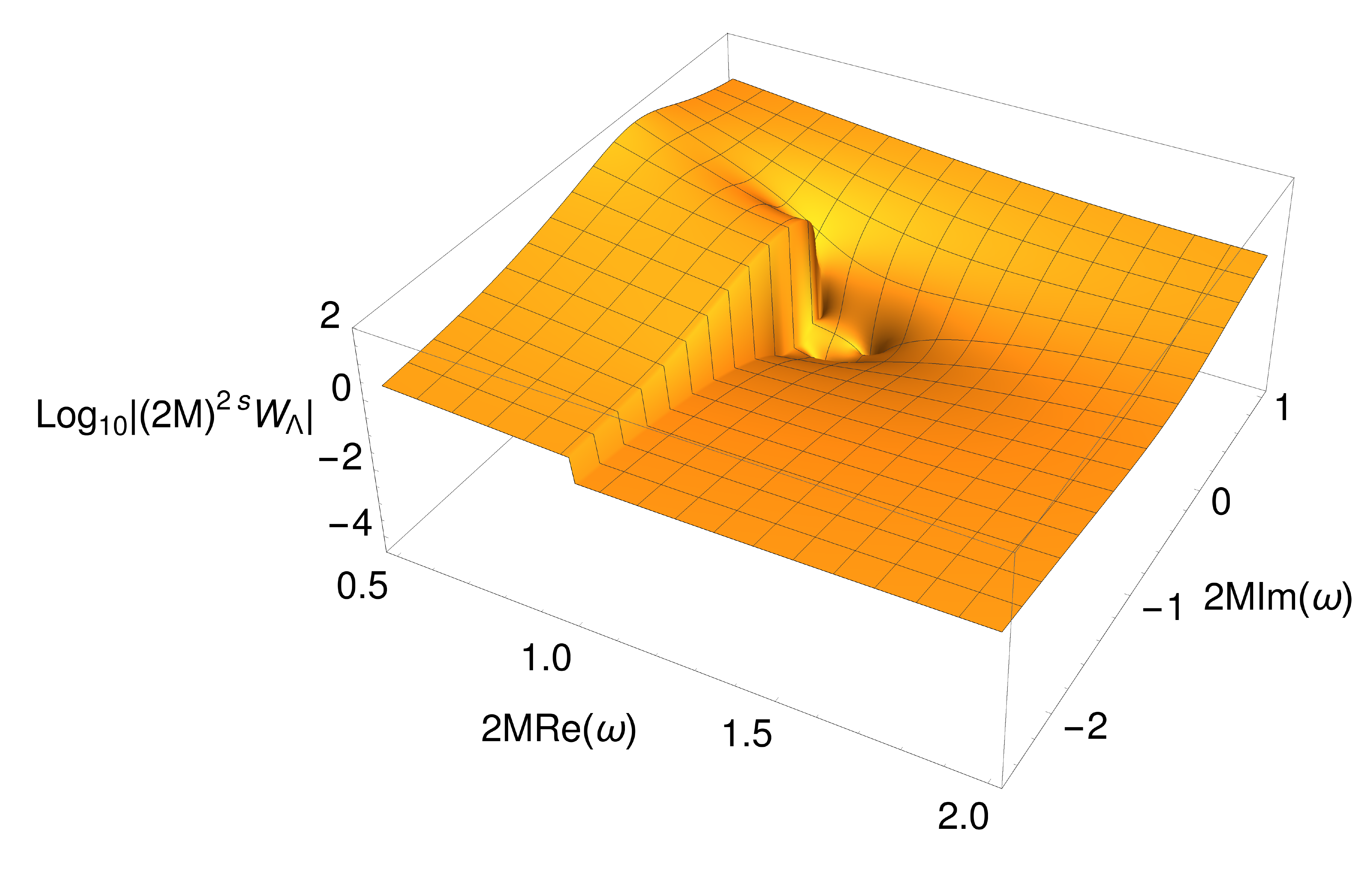}} \quad            
            \subfloat[$s=-2,\ell=2,m=2,a=M$]{
\label{fig3DQNMsm2l2m2aM}
      \includegraphics[width=.55\textwidth]{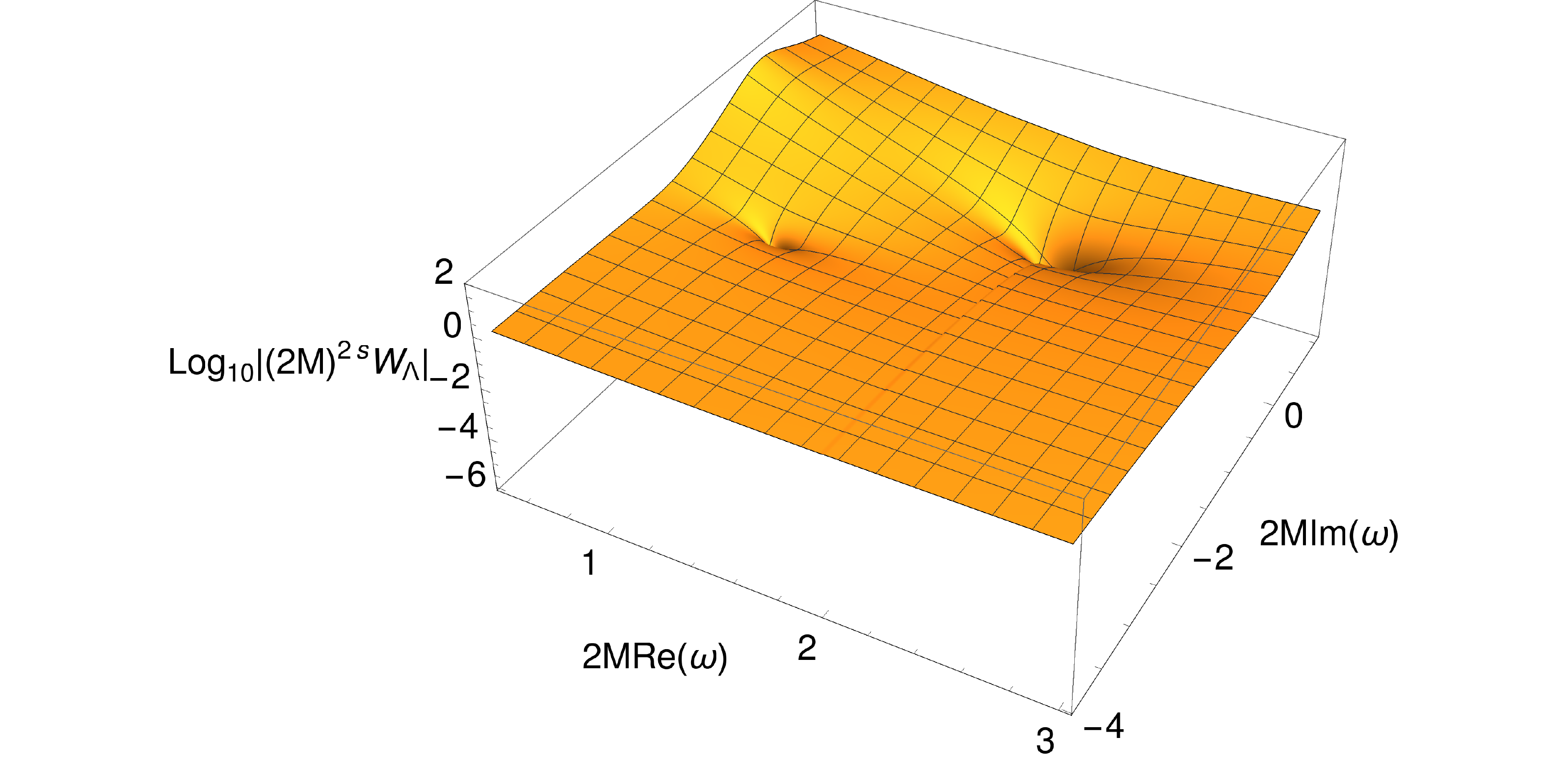}}   
   \end{center}
\caption{
These are 3-D versions of the 
plots
in 
Fig.\ref{figQNMsm2l2m0aM}. 
}
\label{fig:QNM s=-2,l=2,m=0,1,a=M,3D}
\end{figure}

\begin{figure}[hp!]
   \begin{center}
   \subfloat[$s=-2,\ell=2,m=-2,a=M$]{
\label{figQNMsm2l2mm2aM}
   \includegraphics[width=.5\textwidth]{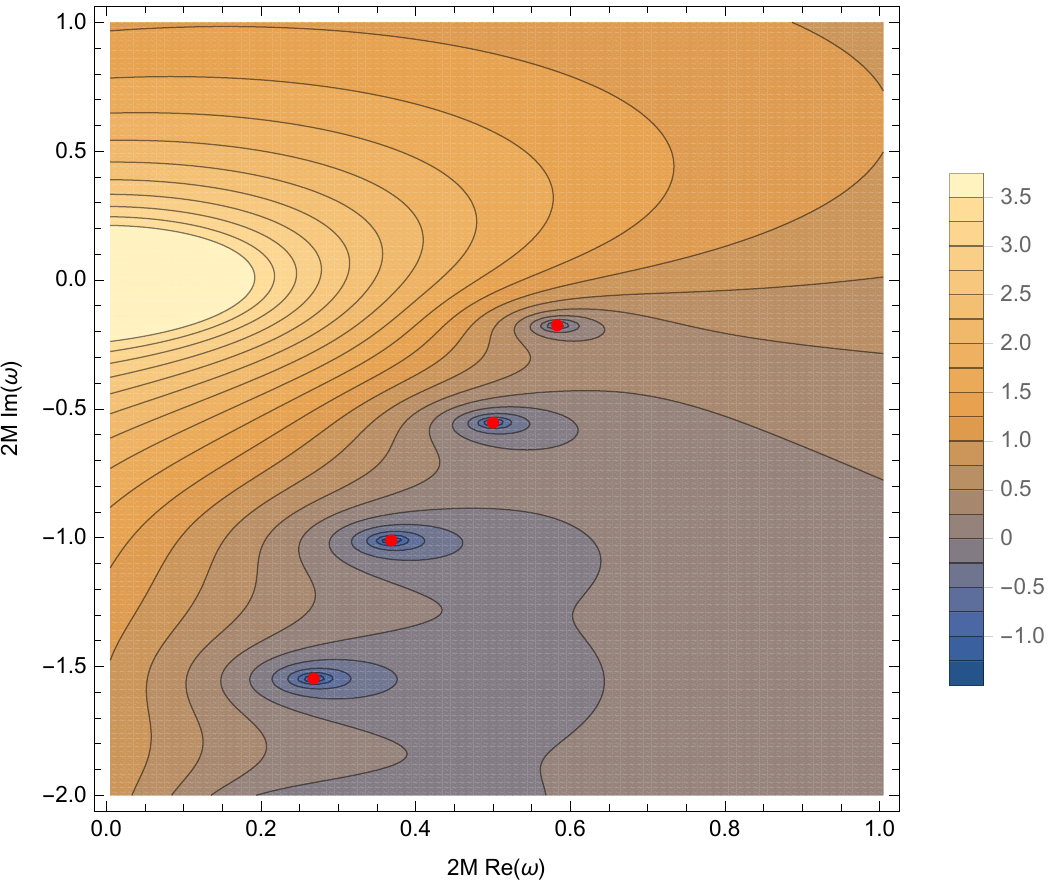}} \quad
      \subfloat[$s=-2,\ell=2,m=-1,a=M$]{
\label{figQNMsm2l2mm1aM}
   \includegraphics[width=.5\textwidth]{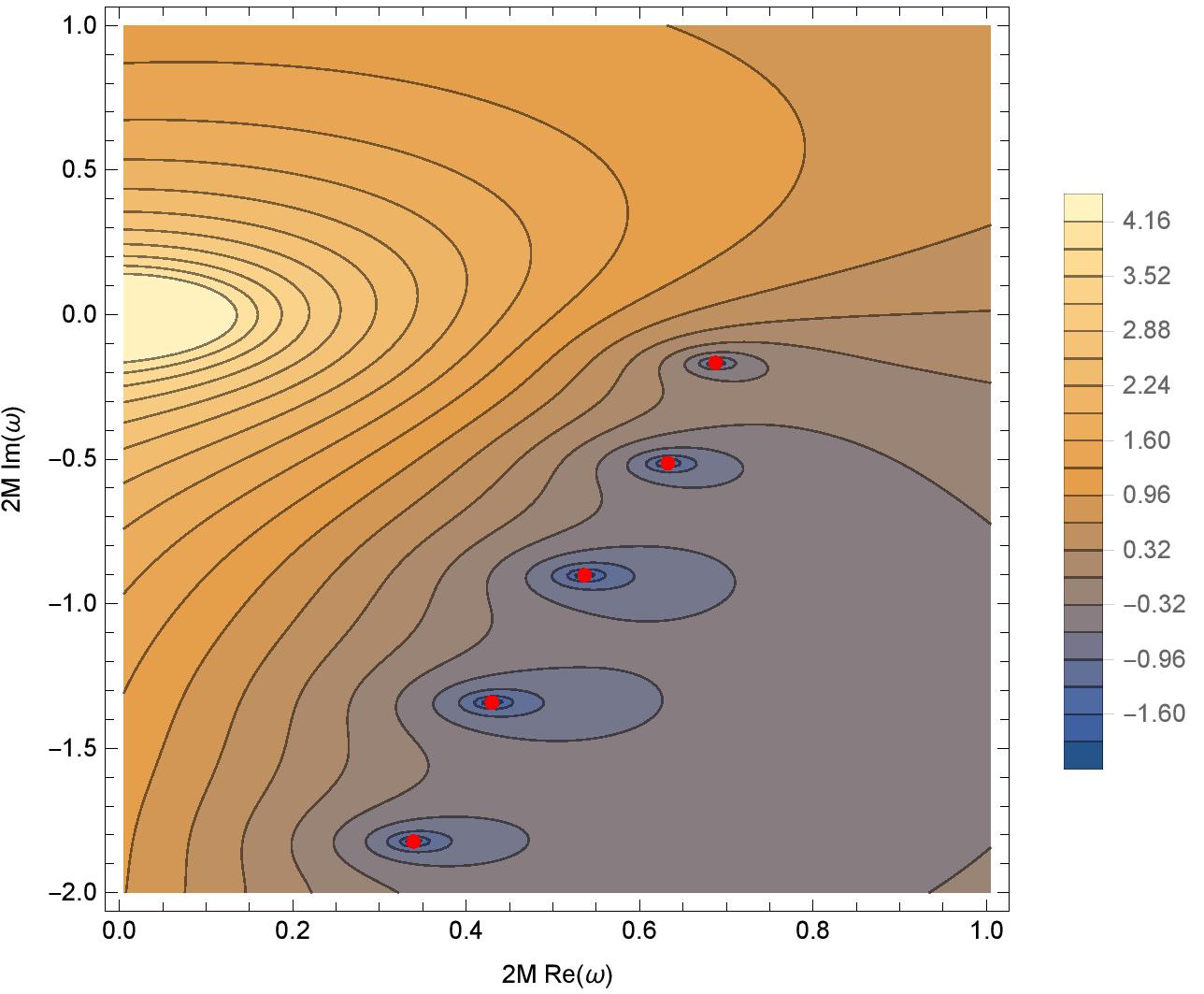}}
   \end{center}
\caption{Contour plots of $\log_{10}\left|\left(2M\right)^{2s}\W\right|$ in the complex-frequency plane in extremal Kerr for $s=-2$, $\ell=2$ and  $m=-1,-2$.
The QNMs 
 in Tables \ref{table:QNMsm2l2mm1a1} and \ref{table:QNMsm2l2mm2a1}  are indicated with red dots. 
Cf. Fig.3 in~\cite{Leaver:1985}.}
\label{figQNMsm2Negatmextrem}
\end{figure}


\subsection{Search for unstable modes}\label{sec:stability}

If the modes of the retarded Green function  possessed a singularity 
on the upper frequency half-plane,
that would indicate a linear instability of the space-time~\cite{detweiler1973stability}.
In principle, such a singularity could be a branch point or a pole (i.e., a mode with no incoming radiation).

Detweiler~\cite{detweiler1973stability} showed that there exist no 
modes with no incoming radiation in the upper plane in the case of scalar perturbations of extremal Kerr.
Later, Aretakis~\cite{aretakis2012decay} proved the 
 decay of the full scalar field in the {\it axisymmetric} case on and outside the horizon; ~\cite{dain2015bounds} proved similarly for axisymmetric gravitational perturbations
 outside the horizon.
On the horizon, it has been shown in~\cite{casals2016horizon,Gralla:2017lto} that the branch point on the superradiant bound frequency leads to the so-called Aretakis phenomenon~\cite{Aretakis:2012ei,Aretakis:2011ha}.
In its turn, the branch point at the origin  leads to a late-time decay in the field~\cite{casals2018perturbations}, as otherwise expected.
It therefore seems that it has not been proven that  the Green function modes with $m\neq 0$ do not possess singularities in the upper complex-frequency plane.

For scalar perturbations and $a\leq M$, Ref.\cite{detweiler1973stability} gives bounds on the real and imaginary parts of the frequencies of potential modes in the upper plane possessing no incoming radiation.
In particular, the real part should lie within the superradiant regime \MC{isn't this also true for general-spin perturbations? explicitly give the bounds on the Im part?}.
Ref.~\cite{Richartz:2015saa} and, in the case of scalar perturbartions, also~\cite{detweiler1973stability}, numerically looked for modes with no incoming radiation in the upper plane of extremal Kerr and could not find any.
Here, we report that we also looked for modes with no incoming radiation in the upper plane of extremal Kerr and did not find any, thus confirming the results in~\cite{detweiler1973stability,Richartz:2015saa}. 
Examples of the search are the already mentioned 
Figs.\ref{figQNMextremes0}--\ref{figQNMsm2Negatmextrem},
where we went up to $2M\text{Im}(\omega)=1$.
These figures also show that there are no ``physical" branch points in the upper plane for these modes, 
thus complementing the investigation in Sec.\ref{sec:search BCs}.


Finally, we note that the result  in~\cite{Andersson:2016epf} that no poles can 
lie on the real axis for $a<M$ together with the fact that there cannot be a pole at  the branch point 
$\oSR$~\cite{casals2016horizon,Richartz:2017qep} 
provides an argument -- although not a rigorous proof -- that no poles cross
the real axis as $a$ increases from $0$ to 
$M$
and so for the absence of poles in the upper plane  in extremal Kerr.



\section{Conclusions}\label{sec:conclusions}

We have carried out a thorough spectroscopic investigation in extremal Kerr space-time as well as in near-extremal Kerr.
We have calculated the superradiant amplification factor and quasi-normal modes and shown the formation of the 
extra superradiant branch cut.
Our results provide a corroboration of the MST method developed in~\cite{casals2018perturbations} and of
the
results
in~\cite{Richartz:2015saa}.
Our tabulated values for  quasi-normal mode frequencies signify an extension in both the accuracy and in the type of
modes tabulated in the literature (\cite{Richartz:2015saa} for extremal Kerr
and~\cite{QNMBertiNew} for near-extremal Kerr), including the astrophysically-relevant quasi-normal  mode for $-s=\ell=m=2$ in Table \ref{table:QNMsm2l3m3a1}.
Furthermore, we searched for both poles and branch points in the upper frequency plane in  extremal Kerr and report that we did not find any,
 providing further support for the mode stability of extremal Kerr off the horizon.
 We also gave an argument for the absence of such poles.
 
A seemingly open issue 
is finding a
condition for existence of quasi-normal modes with a finite imaginary part
 in (near-)extremal Kerr as well as determining the number of such modes.
In particular, the condition Eq.\eqref{eq:cond DM} misses quasi-normal modes  with a finite imaginary part  (the ones which we have called non-standard damped modes),
as seen for the particular cases of $s=-2$ modes with $\ell=m=2$ and $3$.
On the other hand,  we have found no quasi-normal modes for $s=\ell=m=1$ within the large region in the complex plane which we have considered, although of course that is no proof
that there exists no mode outside that region.

In terms of astrophysics,  
 it will be interesting to obtain,
 in the future,
  the gravitational waveform due to a particle inspiraling into an extremal Kerr black hole\footnote{Ref.\cite{sasaki1990gravitational} aimed at doing that but they used an extremal limit of QNM accumulation in NEK instead of directly a BC integral in exactly extremal Kerr.}, and compare 
it
to the results in~\cite{Gralla:2016qfw} for a near-extremal  black hole
and to the results in 
the near-horizon extremal Kerr geometry~\cite{Porfyriadis:2014fja}.
 \MC{other suggestions as to what could be done in the future - TTMs?}


\begin{acknowledgments}
We are grateful to 
Maarten van de Meent,
Adrian C. Ottewill, Maur\'icio Richartz,
Alexei  Starobinsky
and Aaron Zimmerman
 for helpful conversations.
M.C. acknowledges partial financial support by CNPq (Brazil), process number 310200/2017-2. 
L.F. acknowledges financial support by the Coordena\c{c}\~ao de Aperfei\c{c}oamento de Pessoal de N\'ivel Superior - Brasil (CAPES) - Finance Code 001.
\end{acknowledgments}


\appendix

\section{Special frequencies}\label{sec:special freq}

In this appendix we investigate various specific frequencies.
In Sec.\ref{sec:HW} we give evidence that a certain frequency suggested in~\cite{Hartle:Wilkins:1974} to be a branch point is actually not a branch point;
in Sec.\ref{sec:Hod} we give evidence that certain frequencies suggested in~\cite{Hod:2015swa,Hod:2016aoe} to be QNMs are actually not QNMs;
in Sec.\ref{sec:AS} we investigate the so-called algebraically-special frequencies.


\subsection{Hartle-Wilkins frequency}\label{sec:HW}

Hartle and Wilkins~\cite{Hartle:Wilkins:1974} observed
that it was  possible that the frequency
\begin{equation} \label{whw}
\omega_{HW} = 
\oSR
 + i(s-1) \dfrac{r_{\horind}-M}{2Mr_{\horind}}
\end{equation}
is a branch point.
Interestingly, such frequency would lie in the upper-half complex-$\omega$ plane for $s=2$ and, in that case, it might lead to an instability
of the black hole.
We have calculated the Wronskian for $s=\ell=m=2$ 
and the token  value of $a=0.5M$ \MCImp{confirm value of $a$} when going around this frequency: 
$\omega_0=\omega_{HW}$ and $R=|\omega_{HW}|/5$.
We observed no discontinuity in $|\Wsub|$ nor in  $\arg(\Wsub)$,
thus providing
strong evidence that the Hartle-Wilkins frequency is not a branch point 
at least for this mode.


\subsection{Hod frequencies}\label{sec:Hod}

In Sec.\ref{sec:QNMs NEK props} we gave  the generally accepted picture of DMs and ZDMs.
However, it was argued by Hod~\cite{Hod:2015swa,Hod:2016aoe} (based on the analysis in~\cite{detweiler1980black}) that DMs also exist for $\mu\ge \mu_c$ as long as $a$ is sufficiently close to $M$.
For example, for $s=\ell=m=2$, \cite{Hod:2015swa} predicts DMs at
\begin{equation}\label{eq:Hod freq}
\omega\approx 
\oSR
+\left(0.324-0.07i\right)e^{-1.532n},
\end{equation}
where $n\in\mathbb{Z}^+$.
However, Refs.~\cite{Zimmerman:2015rua} and~\cite{Richartz:2015saa}, carried out a numerical search for these DMs 
in, respectively, NEK and extremal Kerr, and did not find any.
Similarly, we  carried out a numerical search in extremal Kerr for  these DMs for  $n=2$  
using a grid of stepsize 
$\Delta\omega_r=\Delta\omega_i=10^{-6}$
 along the directions of both the real and imaginary parts of the frequency and we also
found no evidence of the presence of such DMs.




\subsection{Algebraically-special Frequencies}\label{sec:AS}\MCImp{Is this section worth it given that Fig.7~\cite{onozawa1997detailed} already gives our plot? I think it's only worth keeping it if we do something extra like plot the AS freqs. for various $\ell$, $m$...}

Gravitational ($|s|=2$) perturbations of black holes admit frequencies for which the radial solution for 
spin $s$ (either $s=+2$ or $s=-2$)
is the trivial solution whereas the radial solution for spin ``$-s$" is a  non-trivial solution.
These frequencies are the so-called  algebraically-special (AS) frequencies~\cite{1973JMP....14.1453W,chandrasekhar1984algebraically}.
\MC{say what b.c. the AS modes obey}

The AS frequencies satisfy the following  equation~\cite{chandrasekhar1984algebraically,MaassenvandenBrink:2000ru}:
\begin{align}\label{eq:AS}
&
\eigenAS^2\left(\eigenAS+2\right)^2+8\eigenAS\left(5\eigenAS\left(am\omega-a^2\omega^2\right)+6\left(am\omega+a^2\omega^2\right)\right)+
\nonumber \\ &
36\omega^2+144\left(am\omega-a^2\omega^2\right)^2=0,
\end{align}
where $\eigenAS\equiv\left. \eigenl\right|_{s=-2}$.
The left-hand side of Eq.\eqref{eq:AS} is the so-called Teukolsky-Starobinsky constant, which serves to relate (radial and angular) solutions for spin $s$ to solutions with spin ``$-s$".

The radial solutions at AS frequencies can be found in closed form~\cite{chandrasekhar1984algebraically} and, for $a>0$, they 
correspond to totally transmitted modes (TTMs), i.e., modes with zero reflection coefficient~\cite{MaassenvandenBrink:2000ru}.
For $a=0$, the AS frequencies
 are given by
 ``$-i(\ell-1)(\ell+2)\left((\ell-1)(\ell+2)+2\right)/6$"
 ~\cite{MaassenvandenBrink:2000ru}.
 As $a$ increases, not only the AS frequencies move away from the negative imaginary axis for $m\neq 0$ but also
 a family of QNMs stems off from the AS frequency at $a=0$~\cite{MaassenvandenBrink:2000ru}.
 
\begin{figure}[hp!]
   \includegraphics[width=.4\textwidth]{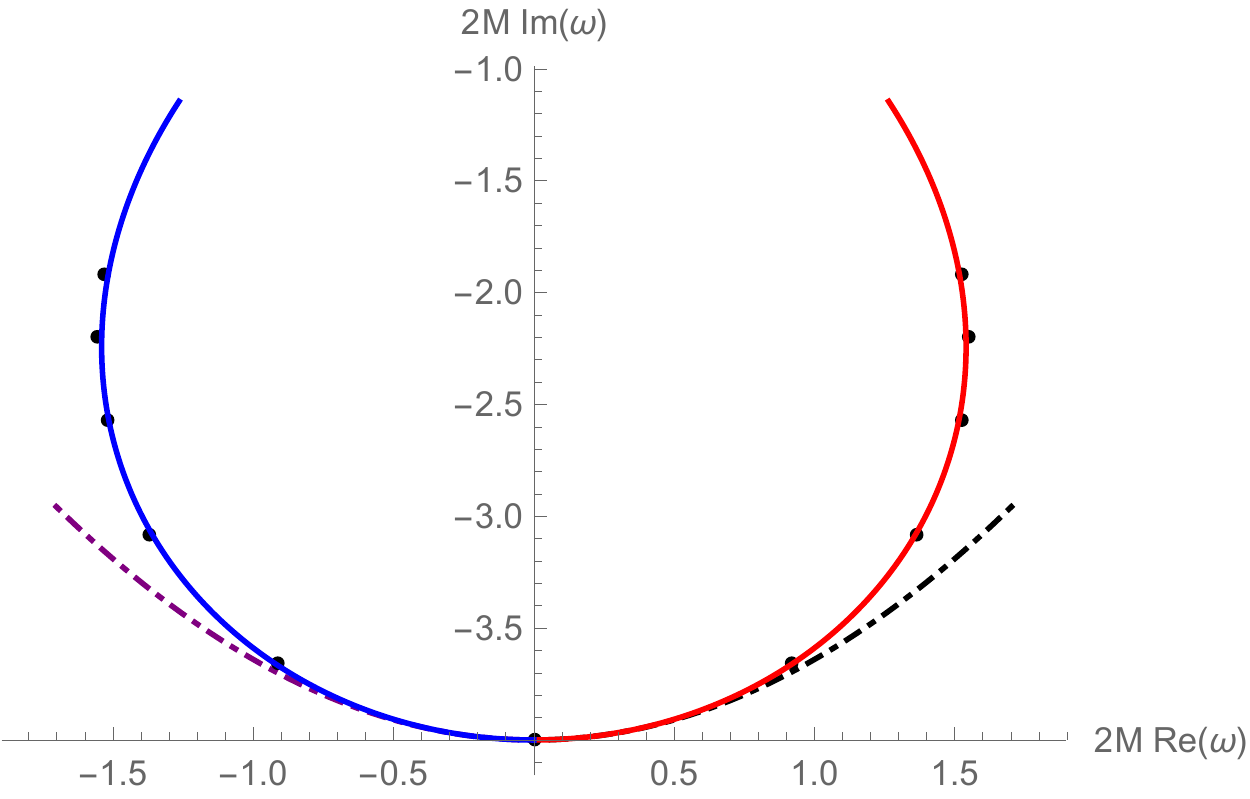} 
\caption{
Algebraically-special frequencies for  $\ell=2$ and $m=+2$ (red) and $m=-2$ (blue).
The solid curves are our  values for the roots of Eq.\eqref{eq:AS} (as $a$ increases, the value of the imaginary part becomes less
negative);  the black dots are the AS frequencies in Table A.1 in~\cite{chandrasekhar1984algebraically} (N.B.: for $a=0.3M$ and $m=-2$ we changed the  sign of the imaginary part of 
the frequency in~\cite{chandrasekhar1984algebraically});
the dashed curves correspond to Eq.7.27~\cite{MaassenvandenBrink:2000ru}.
At $a=0$ the AS frequency is ``$-4i$" and as $a$ increases the AS frequencies move away from it.
}
\label{fig:AS}
\end{figure}

Here we 
 solve Eq.\eqref{eq:AS} numerically and plot the result in Fig.\ref{fig:AS}.
From Eq.\eqref{eq:symm eigen} it follows that the AS frequency for ``$-m$" is equal to that for ``$+m$" after changing the sign of its
real part, as reflected  in Fig.\ref{fig:AS}.
The case for $\ell=2$ was already plotted in Fig.7~\cite{onozawa1997detailed}, with which our Fig.\ref{fig:AS} agrees.
We also compare the numerical values of the AS frequencies with the small $a/M$ expansion in Eq.7.26~\cite{MaassenvandenBrink:2000ru},
\MC{in the end we don't  include plots for extra $\ell=?$ modes?} and here we give the  values of the AS frequencies for extremal Kerr for $s=-2$, $\ell=2$ and $m=\pm 2$:
$2M\omega \approx \pm 1.26369
- 1.14328i$.

\section{Tables of Quasi-Normal Modes in NEK}
\label{sec: app QNMs NEK}

In Tables \ref{table:QNMNEKsm2l2m2}--\ref{table:QNMNEKsm2l2m1} we provide the numerical values of  QNM frequencies in NEK to 16 digits of precision.
We tried to compare our values against those in~\cite{QNMBertiNew} but it seems that some overtones in~\cite{QNMBertiNew} are missing.
As stated in~\cite{QNMBertiNew}, the values there are ``unreliable very close to the Kerr extremal limit (roughly, when $a/M\ge 0.999$)".
Therefore, the comparison between our values and those in ~\cite{QNMBertiNew} is not straight-forward and we only compared a
few  overtones in common, for which we found agreement up to around 7 digits of precision.
We believe that here we provide all QNM frequencies up to the overtone  indicated (and so we provide new frequencies not already
in the literature)
to 16 digits of precision.

The QNMs in Table \ref{table:QNMNEKsm2l2m2}  are plotted in Fig.\ref{figNEKd};
the ones in Tables \ref{table:QNMNEKsm2l2m0}--\ref{table:QNMNEKsm2l2m0cont} and \ref{table:QNMNEKsm2l2m1}  are  in Figs.\ref{figNEKa} and \ref{figNEKc}, respectively.

\begin{table}[hp!] 
\centering
\begin{tabular}{||c c c||c c c||} 
\hline
 n & $\Real(2M\oQNM)$ & $\Imag(2M\oQNM)$  \\  [0.5ex] 
 \hline\hline
 0 & 1.971347096765401 & -6.937346135358345 E-3   \\ 
 1 & 1.971346859396121 & -2.081244162006215 E-2   \\ 
 2 & 1.971345377490081 & -3.468824642756046 E-2   \\ 
 3 & 1.971341924956729 & -4.856346528969809 E-2  \\ 
 4 & 1.971337594793333 & -6.243678169903200 E-2  \\ 
 5 & 1.971334086965111 & -7.630839456258754 E-2  \\ 
 6 & 1.971332336634383 & -9.017958067268572 E-2   \\ 
 7 & 1.971332266312562 & -1.040517407195357 E-1 \\ 
 8 & 1.971333151417415 & -1.179258517311285 E-1  \\ 
 9 & 1.971334044035797 & -1.318023508789842 E-1  \\ 
10 & 1.971334051188986 & -1.456812360708640 E-1   \\ 
11 & 1.971332463130118 & -1.595622216149717 E-1   \\ 
12 & 1.971328783808630 & -1.734448742298788 E-1  \\ 
13 & 1.971322712090932 & -1.873287088529975 E-1  \\   [1ex] 
 \hline
\end{tabular}
\caption{QNM frequencies for the mode $s=-2$, $\ell=m=2$ in NEK with $a=0.9999M$. All modes found are ZDMs.}
\label{table:QNMNEKsm2l2m2}
\end{table}

\begin{table}[hp!] 
\centering
\begin{tabular}{||c c c||c c c||} 
\hline
 n & $\Real(2M\oQNM)$ & $\Imag(2M\oQNM)$  \\  [0.5ex] 
 \hline\hline
 \;\;0*& 8.497014123708619 E-1 &  -1.439808152474526 E-1 \\
 1 & 
  &  -1.908029268511688 E-1 \\
 2 &
  &  -2.549128389211352 E-1 \\
 3 & 
   &  -3.192664571585073 E-1 \\
 4 & 
 &  -3.838535854911746 E-1   \\
 \;\;5* & 8.052661998429733 E-1 &  -4.375466529029460 E-1  \\
 6 &
  &  -4.486623210969344 E-1  \\
 7 &
  &  -5.136796393498003 E-1 \\
 8 & 
  &  -5.788919278697769 E-1   \\
 9 & 
 &  -6.442854493902452 E-1     \\
 10 & 
  &  -7.098467272001537 E-1   \\
 \;\;11*& 7.172155884312021 E-1 &  -7.500069540472439 E-1 \\
 12 &
 &  -7.755628272018845 E-1    \\
 13 & 
 &  -8.414214727066511 E-1   \\
 14 & 
 &  -9.074109470708575 E-1  \\
 15 &
 &  -9.7351989539484129 E-1   \\
 16 & 
 &  -1.039737374647090  \\
 \;\;17*& 5.925271449069068 E-1 &  -1.099229986861354 \\
 18 & 
  &  -1.106053542725389   \\
 19 & 
  &  -1.172460898950187    \\
 20 &
  &  -1.238955080092425    \\
 21 & 
 &  -1.305533662382188 \\
 22 & 
  &  -1.372192244020426 \\
 23 & 
  &  -1.438919629496156  \\
 \;\;24*& 4.529296412578228 E-1 &  -1.500426836467437         \\
 25 &
 &  -1.505696853169629 \\ [1ex] 
 \hline
\end{tabular}
\caption{QNM frequencies for the mode $s=-2$, $\ell=2$, $m=0$  in NEK with $a=0.998M$. DMs are marked with an asterisk next to the overtone number.
We do not include the real part of the ZDM frequencies since these values decreased below the numerical accuracy of our calculation as we kept increasing the accuracy.
}
\label{table:QNMNEKsm2l2m0}
\end{table}

\begin{table}[hp!] 
\centering
\begin{tabular}{||c c c||c c c||} 
\hline
 n & $\Real(2M\oQNM)$ & $\Imag(2M\oQNM)$  \\  [0.5ex] 
 \hline\hline
 26 & 
 &  -1.572504528889339 \\ 
 27 & 
 &  -1.639336645792597 \\
 28 & 
  &  -1.706210683167845 \\
29 & 
 &  -1.773163049029994 \\
30 &
&  -1.840225648392450 \\
31 &
  &  -1.907389617996063 \\
 \;\;32* & 3.274322036264049 E-1 &  -1.948816546673616 \\
  33 & 
     &  -1.974581437611800 \\
  34 &
   &  -2.041695632198213 \\
  35 & 
     &  -2.108681108639191 \\
  36 &
     &  -2.175597100785044 \\
  37 &
    &  -2.242590510389330 \\
  38 &
     &  -2.309831201813220 \\
39 &
 &  -2.377411305259576 \\
\;\;40* & 2.294864986928096 E-1     &  -2.424608143864245 \\
 41 &
          &  -2.445173832897294 \\
 42 &
        &  -2.512687475720510 \\
 43 &
         &  -2.579675360638962 \\
 44 & 
       &  -2.646296875637739 \\ 
 45 & 
      &  -2.712932421578639 \\
 46 &
    &   -2.779995534963594 \\
 47 &
   &  -2.847820373618469 \\
 \;\;48* & 1.573845170494564 E-1    &  -2.912834509300094 \\
 49 &
    &  -2.916492009814639 \\
 50 & 
     &  -2.984858687619585 \\ [1ex] 
 \hline
\end{tabular}
\caption{Continuation of Table \ref{table:QNMNEKsm2l2m0}.}
\label{table:QNMNEKsm2l2m0cont}
\end{table}

\begin{table}[hp!] \label{tableB2}
\centering
\begin{tabular}{||c c c||c c c||} 
\hline
 n & $\Real(2M\oQNM)$ & $\Imag(2M\oQNM)$   \\  [0.5ex] 
 \hline\hline
\;\;0* & 1.159590591695038 & -8.015520552061875 E-2 \\
 1 & 1.012824618998511 & -1.548386076219234 E-1\\
 2 & 1.019199556792002 & -2.218210493975133 E-1\\
  \;\;3* & 1.081234285735362 & -2.414640438496771 E-1\\ 
 4 & 1.018911494244786 & -2.907420089883552 E-1\\ 
 5 & 1.016501421337247 & -3.523120790458689 E-1\\
 6 & 1.015255997696160 & -4.099635096644475 E-1\\
 7 & 1.013393080486841 & -4.654490258922171 E-1\\
 8 & 1.010140571713775 & -5.207860324030137 E-1 \\
 9 & 1.006430702232703 & -5.770692853983539 E-1 \\
 10 & 1.003030985193748 & -6.340464951810719 E-1 \\
 11 & 1.000036494376247 & -6.912846069656605 E-1\\
 12 & 9.973003500319220 E-1 & -7.485746306385129 E-1 \\ [1ex] 
 \hline
\end{tabular}
\caption{QNM frequencies for the mode $s=-2$, $\ell=2$, $m=1$  in NEK with $a=0.998M$. DMs are marked with an asterisk next to the overtone number.
}
\label{table:QNMNEKsm2l2m1}
\end{table}


\section{Tables of Quasi-Normal Modes in extremal Kerr}
\label{sec:app QNMs extreme}

In Tables \ref{table:QNMs0l3m2a1}--\ref{table:QNMsm2l3m3a1} we provide the numerical values of  QNM frequencies in extremal Kerr (i.e., $a=M$) to 16 digits of precision.
We note that, for each table, we provide all the QNM overtones 
which we found  in the analyzed region (which is the region plotted in the figures).
 
Some values here should be compared with values in~\cite{Richartz:2015saa}; we found agreement  to all digits given in~\cite{Richartz:2015saa}, which is up to 7 digits.
Here we also provide new QNM values.

The QNMs in Table \ref{table:QNMs0l3m2a1} 
are plotted in Figs.\ref{figQNMs0l3m2} and \ref{fig:phase s0l3m2}  (see also Fig.\ref{figBCa1});
the ones in Table \ref{table:QNMsm2l2m0a1} are in Figs.\ref{fig:phase sm2l2m0} and \ref{figQNMsm2l2m0aM} (see also Fig.\ref{fig3DQNMsm2l2m0aM});
the ones in Table \ref{table:QNMsm2l2m1a1} are in Fig.\ref{figQNMsm2l2m1aM} (see also Fig.\ref{fig3DQNMsm2l2m1aM});
the one in Table \ref{table:QNMsm2l2m2a1} is in Fig.\ref{figQNMsm2l2m2aM};
the ones in Table \ref{table:QNMsm2l2mm1a1} are in Figs.\ref{fig:phase sm2l2mm1} and \ref{figQNMsm2l2mm1aM};
the ones in Table \ref{table:QNMsm2l2mm2a1} are in Fig.\ref{figQNMsm2l2mm2aM};
the one in Table \ref{table:QNMsm2l3m3a1} is in Fig.\ref{figQNMsm2extreml3}.

\begin{table}[hp!] 
\centering
\begin{tabular}{||c c c||} 
\hline
 n & $\Real(2M\oQNM)$ & $\Imag(2M\oQNM)$   \\  [0.5ex] 
 \hline\hline
 0 & 2.143189451642680 & -6.447596182703441 E-2   \\ 
 1 & 2.065635617386494 & -1.936077608481560 E-1  \\  [1ex] 
 \hline
\end{tabular}
\caption{QNM frequencies for the mode $s=0$, $\ell=3$, $m=2$  in extremal Kerr. Only two QNMs were found in the analyzed region.}
\label{table:QNMs0l3m2a1}
\end{table}

\begin{table}[hp!]
\centering
\begin{tabular}{||c c c||}  
\hline
 n & $\Real(2M\oQNM)$ & $\Imag(2M\oQNM)$ \\  [0.5ex] 
 \hline\hline
0 & 8.502902182451604 E-1 & -1.436123679332644 E-1  \\ 
 1 & 8.054871834090764 E-1 & -4.365659779474246 E-1   \\ 
 2 & 7.168565309137984 E-1 & -7.487455716919170 E-1 \\ 
 3 & 5.916519687840184 E-1 & -1.098077974425383  \\ 
 4 & 4.518396119844236 E-1 & -1.499593272060030  \\ 
 5 & 3.264000679767349 E-1 & -1.948244495660487  \\ 
 6 & 2.286126533997227 E-1 & -2.424170217742298  \\   [1ex] 
 \hline
\end{tabular}
\caption{QNM frequencies for the mode $s=-2$, $\ell=2$, $m=0$ in extremal Kerr. }
\label{table:QNMsm2l2m0a1}
\end{table}

\begin{table}[hp!] 
\centering
\begin{tabular}{||c c c||} 
\hline
 n & $\Real(2M\oQNM)$ & $\Imag(2M\oQNM)$ \\  [0.5ex] 
 \hline\hline
 0 & 1.162866404907267 & -7.651091051516613 E-2  \\ 
 1 & 1.077709328395424 & -2.372558086777354 E-1   \\   [1ex] 
 \hline
\end{tabular}
\caption{QNM frequencies for the mode $s=-2$, $\ell=2$, $m=1$ in  extremal Kerr.
}
\label{table:QNMsm2l2m1a1}
\end{table}

\begin{table}[hp!] 
\centering
\begin{tabular}{||c c c||} 
\hline
 n & $\Real(2M\oQNM)$ & $\Imag(2M\oQNM)$   \\  [0.5ex] 
 \hline\hline
 0 & 1.006919901580225 & -1.414797640581692 E-1
   \\  [1ex] 
 \hline
\end{tabular}
\caption{QNM frequency for the mode $s=-2$, $\ell=m=2$  in extremal Kerr. }
\label{table:QNMsm2l2m2a1}
\end{table}

\begin{table}[hp!] 
\centering
\begin{tabular}{||c c c||} 
\hline
 n & $\Real(2M\oQNM)$ & $\Imag(2M\oQNM)$ \\  [0.5ex] 
 \hline\hline
0 & 6.877231140851841 E-1 & -1.667681875043246 E-1 \\
1 & 6.326105018403029 E-1 & -5.140477960506172 E-1\\
2 & 5.364531452887850 E-1 & -9.014823440124349 E-1\\
3 & 4.295419740535755 E-1 & -1.341635185721246 \\
4 & 3.391588469653389 E-1 & -1.822506293258611 \\  [1ex] 
 \hline
\end{tabular}
\caption{QNM frequencies for the mode $s=-2$, $\ell=2$, $m=-1$ in  extremal Kerr. 
}
\label{table:QNMsm2l2mm1a1}
\end{table}

\begin{table}[hp!]
\centering
\begin{tabular}{||c c c||} 
\hline
 n & $\Real(2M\oQNM)$ & $\Imag(2M\oQNM)$ \\  [0.5ex] 
 \hline\hline
0 & 5.831069287317494 E-1& -1.760516746384491 E-1 \\
1 & 5.002925334514074 E-1& -5.534767606703243 E-1\\
2 & 3.684811261106501 E-1& -1.010929783409981 \\
3 & 2.681101698130941 E-1& -1.546682699292125 \\   [1ex] 
 \hline
\end{tabular}
\caption{QNM frequencies for the mode $s=-2$, $\ell=2$, $m=-2$ in  extremal Kerr.
}
\label{table:QNMsm2l2mm2a1}
\end{table}

\begin{table}[hp!] 
\centering
\begin{tabular}{||c c c||} 
\hline
 n & $\Real(2M\oQNM)$ & $\Imag(2M\oQNM)$   \\  [0.5ex] 
 \hline\hline
 0 & 2.726653450670023 & -2.304765481816845       
   \\  [1ex] 
 \hline
\end{tabular}
\caption{QNM frequency for the mode $s=-2$, $\ell=m=3$  in extremal Kerr. }
\label{table:QNMsm2l3m3a1}
\end{table}



\begin{thebibliography}{87}
\expandafter\ifx\csname natexlab\endcsname\relax\def\natexlab#1{#1}\fi
\expandafter\ifx\csname bibnamefont\endcsname\relax
  \def\bibnamefont#1{#1}\fi
\expandafter\ifx\csname bibfnamefont\endcsname\relax
  \def\bibfnamefont#1{#1}\fi
\expandafter\ifx\csname citenamefont\endcsname\relax
  \def\citenamefont#1{#1}\fi
\expandafter\ifx\csname url\endcsname\relax
  \def\url#1{\texttt{#1}}\fi
\expandafter\ifx\csname urlprefix\endcsname\relax\def\urlprefix{URL }\fi
\providecommand{\bibinfo}[2]{#2}
\providecommand{\eprint}[2][]{\url{#2}}

\bibitem[{\citenamefont{Strominger and Vafa}(1996)}]{Strominger:1996sh}
\bibinfo{author}{\bibfnamefont{A.}~\bibnamefont{Strominger}} \bibnamefont{and}
  \bibinfo{author}{\bibfnamefont{C.}~\bibnamefont{Vafa}},
  \bibinfo{journal}{Phys. Lett.} \textbf{\bibinfo{volume}{B379}},
  \bibinfo{pages}{99} (\bibinfo{year}{1996}), \eprint{hep-th/9601029}.

\bibitem[{\citenamefont{Bardeen and Horowitz}(1999)}]{Bardeen:1999px}
\bibinfo{author}{\bibfnamefont{J.~M.} \bibnamefont{Bardeen}} \bibnamefont{and}
  \bibinfo{author}{\bibfnamefont{G.~T.} \bibnamefont{Horowitz}},
  \bibinfo{journal}{Phys. Rev.} \textbf{\bibinfo{volume}{D60}},
  \bibinfo{pages}{104030} (\bibinfo{year}{1999}), \eprint{hep-th/9905099}.

\bibitem[{\citenamefont{Guica et~al.}(2009)\citenamefont{Guica, Hartman, Song,
  and Strominger}}]{PhysRevD.80.124008}
\bibinfo{author}{\bibfnamefont{M.}~\bibnamefont{Guica}},
  \bibinfo{author}{\bibfnamefont{T.}~\bibnamefont{Hartman}},
  \bibinfo{author}{\bibfnamefont{W.}~\bibnamefont{Song}}, \bibnamefont{and}
  \bibinfo{author}{\bibfnamefont{A.}~\bibnamefont{Strominger}},
  \bibinfo{journal}{Phys. Rev. D} \textbf{\bibinfo{volume}{80}},
  \bibinfo{pages}{124008} (\bibinfo{year}{2009}),
  \urlprefix\url{https://link.aps.org/doi/10.1103/PhysRevD.80.124008}.

\bibitem[{\citenamefont{Penrose}(1969)}]{penrose1969gravitational}
\bibinfo{author}{\bibfnamefont{R.}~\bibnamefont{Penrose}}, \bibinfo{type}{Tech.
  Rep.}, \bibinfo{institution}{Birkbeck Coll., London} (\bibinfo{year}{1969}).

\bibitem[{\citenamefont{Gou et~al.}(2014)\citenamefont{Gou, McClintock,
  Remillard, Steiner, Reid, Orosz, Narayan, Hanke, and Garc'a}}]{Gou:2013dna}
\bibinfo{author}{\bibfnamefont{L.}~\bibnamefont{Gou}},
  \bibinfo{author}{\bibfnamefont{J.~E.} \bibnamefont{McClintock}},
  \bibinfo{author}{\bibfnamefont{R.~A.} \bibnamefont{Remillard}},
  \bibinfo{author}{\bibfnamefont{J.~F.} \bibnamefont{Steiner}},
  \bibinfo{author}{\bibfnamefont{M.~J.} \bibnamefont{Reid}},
  \bibinfo{author}{\bibfnamefont{J.~A.} \bibnamefont{Orosz}},
  \bibinfo{author}{\bibfnamefont{R.}~\bibnamefont{Narayan}},
  \bibinfo{author}{\bibfnamefont{M.}~\bibnamefont{Hanke}}, \bibnamefont{and}
  \bibinfo{author}{\bibfnamefont{J.}~\bibnamefont{Garc'a}},
  \bibinfo{journal}{Astrophys. J.} \textbf{\bibinfo{volume}{790}},
  \bibinfo{pages}{29} (\bibinfo{year}{2014}), \eprint{1308.4760}.

\bibitem[{\citenamefont{McClintock et~al.}(2006)\citenamefont{McClintock,
  Shafee, Narayan, Remillard, Davis, and Li}}]{McClintock:2006xd}
\bibinfo{author}{\bibfnamefont{J.~E.} \bibnamefont{McClintock}},
  \bibinfo{author}{\bibfnamefont{R.}~\bibnamefont{Shafee}},
  \bibinfo{author}{\bibfnamefont{R.}~\bibnamefont{Narayan}},
  \bibinfo{author}{\bibfnamefont{R.~A.} \bibnamefont{Remillard}},
  \bibinfo{author}{\bibfnamefont{S.~W.} \bibnamefont{Davis}}, \bibnamefont{and}
  \bibinfo{author}{\bibfnamefont{L.-X.} \bibnamefont{Li}},
  \bibinfo{journal}{Astrophys. J.} \textbf{\bibinfo{volume}{652}},
  \bibinfo{pages}{518} (\bibinfo{year}{2006}), \eprint{astro-ph/0606076}.

\bibitem[{\citenamefont{Reynolds}(2013)}]{0264-9381-30-24-244004}
\bibinfo{author}{\bibfnamefont{C.~S.} \bibnamefont{Reynolds}},
  \bibinfo{journal}{Classical and Quantum Gravity}
  \textbf{\bibinfo{volume}{30}}, \bibinfo{pages}{244004}
  (\bibinfo{year}{2013}),
  \urlprefix\url{http://stacks.iop.org/0264-9381/30/i=24/a=244004}.

\bibitem[{\citenamefont{Gralla et~al.}(2016{\natexlab{a}})\citenamefont{Gralla,
  Hughes, and Warburton}}]{Gralla:2016qfw}
\bibinfo{author}{\bibfnamefont{S.~E.} \bibnamefont{Gralla}},
  \bibinfo{author}{\bibfnamefont{S.~A.} \bibnamefont{Hughes}},
  \bibnamefont{and}
  \bibinfo{author}{\bibfnamefont{N.}~\bibnamefont{Warburton}},
  \bibinfo{journal}{Class. Quant. Grav.} \textbf{\bibinfo{volume}{33}},
  \bibinfo{pages}{155002} (\bibinfo{year}{2016}{\natexlab{a}}),
  \eprint{1603.01221}.

\bibitem[{\citenamefont{Hadar et~al.}(2014)\citenamefont{Hadar, Porfyriadis,
  and Strominger}}]{Hadar:2014dpa}
\bibinfo{author}{\bibfnamefont{S.}~\bibnamefont{Hadar}},
  \bibinfo{author}{\bibfnamefont{A.~P.} \bibnamefont{Porfyriadis}},
  \bibnamefont{and}
  \bibinfo{author}{\bibfnamefont{A.}~\bibnamefont{Strominger}},
  \bibinfo{journal}{Phys. Rev.} \textbf{\bibinfo{volume}{D90}},
  \bibinfo{pages}{064045} (\bibinfo{year}{2014}), \eprint{1403.2797}.

\bibitem[{\citenamefont{Compre et~al.}(2018)\citenamefont{Compre, Fransen,
  Hertog, and Long}}]{Compere:2017hsi}
\bibinfo{author}{\bibfnamefont{G.}~\bibnamefont{Compre}},
  \bibinfo{author}{\bibfnamefont{K.}~\bibnamefont{Fransen}},
  \bibinfo{author}{\bibfnamefont{T.}~\bibnamefont{Hertog}}, \bibnamefont{and}
  \bibinfo{author}{\bibfnamefont{J.}~\bibnamefont{Long}},
  \bibinfo{journal}{Class. Quant. Grav.} \textbf{\bibinfo{volume}{35}},
  \bibinfo{pages}{104002} (\bibinfo{year}{2018}), \eprint{1712.07130}.

\bibitem[{\citenamefont{Vishveshwara}(1970)}]{Vishveshwara:1970zz}
\bibinfo{author}{\bibfnamefont{C.~V.} \bibnamefont{Vishveshwara}},
  \bibinfo{journal}{Nature} \textbf{\bibinfo{volume}{227}},
  \bibinfo{pages}{936} (\bibinfo{year}{1970}).

\bibitem[{\citenamefont{Berti et~al.}(2009)\citenamefont{Berti, Cardoso, and
  Starinets}}]{Berti:2009kk}
\bibinfo{author}{\bibfnamefont{E.}~\bibnamefont{Berti}},
  \bibinfo{author}{\bibfnamefont{V.}~\bibnamefont{Cardoso}}, \bibnamefont{and}
  \bibinfo{author}{\bibfnamefont{A.~O.} \bibnamefont{Starinets}},
  \bibinfo{journal}{Class. Quant. Grav.} \textbf{\bibinfo{volume}{26}},
  \bibinfo{pages}{163001} (\bibinfo{year}{2009}), \eprint{0905.2975}.

\bibitem[{\citenamefont{Abbott et~al.}(2016)\citenamefont{Abbott, Abbott,
  Abbott, Abernathy, Acernese, Ackley, Adams, Adams, Addesso, Adhikari
  et~al.}}]{PhysRevLett.116.061102}
\bibinfo{author}{\bibfnamefont{B.~P.} \bibnamefont{Abbott}},
  \bibinfo{author}{\bibfnamefont{R.}~\bibnamefont{Abbott}},
  \bibinfo{author}{\bibfnamefont{T.~D.} \bibnamefont{Abbott}},
  \bibinfo{author}{\bibfnamefont{M.~R.} \bibnamefont{Abernathy}},
  \bibinfo{author}{\bibfnamefont{F.}~\bibnamefont{Acernese}},
  \bibinfo{author}{\bibfnamefont{K.}~\bibnamefont{Ackley}},
  \bibinfo{author}{\bibfnamefont{C.}~\bibnamefont{Adams}},
  \bibinfo{author}{\bibfnamefont{T.}~\bibnamefont{Adams}},
  \bibinfo{author}{\bibfnamefont{P.}~\bibnamefont{Addesso}},
  \bibinfo{author}{\bibfnamefont{R.~X.} \bibnamefont{Adhikari}},
  \bibnamefont{et~al.} (\bibinfo{collaboration}{LIGO Scientific Collaboration
  and Virgo Collaboration}), \bibinfo{journal}{Phys. Rev. Lett.}
  \textbf{\bibinfo{volume}{116}}, \bibinfo{pages}{061102}
  (\bibinfo{year}{2016}),
  \urlprefix\url{http://link.aps.org/doi/10.1103/PhysRevLett.116.061102}.

\bibitem[{\citenamefont{Detweiler}(1980{\natexlab{a}})}]{detweiler1980black}
\bibinfo{author}{\bibfnamefont{S.}~\bibnamefont{Detweiler}},
  \bibinfo{journal}{The Astrophysical Journal} \textbf{\bibinfo{volume}{239}},
  \bibinfo{pages}{292} (\bibinfo{year}{1980}{\natexlab{a}}).

\bibitem[{\citenamefont{Glampedakis and Andersson}(2001)}]{PhysRevD.64.104021}
\bibinfo{author}{\bibfnamefont{K.}~\bibnamefont{Glampedakis}} \bibnamefont{and}
  \bibinfo{author}{\bibfnamefont{N.}~\bibnamefont{Andersson}},
  \bibinfo{journal}{Phys. Rev. D} \textbf{\bibinfo{volume}{64}},
  \bibinfo{pages}{104021} (\bibinfo{year}{2001}),
  \urlprefix\url{http://link.aps.org/doi/10.1103/PhysRevD.64.104021}.

\bibitem[{\citenamefont{Casals and Zimmerman}(2018)}]{casals2018perturbations}
\bibinfo{author}{\bibfnamefont{M.}~\bibnamefont{Casals}} \bibnamefont{and}
  \bibinfo{author}{\bibfnamefont{P.}~\bibnamefont{Zimmerman}},
  \bibinfo{journal}{arXiv preprint arXiv:1801.05830}  (\bibinfo{year}{2018}).

\bibitem[{\citenamefont{Starobinskii}(1973)}]{starobinskii1973amplification}
\bibinfo{author}{\bibfnamefont{A.}~\bibnamefont{Starobinskii}},
  \bibinfo{journal}{Zh. Eksp. Teor. Fiz} \textbf{\bibinfo{volume}{64}},
  \bibinfo{pages}{48} (\bibinfo{year}{1973}).

\bibitem[{\citenamefont{Zel'Dovich}(1971)}]{zel1971generation}
\bibinfo{author}{\bibfnamefont{Y.~B.} \bibnamefont{Zel'Dovich}},
  \bibinfo{journal}{ZhETF Pisma Redaktsiiu} \textbf{\bibinfo{volume}{14}},
  \bibinfo{pages}{270} (\bibinfo{year}{1971}).

\bibitem[{\citenamefont{Leaver}(1986{\natexlab{a}})}]{Leaver:1986}
\bibinfo{author}{\bibfnamefont{E.~W.} \bibnamefont{Leaver}},
  \bibinfo{journal}{Phys. Rev. D} \textbf{\bibinfo{volume}{34}},
  \bibinfo{pages}{384} (\bibinfo{year}{1986}{\natexlab{a}}).

\bibitem[{\citenamefont{Casals and Ottewill}(2015)}]{Casals:Ottewill:2015}
\bibinfo{author}{\bibfnamefont{M.}~\bibnamefont{Casals}} \bibnamefont{and}
  \bibinfo{author}{\bibfnamefont{A.}~\bibnamefont{Ottewill}},
  \bibinfo{journal}{Phys. Rev. D} \textbf{\bibinfo{volume}{92}},
  \bibinfo{pages}{124055} (\bibinfo{year}{2015}),
  \urlprefix\url{http://link.aps.org/doi/10.1103/PhysRevD.92.124055}.

\bibitem[{\citenamefont{Casals et~al.}(2013)\citenamefont{Casals, Dolan,
  Ottewill, and Wardell}}]{CDOW13}
\bibinfo{author}{\bibfnamefont{M.}~\bibnamefont{Casals}},
  \bibinfo{author}{\bibfnamefont{S.}~\bibnamefont{Dolan}},
  \bibinfo{author}{\bibfnamefont{A.~C.} \bibnamefont{Ottewill}},
  \bibnamefont{and} \bibinfo{author}{\bibfnamefont{B.}~\bibnamefont{Wardell}},
  \bibinfo{journal}{Phys. Rev. D} \textbf{\bibinfo{volume}{88}},
  \bibinfo{pages}{044022} (\bibinfo{year}{2013}),
  \urlprefix\url{http://link.aps.org/doi/10.1103/PhysRevD.88.044022}.

\bibitem[{\citenamefont{Casals et~al.}(2016{\natexlab{a}})\citenamefont{Casals,
  Kavanagh, and Ottewill}}]{PhysRevD.94.124053}
\bibinfo{author}{\bibfnamefont{M.}~\bibnamefont{Casals}},
  \bibinfo{author}{\bibfnamefont{C.}~\bibnamefont{Kavanagh}}, \bibnamefont{and}
  \bibinfo{author}{\bibfnamefont{A.~C.} \bibnamefont{Ottewill}},
  \bibinfo{journal}{Phys. Rev. D} \textbf{\bibinfo{volume}{94}},
  \bibinfo{pages}{124053} (\bibinfo{year}{2016}{\natexlab{a}}),
  \urlprefix\url{http://link.aps.org/doi/10.1103/PhysRevD.94.124053}.

\bibitem[{\citenamefont{Casals and Ottewill}(2012)}]{PhysRevLett.109.111101}
\bibinfo{author}{\bibfnamefont{M.}~\bibnamefont{Casals}} \bibnamefont{and}
  \bibinfo{author}{\bibfnamefont{A.}~\bibnamefont{Ottewill}},
  \bibinfo{journal}{Phys. Rev. Lett.} \textbf{\bibinfo{volume}{109}},
  \bibinfo{pages}{111101} (\bibinfo{year}{2012}),
  \urlprefix\url{http://link.aps.org/doi/10.1103/PhysRevLett.109.111101}.

\bibitem[{\citenamefont{Casals and Ottewill}(2013)}]{Casals:2012ng}
\bibinfo{author}{\bibfnamefont{M.}~\bibnamefont{Casals}} \bibnamefont{and}
  \bibinfo{author}{\bibfnamefont{A.~C.} \bibnamefont{Ottewill}},
  \bibinfo{journal}{Phys.Rev.} \textbf{\bibinfo{volume}{D87}},
  \bibinfo{pages}{064010} (\bibinfo{year}{2013}), \eprint{1210.0519}.

\bibitem[{\citenamefont{Price}(1972)}]{Price:1971fb}
\bibinfo{author}{\bibfnamefont{R.~H.} \bibnamefont{Price}},
  \bibinfo{journal}{Phys. Rev.} \textbf{\bibinfo{volume}{D5}},
  \bibinfo{pages}{2419} (\bibinfo{year}{1972}).

\bibitem[{\citenamefont{Damour et~al.}(1976)\citenamefont{Damour, Deruelle, and
  Ruffini}}]{damour1976quantum}
\bibinfo{author}{\bibfnamefont{T.}~\bibnamefont{Damour}},
  \bibinfo{author}{\bibfnamefont{N.}~\bibnamefont{Deruelle}}, \bibnamefont{and}
  \bibinfo{author}{\bibfnamefont{R.}~\bibnamefont{Ruffini}},
  \bibinfo{journal}{Lettere Al Nuovo Cimento Series 2}
  \textbf{\bibinfo{volume}{15}}, \bibinfo{pages}{257} (\bibinfo{year}{1976}).

\bibitem[{\citenamefont{Zouros and Eardley}(1979)}]{zouros1979instabilities}
\bibinfo{author}{\bibfnamefont{T.~J.} \bibnamefont{Zouros}} \bibnamefont{and}
  \bibinfo{author}{\bibfnamefont{D.~M.} \bibnamefont{Eardley}},
  \bibinfo{journal}{Annals of physics} \textbf{\bibinfo{volume}{118}},
  \bibinfo{pages}{139} (\bibinfo{year}{1979}).

\bibitem[{\citenamefont{Detweiler}(1980{\natexlab{b}})}]{PhysRevD.22.2323}
\bibinfo{author}{\bibfnamefont{S.}~\bibnamefont{Detweiler}},
  \bibinfo{journal}{Phys. Rev. D} \textbf{\bibinfo{volume}{22}},
  \bibinfo{pages}{2323} (\bibinfo{year}{1980}{\natexlab{b}}),
  \urlprefix\url{http://link.aps.org/doi/10.1103/PhysRevD.22.2323}.

\bibitem[{\citenamefont{Press and Teukolsky}(1972)}]{Press:1972zz}
\bibinfo{author}{\bibfnamefont{W.~H.} \bibnamefont{Press}} \bibnamefont{and}
  \bibinfo{author}{\bibfnamefont{S.~A.} \bibnamefont{Teukolsky}},
  \bibinfo{journal}{Nature} \textbf{\bibinfo{volume}{238}},
  \bibinfo{pages}{211} (\bibinfo{year}{1972}).

\bibitem[{\citenamefont{Cardoso and Dias}(2004)}]{PhysRevD.70.084011}
\bibinfo{author}{\bibfnamefont{V.}~\bibnamefont{Cardoso}} \bibnamefont{and}
  \bibinfo{author}{\bibfnamefont{O.~J.~C.} \bibnamefont{Dias}},
  \bibinfo{journal}{Phys. Rev. D} \textbf{\bibinfo{volume}{70}},
  \bibinfo{pages}{084011} (\bibinfo{year}{2004}),
  \urlprefix\url{http://link.aps.org/doi/10.1103/PhysRevD.70.084011}.

\bibitem[{\citenamefont{Whiting}(1989)}]{whiting1989mode}
\bibinfo{author}{\bibfnamefont{B.}~\bibnamefont{Whiting}},
  \bibinfo{journal}{Journal of Mathematical Physics}
  \textbf{\bibinfo{volume}{30}}, \bibinfo{pages}{1301} (\bibinfo{year}{1989}).

\bibitem[{\citenamefont{Dafermos et~al.}(2016)\citenamefont{Dafermos,
  Rodnianski, and Shlapentokh-Rothman}}]{dafermos2016decay}
\bibinfo{author}{\bibfnamefont{M.}~\bibnamefont{Dafermos}},
  \bibinfo{author}{\bibfnamefont{I.}~\bibnamefont{Rodnianski}},
  \bibnamefont{and}
  \bibinfo{author}{\bibfnamefont{Y.}~\bibnamefont{Shlapentokh-Rothman}},
  \bibinfo{journal}{annals of Mathematics} pp. \bibinfo{pages}{787--913}
  (\bibinfo{year}{2016}).

\bibitem[{\citenamefont{Aretakis}(2015)}]{Aretakis:2012ei}
\bibinfo{author}{\bibfnamefont{S.}~\bibnamefont{Aretakis}},
  \bibinfo{journal}{Adv. Theor. Math. Phys.} \textbf{\bibinfo{volume}{19}},
  \bibinfo{pages}{507} (\bibinfo{year}{2015}), \eprint{1206.6598}.

\bibitem[{\citenamefont{Aretakis}(2011)}]{Aretakis:2011ha}
\bibinfo{author}{\bibfnamefont{S.}~\bibnamefont{Aretakis}},
  \bibinfo{journal}{Commun. Math. Phys.} \textbf{\bibinfo{volume}{307}},
  \bibinfo{pages}{17} (\bibinfo{year}{2011}), \eprint{1110.2007}.

\bibitem[{\citenamefont{Lucietti and Reall}(2012)}]{lucietti2012gravitational}
\bibinfo{author}{\bibfnamefont{J.}~\bibnamefont{Lucietti}} \bibnamefont{and}
  \bibinfo{author}{\bibfnamefont{H.~S.} \bibnamefont{Reall}},
  \bibinfo{journal}{Physical Review D} \textbf{\bibinfo{volume}{86}},
  \bibinfo{pages}{104030} (\bibinfo{year}{2012}).

\bibitem[{\citenamefont{Casals et~al.}(2016{\natexlab{b}})\citenamefont{Casals,
  Gralla, and Zimmerman}}]{casals2016horizon}
\bibinfo{author}{\bibfnamefont{M.}~\bibnamefont{Casals}},
  \bibinfo{author}{\bibfnamefont{S.~E.} \bibnamefont{Gralla}},
  \bibnamefont{and}
  \bibinfo{author}{\bibfnamefont{P.}~\bibnamefont{Zimmerman}},
  \bibinfo{journal}{Physical Review D} \textbf{\bibinfo{volume}{94}},
  \bibinfo{pages}{064003} (\bibinfo{year}{2016}{\natexlab{b}}).

\bibitem[{\citenamefont{Gralla and Zimmerman}(2018)}]{Gralla:2017lto}
\bibinfo{author}{\bibfnamefont{S.~E.} \bibnamefont{Gralla}} \bibnamefont{and}
  \bibinfo{author}{\bibfnamefont{P.}~\bibnamefont{Zimmerman}},
  \bibinfo{journal}{Class. Quant. Grav.} \textbf{\bibinfo{volume}{35}},
  \bibinfo{pages}{095002} (\bibinfo{year}{2018}), \eprint{1711.00855}.

\bibitem[{\citenamefont{Aretakis}(2012)}]{aretakis2012decay}
\bibinfo{author}{\bibfnamefont{S.}~\bibnamefont{Aretakis}},
  \bibinfo{journal}{Journal of Functional Analysis}
  \textbf{\bibinfo{volume}{263}}, \bibinfo{pages}{2770} (\bibinfo{year}{2012}).

\bibitem[{\citenamefont{Dain and de~Austria}(2015)}]{dain2015bounds}
\bibinfo{author}{\bibfnamefont{S.}~\bibnamefont{Dain}} \bibnamefont{and}
  \bibinfo{author}{\bibfnamefont{I.~G.} \bibnamefont{de~Austria}},
  \bibinfo{journal}{Classical and Quantum Gravity}
  \textbf{\bibinfo{volume}{32}}, \bibinfo{pages}{135010}
  (\bibinfo{year}{2015}).

\bibitem[{\citenamefont{Burko and Khanna}(2017)}]{Burko:2017eky}
\bibinfo{author}{\bibfnamefont{L.~M.} \bibnamefont{Burko}} \bibnamefont{and}
  \bibinfo{author}{\bibfnamefont{G.}~\bibnamefont{Khanna}}
  (\bibinfo{year}{2017}), \eprint{1709.10155}.

\bibitem[{\citenamefont{Detweiler and Ipser}(1973)}]{detweiler1973stability}
\bibinfo{author}{\bibfnamefont{S.~L.} \bibnamefont{Detweiler}}
  \bibnamefont{and} \bibinfo{author}{\bibfnamefont{J.~R.} \bibnamefont{Ipser}},
  \bibinfo{journal}{The Astrophysical Journal} \textbf{\bibinfo{volume}{185}},
  \bibinfo{pages}{675} (\bibinfo{year}{1973}).

\bibitem[{\citenamefont{Richartz}(2016)}]{Richartz:2015saa}
\bibinfo{author}{\bibfnamefont{M.}~\bibnamefont{Richartz}},
  \bibinfo{journal}{Phys. Rev.} \textbf{\bibinfo{volume}{D93}},
  \bibinfo{pages}{064062} (\bibinfo{year}{2016}), \eprint{1509.04260}.
  
  \bibitem[{\citenamefont{Cook}(2014)}]{Cook:2014cta}
\bibinfo{author}{\bibfnamefont{G.~B.}~\bibnamefont{Cook}},
\bibinfo{author}{\bibfnamefont{M.}~\bibnamefont{Zalutskiy}},
  \bibinfo{journal}{Phys. Rev.} \textbf{\bibinfo{volume}{D90}},
  \bibinfo{pages}{124021} (\bibinfo{year}{2014}), \eprint{1410.7698}.


\bibitem[{QNM()}]{QNMBertiNew}
\bibinfo{howpublished}{\url{https://pages.jh.edu/~eberti2/ringdown/},
  \url{https://centra.tecnico.ulisboa.pt/network/grit/files/ringdown/}}.

\bibitem[{\citenamefont{Yang et~al.}(2013{\natexlab{a}})\citenamefont{Yang,
  Zhang, Zimmerman, Nichols, Berti, and Chen}}]{Yang:2012pj}
\bibinfo{author}{\bibfnamefont{H.}~\bibnamefont{Yang}},
  \bibinfo{author}{\bibfnamefont{F.}~\bibnamefont{Zhang}},
  \bibinfo{author}{\bibfnamefont{A.}~\bibnamefont{Zimmerman}},
  \bibinfo{author}{\bibfnamefont{D.~A.} \bibnamefont{Nichols}},
  \bibinfo{author}{\bibfnamefont{E.}~\bibnamefont{Berti}}, \bibnamefont{and}
  \bibinfo{author}{\bibfnamefont{Y.}~\bibnamefont{Chen}},
  \bibinfo{journal}{Phys. Rev.} \textbf{\bibinfo{volume}{D87}},
  \bibinfo{pages}{041502} (\bibinfo{year}{2013}{\natexlab{a}}),
  \eprint{1212.3271}.

\bibitem[{\citenamefont{Hod}(2008)}]{hod2008slow}
\bibinfo{author}{\bibfnamefont{S.}~\bibnamefont{Hod}},
  \bibinfo{journal}{Physical Review D} \textbf{\bibinfo{volume}{78}},
  \bibinfo{pages}{084035} (\bibinfo{year}{2008}).

\bibitem[{\citenamefont{Yang et~al.}(2013{\natexlab{b}})\citenamefont{Yang,
  Zimmerman, Zengino{\u{g}}lu, Zhang, Berti, and Chen}}]{yang2013quasinormal}
\bibinfo{author}{\bibfnamefont{H.}~\bibnamefont{Yang}},
  \bibinfo{author}{\bibfnamefont{A.}~\bibnamefont{Zimmerman}},
  \bibinfo{author}{\bibfnamefont{A.}~\bibnamefont{Zengino{\u{g}}lu}},
  \bibinfo{author}{\bibfnamefont{F.}~\bibnamefont{Zhang}},
  \bibinfo{author}{\bibfnamefont{E.}~\bibnamefont{Berti}}, \bibnamefont{and}
  \bibinfo{author}{\bibfnamefont{Y.}~\bibnamefont{Chen}},
  \bibinfo{journal}{Physical Review D} \textbf{\bibinfo{volume}{88}},
  \bibinfo{pages}{044047} (\bibinfo{year}{2013}{\natexlab{b}}).

\bibitem[{\citenamefont{Zimmerman et~al.}(2015)\citenamefont{Zimmerman, Yang,
  Zhang, Nichols, Berti, and Chen}}]{Zimmerman:2015rua}
\bibinfo{author}{\bibfnamefont{A.}~\bibnamefont{Zimmerman}},
  \bibinfo{author}{\bibfnamefont{H.}~\bibnamefont{Yang}},
  \bibinfo{author}{\bibfnamefont{F.}~\bibnamefont{Zhang}},
  \bibinfo{author}{\bibfnamefont{D.~A.} \bibnamefont{Nichols}},
  \bibinfo{author}{\bibfnamefont{E.}~\bibnamefont{Berti}}, \bibnamefont{and}
  \bibinfo{author}{\bibfnamefont{Y.}~\bibnamefont{Chen}}
  (\bibinfo{year}{2015}), \eprint{1510.08159}.

\bibitem[{\citenamefont{Starobinskii and
  Churilov}(1974)}]{Starobinskil:1974nkd}
\bibinfo{author}{\bibfnamefont{A.~A.} \bibnamefont{Starobinskii}}
  \bibnamefont{and} \bibinfo{author}{\bibfnamefont{S.~M.}
  \bibnamefont{Churilov}}, \bibinfo{journal}{Sov. Phys. JETP}
  \textbf{\bibinfo{volume}{65}}, \bibinfo{pages}{1} (\bibinfo{year}{1974}).

\bibitem[{\citenamefont{Mano et~al.}(1996{\natexlab{a}})\citenamefont{Mano,
  Suzuki, and Takasugi}}]{Mano:Suzuki:Takasugi:1996}
\bibinfo{author}{\bibfnamefont{S.}~\bibnamefont{Mano}},
  \bibinfo{author}{\bibfnamefont{H.}~\bibnamefont{Suzuki}}, \bibnamefont{and}
  \bibinfo{author}{\bibfnamefont{E.}~\bibnamefont{Takasugi}},
  \bibinfo{journal}{Prog. Theor. Phys.} \textbf{\bibinfo{volume}{95}},
  \bibinfo{pages}{1079} (\bibinfo{year}{1996}{\natexlab{a}}).

\bibitem[{\citenamefont{Mano et~al.}(1996{\natexlab{b}})\citenamefont{Mano,
  Suzuki, and Takasugi}}]{Mano:1996mf}
\bibinfo{author}{\bibfnamefont{S.}~\bibnamefont{Mano}},
  \bibinfo{author}{\bibfnamefont{H.}~\bibnamefont{Suzuki}}, \bibnamefont{and}
  \bibinfo{author}{\bibfnamefont{E.}~\bibnamefont{Takasugi}},
  \bibinfo{journal}{Prog. Theor. Phys.} \textbf{\bibinfo{volume}{96}},
  \bibinfo{pages}{549} (\bibinfo{year}{1996}{\natexlab{b}}),
  \eprint{gr-qc/9605057}.

\bibitem[{\citenamefont{Sasaki and Tagoshi}(2003)}]{Sasaki:2003xr}
\bibinfo{author}{\bibfnamefont{M.}~\bibnamefont{Sasaki}} \bibnamefont{and}
  \bibinfo{author}{\bibfnamefont{H.}~\bibnamefont{Tagoshi}},
  \bibinfo{journal}{Living Rev. Rel.} \textbf{\bibinfo{volume}{6}},
  \bibinfo{pages}{6} (\bibinfo{year}{2003}), \eprint{gr-qc/0306120}.

\bibitem[{\citenamefont{Zhang et~al.}(2013)\citenamefont{Zhang, Berti, and
  Cardoso}}]{zhang2013quasinormal}
\bibinfo{author}{\bibfnamefont{Z.}~\bibnamefont{Zhang}},
  \bibinfo{author}{\bibfnamefont{E.}~\bibnamefont{Berti}}, \bibnamefont{and}
  \bibinfo{author}{\bibfnamefont{V.}~\bibnamefont{Cardoso}},
  \bibinfo{journal}{Physical Review D} \textbf{\bibinfo{volume}{88}},
  \bibinfo{pages}{044018} (\bibinfo{year}{2013}).

\bibitem[{\citenamefont{Leaver}(1985)}]{Leaver:1985}
\bibinfo{author}{\bibfnamefont{E.~W.} \bibnamefont{Leaver}},
  \bibinfo{journal}{Proc. Roy. Soc. Lond. A} \textbf{\bibinfo{volume}{402}},
  \bibinfo{pages}{285} (\bibinfo{year}{1985}).

\bibitem[{\citenamefont{{Wald}}(1973)}]{1973JMP....14.1453W}
\bibinfo{author}{\bibfnamefont{R.~M.} \bibnamefont{{Wald}}},
  \bibinfo{journal}{Journal of Mathematical Physics}
  \textbf{\bibinfo{volume}{14}}, \bibinfo{pages}{1453} (\bibinfo{year}{1973}).

\bibitem[{\citenamefont{Chandrasekhar}(1984)}]{chandrasekhar1984algebraically}
\bibinfo{author}{\bibfnamefont{S.}~\bibnamefont{Chandrasekhar}},
  \bibinfo{journal}{Proceedings of the Royal Society of London. Series A,
  Mathematical and Physical Sciences} pp. \bibinfo{pages}{1--13}
  (\bibinfo{year}{1984}).

\bibitem[{\citenamefont{Teukolsky}(1973)}]{Teukolsky:1973ha}
\bibinfo{author}{\bibfnamefont{S.~A.} \bibnamefont{Teukolsky}},
  \bibinfo{journal}{Astrophys. J.} \textbf{\bibinfo{volume}{185}},
  \bibinfo{pages}{635} (\bibinfo{year}{1973}).

\bibitem[{\citenamefont{Berti et~al.}(2006)\citenamefont{Berti, Cardoso, and
  Casals}}]{Berti:2005gp}
\bibinfo{author}{\bibfnamefont{E.}~\bibnamefont{Berti}},
  \bibinfo{author}{\bibfnamefont{V.}~\bibnamefont{Cardoso}}, \bibnamefont{and}
  \bibinfo{author}{\bibfnamefont{M.}~\bibnamefont{Casals}},
  \bibinfo{journal}{Phys. Rev.} \textbf{\bibinfo{volume}{D73}},
  \bibinfo{pages}{024013} (\bibinfo{year}{2006}), \eprint{gr-qc/0511111}.

\bibitem[{\citenamefont{Yang et~al.}(2014)\citenamefont{Yang, Zhang, Zimmerman,
  and Chen}}]{Yang:2013shb}
\bibinfo{author}{\bibfnamefont{H.}~\bibnamefont{Yang}},
  \bibinfo{author}{\bibfnamefont{F.}~\bibnamefont{Zhang}},
  \bibinfo{author}{\bibfnamefont{A.}~\bibnamefont{Zimmerman}},
  \bibnamefont{and} \bibinfo{author}{\bibfnamefont{Y.}~\bibnamefont{Chen}},
  \bibinfo{journal}{Phys.Rev.} \textbf{\bibinfo{volume}{D89}},
  \bibinfo{pages}{064014} (\bibinfo{year}{2014}), \eprint{1311.3380}.

\bibitem[{\citenamefont{Leaver}(1986{\natexlab{b}})}]{Leaver:1986a}
\bibinfo{author}{\bibfnamefont{E.~W.} \bibnamefont{Leaver}},
  \bibinfo{journal}{J.\ Math.\ Phys.} \textbf{\bibinfo{volume}{27}},
  \bibinfo{pages}{1238} (\bibinfo{year}{1986}{\natexlab{b}}).

\bibitem[{\citenamefont{Hod}(2000)}]{PhysRevLett.84.10}
\bibinfo{author}{\bibfnamefont{S.}~\bibnamefont{Hod}}, \bibinfo{journal}{Phys.
  Rev. Lett.} \textbf{\bibinfo{volume}{84}}, \bibinfo{pages}{10}
  (\bibinfo{year}{2000}),
  \urlprefix\url{http://link.aps.org/doi/10.1103/PhysRevLett.84.10}.

\bibitem[{\citenamefont{Oguchi}(1970)}]{oguchi1970eigenvalues}
\bibinfo{author}{\bibfnamefont{T.}~\bibnamefont{Oguchi}},
  \bibinfo{journal}{Radio Science} \textbf{\bibinfo{volume}{5}},
  \bibinfo{pages}{1207} (\bibinfo{year}{1970}).

\bibitem[{\citenamefont{Barrowes et~al.}(2004)\citenamefont{Barrowes, O'Neill,
  M., and Kong}}]{BONGK:2004}
\bibinfo{author}{\bibfnamefont{B.~E.} \bibnamefont{Barrowes}},
  \bibinfo{author}{\bibfnamefont{K.}~\bibnamefont{O'Neill}},
  \bibinfo{author}{\bibfnamefont{G.~T.} \bibnamefont{M.}}, \bibnamefont{and}
  \bibinfo{author}{\bibfnamefont{J.~A.} \bibnamefont{Kong}},
  \bibinfo{journal}{Studies in Applied Mathematics}
  \textbf{\bibinfo{volume}{113}}, \bibinfo{pages}{271} (\bibinfo{year}{2004}).

\bibitem[{\citenamefont{Hartle and Wilkins}(1974)}]{Hartle:Wilkins:1974}
\bibinfo{author}{\bibfnamefont{J.~B.} \bibnamefont{Hartle}} \bibnamefont{and}
  \bibinfo{author}{\bibfnamefont{D.~C.} \bibnamefont{Wilkins}},
  \bibinfo{journal}{Commun. Math. Phys.} \textbf{\bibinfo{volume}{38}},
  \bibinfo{pages}{47} (\bibinfo{year}{1974}).

\bibitem[{\citenamefont{Castro et~al.}(2013)\citenamefont{Castro, Lapan,
  Maloney, and Rodriguez}}]{castro2013black}
\bibinfo{author}{\bibfnamefont{A.}~\bibnamefont{Castro}},
  \bibinfo{author}{\bibfnamefont{J.~M.} \bibnamefont{Lapan}},
  \bibinfo{author}{\bibfnamefont{A.}~\bibnamefont{Maloney}}, \bibnamefont{and}
  \bibinfo{author}{\bibfnamefont{M.~J.} \bibnamefont{Rodriguez}},
  \bibinfo{journal}{Classical and Quantum Gravity}
  \textbf{\bibinfo{volume}{30}}, \bibinfo{pages}{165005}
  (\bibinfo{year}{2013}).

\bibitem[{\citenamefont{Gralla et~al.}(2015)\citenamefont{Gralla, Porfyriadis,
  and Warburton}}]{Gralla:2015rpa}
\bibinfo{author}{\bibfnamefont{S.~E.} \bibnamefont{Gralla}},
  \bibinfo{author}{\bibfnamefont{A.~P.} \bibnamefont{Porfyriadis}},
  \bibnamefont{and}
  \bibinfo{author}{\bibfnamefont{N.}~\bibnamefont{Warburton}},
  \bibinfo{journal}{Phys. Rev.} \textbf{\bibinfo{volume}{D92}},
  \bibinfo{pages}{064029} (\bibinfo{year}{2015}), \eprint{1506.08496}.

\bibitem[{\citenamefont{Fujita and Tagoshi}(2005)}]{fujita2005new}
\bibinfo{author}{\bibfnamefont{R.}~\bibnamefont{Fujita}} \bibnamefont{and}
  \bibinfo{author}{\bibfnamefont{H.}~\bibnamefont{Tagoshi}},
  \bibinfo{journal}{Progress of theoretical physics}
  \textbf{\bibinfo{volume}{113}}, \bibinfo{pages}{1165} (\bibinfo{year}{2005}).

\bibitem[{\citenamefont{Rodriguez}()}]{monodromysite}
\bibinfo{author}{\bibfnamefont{M.~J.} \bibnamefont{Rodriguez}},
  \bibinfo{howpublished}{\url{https://sites.google.com/site/justblackholes/techy-zone}}.

\bibitem[{BHP()}]{BHPToolkit}
\emph{\bibinfo{title}{{Black Hole Perturbation Toolkit}}},
  \bibinfo{howpublished}{\url{bhptoolkit.org}}.

\bibitem[{\citenamefont{Teukolsky and
  Press}(1974)}]{teukolsky1974perturbations}
\bibinfo{author}{\bibfnamefont{S.~A.} \bibnamefont{Teukolsky}}
  \bibnamefont{and} \bibinfo{author}{\bibfnamefont{W.}~\bibnamefont{Press}},
  \bibinfo{journal}{The Astrophysical Journal} \textbf{\bibinfo{volume}{193}},
  \bibinfo{pages}{443} (\bibinfo{year}{1974}).

\bibitem[{\citenamefont{Chandrasekhar}(1983)}]{Chandrasekhar}
\bibinfo{author}{\bibfnamefont{S.}~\bibnamefont{Chandrasekhar}},
  \emph{\bibinfo{title}{The Mathematical Theory of Black Holes}}
  (\bibinfo{publisher}{Oxford University Press}, \bibinfo{address}{New York},
  \bibinfo{year}{1983}).

\bibitem[{\citenamefont{Maassen van~den
  Brink}(2000)}]{MaassenvandenBrink:2000ru}
\bibinfo{author}{\bibfnamefont{A.}~\bibnamefont{Maassen van~den Brink}},
  \bibinfo{journal}{Phys. Rev.} \textbf{\bibinfo{volume}{D62}},
  \bibinfo{pages}{064009} (\bibinfo{year}{2000}), \eprint{gr-qc/0001032}.

\bibitem[{\citenamefont{Rudin}(1987)}]{Rudin}
\bibinfo{author}{\bibfnamefont{W.}~\bibnamefont{Rudin}},
  \emph{\bibinfo{title}{Real and Complex Analysis}}
  (\bibinfo{publisher}{McGraw-Hill Book Co., Singapore}, \bibinfo{year}{1987}).

\bibitem[{\citenamefont{Nelder and Mead}(1965)}]{NM}
\bibinfo{author}{\bibfnamefont{J.~A.} \bibnamefont{Nelder}} \bibnamefont{and}
  \bibinfo{author}{\bibfnamefont{R.}~\bibnamefont{Mead}}, \bibinfo{journal}{The
  Computer Journal} \textbf{\bibinfo{volume}{7}}, \bibinfo{pages}{308}
  (\bibinfo{year}{1965}).

\bibitem[{\citenamefont{Brito et~al.}(2015)\citenamefont{Brito, Cardoso, and
  Pani}}]{brito2015superradiance}
\bibinfo{author}{\bibfnamefont{R.}~\bibnamefont{Brito}},
  \bibinfo{author}{\bibfnamefont{V.}~\bibnamefont{Cardoso}}, \bibnamefont{and}
  \bibinfo{author}{\bibfnamefont{P.}~\bibnamefont{Pani}},
  \bibinfo{journal}{arXiv preprint arXiv:1501.06570}  (\bibinfo{year}{2015}).

\bibitem[{\citenamefont{Mashhoon}(1985)}]{PhysRevD.31.290}
\bibinfo{author}{\bibfnamefont{B.}~\bibnamefont{Mashhoon}},
  \bibinfo{journal}{Phys. Rev. D} \textbf{\bibinfo{volume}{31}},
  \bibinfo{pages}{290} (\bibinfo{year}{1985}),
  \urlprefix\url{http://link.aps.org/doi/10.1103/PhysRevD.31.290}.

\bibitem[{\citenamefont{Hod}(2013)}]{hod2013purely}
\bibinfo{author}{\bibfnamefont{S.}~\bibnamefont{Hod}},
  \bibinfo{journal}{Physical Review D} \textbf{\bibinfo{volume}{88}},
  \bibinfo{pages}{084018} (\bibinfo{year}{2013}).

\bibitem[{\citenamefont{Cot\ifmmode~\u{a}\else
  \u{a}\fi{}escu}(1999)}]{PhysRevD.60.107504}
\bibinfo{author}{\bibfnamefont{I.~I.} \bibnamefont{Cot\ifmmode~\u{a}\else
  \u{a}\fi{}escu}}, \bibinfo{journal}{Phys. Rev. D}
  \textbf{\bibinfo{volume}{60}}, \bibinfo{pages}{107504}
  (\bibinfo{year}{1999}),
  \urlprefix\url{http://link.aps.org/doi/10.1103/PhysRevD.60.107504}.

\bibitem[{\citenamefont{Andersson et~al.}(2017)\citenamefont{Andersson, Ma,
  Paganini, and Whiting}}]{Andersson:2016epf}
\bibinfo{author}{\bibfnamefont{L.}~\bibnamefont{Andersson}},
  \bibinfo{author}{\bibfnamefont{S.}~\bibnamefont{Ma}},
  \bibinfo{author}{\bibfnamefont{C.}~\bibnamefont{Paganini}}, \bibnamefont{and}
  \bibinfo{author}{\bibfnamefont{B.~F.} \bibnamefont{Whiting}},
  \bibinfo{journal}{J. Math. Phys.} \textbf{\bibinfo{volume}{58}},
  \bibinfo{pages}{072501} (\bibinfo{year}{2017}), \eprint{1607.02759}.

\bibitem[{\citenamefont{Gralla et~al.}(2016{\natexlab{b}})\citenamefont{Gralla,
  Zimmerman, and Zimmerman}}]{gralla2016transient}
\bibinfo{author}{\bibfnamefont{S.~E.} \bibnamefont{Gralla}},
  \bibinfo{author}{\bibfnamefont{A.}~\bibnamefont{Zimmerman}},
  \bibnamefont{and}
  \bibinfo{author}{\bibfnamefont{P.}~\bibnamefont{Zimmerman}},
  \bibinfo{journal}{Physical Review D} \textbf{\bibinfo{volume}{94}},
  \bibinfo{pages}{084017} (\bibinfo{year}{2016}{\natexlab{b}}).

\bibitem[{\citenamefont{Keshet and Neitzke}(2008)}]{Keshet:2007be}
\bibinfo{author}{\bibfnamefont{U.}~\bibnamefont{Keshet}} \bibnamefont{and}
  \bibinfo{author}{\bibfnamefont{A.}~\bibnamefont{Neitzke}},
  \bibinfo{journal}{Phys. Rev.} \textbf{\bibinfo{volume}{D78}},
  \bibinfo{pages}{044006} (\bibinfo{year}{2008}), \eprint{0709.1532}.

\bibitem[{\citenamefont{Longo}(2018)}]{MSc-Longo}
\bibinfo{author}{\bibfnamefont{L.~F.} \bibnamefont{Longo}}, Master's thesis,
  \bibinfo{school}{Centro Brasileiro de Pesquisas F\'isicas}
  (\bibinfo{year}{2018}).

\bibitem[{\citenamefont{Richartz et~al.}(2017)\citenamefont{Richartz, Herdeiro,
  and Berti}}]{Richartz:2017qep}
\bibinfo{author}{\bibfnamefont{M.}~\bibnamefont{Richartz}},
  \bibinfo{author}{\bibfnamefont{C.~A.~R.} \bibnamefont{Herdeiro}},
  \bibnamefont{and} \bibinfo{author}{\bibfnamefont{E.}~\bibnamefont{Berti}},
  \bibinfo{journal}{Phys. Rev.} \textbf{\bibinfo{volume}{D96}},
  \bibinfo{pages}{044034} (\bibinfo{year}{2017}), \eprint{1706.01112}.

\bibitem[{\citenamefont{Sasaki and Nakamura}(1990)}]{sasaki1990gravitational}
\bibinfo{author}{\bibfnamefont{M.}~\bibnamefont{Sasaki}} \bibnamefont{and}
  \bibinfo{author}{\bibfnamefont{T.}~\bibnamefont{Nakamura}},
  \bibinfo{journal}{General Relativity and Gravitation}
  \textbf{\bibinfo{volume}{22}}, \bibinfo{pages}{1351} (\bibinfo{year}{1990}).

\bibitem[{\citenamefont{Porfyriadis and
  Strominger}(2014)}]{Porfyriadis:2014fja}
\bibinfo{author}{\bibfnamefont{A.~P.} \bibnamefont{Porfyriadis}}
  \bibnamefont{and}
  \bibinfo{author}{\bibfnamefont{A.}~\bibnamefont{Strominger}},
  \bibinfo{journal}{Phys. Rev.} \textbf{\bibinfo{volume}{D90}},
  \bibinfo{pages}{044038} (\bibinfo{year}{2014}), \eprint{1401.3746}.

\bibitem[{\citenamefont{Hod}(2015)}]{Hod:2015swa}
\bibinfo{author}{\bibfnamefont{S.}~\bibnamefont{Hod}}, \bibinfo{journal}{Eur.
  Phys. J.} \textbf{\bibinfo{volume}{C75}}, \bibinfo{pages}{520}
  (\bibinfo{year}{2015}), \eprint{1510.05604}.

\bibitem[{\citenamefont{Hod}(2016)}]{Hod:2016aoe}
\bibinfo{author}{\bibfnamefont{S.}~\bibnamefont{Hod}}, \bibinfo{journal}{JCAP}
  \textbf{\bibinfo{volume}{1608}}, \bibinfo{pages}{066} (\bibinfo{year}{2016}),
  \eprint{1602.05730}.

\bibitem[{\citenamefont{Onozawa}(1997)}]{onozawa1997detailed}
\bibinfo{author}{\bibfnamefont{H.}~\bibnamefont{Onozawa}},
  \bibinfo{journal}{Physical Review D} \textbf{\bibinfo{volume}{55}},
  \bibinfo{pages}{3593} (\bibinfo{year}{1997}).

\end{thebibliography}


\end{document}